# Modulation of the Octahedral Structure and Potential Superconductivity of $La_3Ni_2O_7$ through Strain Engineering

Zihao Huo[1,‡], Zhihui Luo[3,‡], Peng Zhang[2], Aiqin Yang[2], Zhengtao Liu[1], Xiangru Tao[2], Zihan Zhang[1], Shumin Guo[1], Qiwen Jiang[1], Wenxuan Chen[1], Dao-Xin Yao[3,*], Defang Duan[1,*],Tian Cui[4,1,*]

[1]*Key Laboratory of Material Simulation Methods & Software of Ministry of Education, State Key Laboratory of Superhard Materials, College of Physics, Jilin University, Changchun 130012, China*

[2]*MOE Key Laboratory for Non-equilibrium Synthesis and Modulation of Condensed Matter, Shaanxi Province Key Laboratory of Advanced Functional Materials and Mesoscopic Physics, School of Physics, Xi'an Jiaotong University, Xi'an 710049, China*

[3]*Center for Neutron Science and Technology, Guangdong Provincial Key Laboratory of Magnetoelectric Physics and Devices, State Key Laboratory of Optoelectronic Materials and Technologies, School of Physics, Sun Yat-Sen University, Guangzhou, 510275, China*

[4]*Institute of High Pressure Physics, School of Physical Science and Technology, Ningbo University, Ningbo 315211, China*

*Corresponding author: duandf@jlu.edu.cn (D. D.), yaodaox@mail.sysu.edu.cn (D. Y.), cuitian@nbu.edu.cn (T.C.)

‡These authors contributed equally: Zihao Huo, Zhihui Luo

**Abstract:**

The recent transport measurement of $La_3Ni_2O_7$ uncover a "right-triangle" shape of the superconducting dome in the pressure-temperature (P-T) phase diagram. Motivated by this, we perform theoretical first-principles studies of $La_3Ni_2O_7$ with the pressure ranging from 0 to 100 GPa. Notably, we reveal a pressure dependence of the Ni-$d_{z^2}$ electron density at the Fermi energy ($n_z^{E_F}$) that highly coincides with such shape. On this basis, we further explore the electronic structure under uniaxial stress. By tracking the stress response of $n_z^{E_F}$, we propose that superconductivity can be achieved by applying only ~ 2 GPa of compression along the *c* axis. The idea is further exemplified from the perspectives of lattice distortion, band structure, Fermi surface and superconducting phase coherence. We also discuss the possible charge modulation under the stress and provide an insight to the relation between $n_z^{E_F}$ and the superconducting $T_c$ in $La_3Ni_2O_7$ system. Our study provides a helpful guide to the future experiment.

The high-$T_c$ superconductivity (HTSC) has long been a hot topic in the scientific community. Recently, Sun *et al.* reported the signature of superconductivity with $T_c$ = 80 K at pressures of ~ 14 GPa in the bilayer Ruddlesden-Popper (RP) phase of $La_3Ni_2O_7$ [1], heralding a new HTSC family after the discovery of cuprates [2-6], iron-based [7-12], and infinite-layer nickelates [13-17] superconductors. The discovery was later confirmed by several other groups [18-21], which show that the emergence of HTSC is accompanied by a structural transition from the *Amam* phase to *I*4/*mmm* phase under pressure. In a recent resistance measurement, the hydrostatic pressure is applied up to 104 GPa, which unveils a "right-triangle" shape of the superconducting region in the pressure-temperature (P-T) phase diagram [22]. Such right-triangle shape may suggest a linear dependence of the superconducting $T_c$ on some critical characteristic scales at the high-pressure side of $La_3Ni_2O_7$.

The related superconductivity mechanism of $La_3Ni_2O_7$ has been intensively discussed [23-45]. Density functional theory (DFT) calculations show that the occurrence of the Ni-$d_{z^2}$ spectrum at the Fermi energy ($E_F$) is vital for the observed HTSC [1]. Meanwhile, strongly-correlated analysis indicates the crucial apical Ni-$d_{z^2}$-O-$p_z$-Ni-$d_{z^2}$ bonding structure for the formation of the inter-layer superexchange, which can further drive an $s_\pm$-wave pairing symmetry [28, 35, 40]. These proposals are in general agree with the experimental observations [46, 47], which highlight a different superconducting situation in $La_3Ni_2O_7$ as compared to the cuprates. In light of these, there are several theoretical efforts on the search of other superconductors in RP nickelates [48, 49]. In experiment, a Pr-doped *I*4/*mmm*-$La_2PrNi_2O_7$ have been successfully synthesized, which exhibits a $T_c$ of 78 K at 15 GPa [50]. This compound bears striking similarity to $La_3Ni_2O_7$ in the many aspects, such as lattice, electronic structure, and even the pressure dependence of $T_c$. Notably, this compound bears striking similarity to $La_3Ni_2O_7$ in many aspects, such as lattice, electronic structure, and even the pressure dependence of $T_c$. Also, the trilayer $La_4Ni_3O_{10}$ is confirmed to host superconductivity with $T_c$ ~ 30 K under pressure [51-55]. Theoretical studies on $La_4Ni_3O_7$ suggest a similar mechanism of HTSC and attribute the rapid drop of $T_c$ to the appearance of frustrated superexchanges in the system [56-58].

To gain insights into the superconductivity in RP nickelates, it is helpful to seek novel experimental techniques. Strain engineering is an effective and powerful method in materials design with the ability to regulate electronic properties at the atomic level [59-61]. Previous studies show that, for example, the $T_c$ of crystalline $Sr_2RuO_4$ can be enhanced from 1.5 to 3.4 K by inducing uniaxial strain using a strain cell [62, 63]. Also, in the $YBa_2Cu_3O_{7-\delta}$ system, the weak pressure dependence of $T_c$ is ascribed to a large cancellation of uniaxial pressures in the *a* and *b* axes [64, 65]. In this system, uniaxial pressure can further tune competing orders via charge modulation [66]. These greatly inspire us to apply the same technique in RP nickelates. We can also decompose the unified resistance response of $La_3Ni_2O_7$ along uniaxial directions so as to gain deeper understandings of how the superconductivity develops with the depressions of lattice distortion and charge density wave [19, 21]. Motivated by this, in this paper, we perform the first-principal studies of electronic properties under hydrostatic and uniaxial pressures. For the hydrostatic pressure, we investigate a wide range of pressure from 0 to 100 GPa to cover the experimentally detected right-triangle shape, from which we observe an alignment of the Ni-$d_{z^2}$ DOS at the $E_F$ with such shape. For the uniaxial pressure, we investigate the stress response of electronic structures along three axes. Our results suggest that superconductivity may also be achieved via an equivalent 2 GPa of uniaxial compression along *c* axis, which is much smaller than that is required under the hydrostatic pressure (15 GPa).

The $La_3Ni_2O_7$ structure hosts an orthorhombic symmetry (space group: *Amam*) at ambient pressure [Fig. S1(a) of Supplemental Material (SM)], which is formed by the inter-growth of two planes of $NiO_6$ octahedra and La-O fluorite-type layers stacked along the *c* axis. The in-plane Ni-O bonds exhibit two different lengths of 1.93 and 1.91 Å, while the two out-of-plane Ni-O bonds exhibit various lengths, with 2.3 Å for the outer and 1.97 Å for the inner apical Ni-O bonds. To describe the shape of the $NiO_6$ octahedra, we define the octahedral regularity parameter $R$ [67]:

$$R = \frac{d_z - d_{xy}}{d_z + d_{xy}},$$

where $d_{xy}$ is the average in-plane O-Ni-O distance and $d_z$ is the average out-of-plane O-Ni-O distance. This parameter is a measure of the apical versus basal extension of the octahedra, in which, in extreme cases, $R$ = 1 (-1) corresponds to a fully elongated (compressed) octahedra.

In Fig. 1, we show the DOS of Ni-$d_{z^2}$ and $d_{x^2-y^2}$ orbitals at the $E_F$ as a function of pressure, which we denote as $n_z^{E_F}$ and $n_x^{E_F}$ by the black and red lines respectively. For comparison, we overlay on the experimentally measured $T_c$s [22] by the blue dots. It's striking that the pressure dependence of $n_z^{E_F}$ shows a general consistence with that of the $T_c$ over the whole pressure range. As one can see, when 0 < P < 15 GPa, which is featured by *Amam* symmetry, the $n_z^{E_F}$ curve exhibits an exponential-like increase. This coincides with the rapid increase of the $T_c$ data points in the range of 10 ~ 15 GPa. When P > ~ 15 GPa, both $n_z^{E_F}$ and $T_c$ reaches the maxima and then starts to decrease. But to our surprise, they again follow a similar tendency. On the other hand, we find the DOS of Ni-$d_{x^2-y^2}$ orbital ($n_x^{E_F}$) keeps showing decrease with pressure. The agreement of the $n_z^{E_F}$ curves to the right-triangle shape is in fact embodied in its special electronic structure. Due to the relative localized environment, Ni-$d_{z^2}$ electrons can form a rather flat bonding state around -50 meV [68] and can cause drastic change of the density at the $E_F$ under moderate pressure. Further increasing pressure would push the flat band above the $E_F$, thus causing an opposite decrease of $n_z^{E_F}$. However, since the lattice collapse slows down as keep increasing the pressure, the decrease of $n_z^{E_\mathrm{F}}$ becomes less intensive, which eventually outlines the right-triangle shape.

Now we further study the electronic structure under uniaxial pressures. The stress responses of the *Amam* phase to compression strain $\varepsilon$ along the [100], [010], and [001] directions at ambient pressures are presented in Fig. S2(a) of SM. The slope of the stress-strain curve along different directions is almost constant, indicating that the *Amam* phase does not undergo plastic deformation in the range of compressive strain $\varepsilon$ that we study. In Fig. S2(b) of SM we present $n_z^{E_\mathrm{F}}$ as a function of $\varepsilon$ at ambient pressure. Also, since the evolution of $n_z^{E_\mathrm{F}}$ is similar along the [100] and [010] directions, we only present the results along the [100] and [001] directions in the main text.

In Fig. 2(a), when stress is applied along the [100] direction at ambient pressure, $d_{xy}$ decreases and $d_z$ increases, thereby causing an increase of $R$ from 5.3% to 6% in the range of $0 < \varepsilon < 0.03$. Here we see that the band structure is pushed away from the $E_\mathrm{F}$, as shown in Fig. 2(c) for $\varepsilon = 0.02$. We therefore concludes that applying stress along the [100]/[010] direction should disfavor the superconductivity. In Fig. 2(b), when applying stress along the [001] direction, $d_z$ decreases and $d_{xy}$ increases, thus leads to the decrease in $R$. Note that when $\varepsilon = 0.02$ (equivalent to 2.06 GPa of stress), the octahedra hosts an $R = 3.9\%$ that is close to that under 15 GPa of hydrostatic pressure. In this case, we find the band structure acquires a huge enhancement of the Ni-$d_{z^2}$ spectrum at the $E_\mathrm{F}$ as shown in Fig. 2(d). This may suggest that superconductivity can be promoted under ~ 2 GPa of uniaxial pressure along the c axis, which is much smaller than that is required under the hydrostatic pressure. In Fig. S2(b) of SM we further demonstrate that the evolution of $n_z^{E_\mathrm{F}}$ under *c* axis compression is similar to that in Fig. 1, where it gains an abrupt increase at $\varepsilon = 0.005$ and then slowly decreases when $\varepsilon$ exceeds 0.02. In regard to the self-doping effect, however, there are qualitative difference between the hydrostatic and uniaxial pressures. For the hydrostatic pressure, the self-doping holes mainly come from the Ni-$t_{2g}$ orbitals or from other elements due to the same upward trend of the two Ni-$e_g$ energy levels (Fig. S3 and S22 of SM). But for the uniaxial pressures, the self-doping effect mainly occurs within two Ni-$e_g$ orbitals, which is well illustrated in Fig. 2(e) by the relative variation of the energy levels under different type of stresses. On the other hand, we note the application of uniaxial pressure would not trigger the structural transition but maintains *Amam* symmetry, which results in the simple upward shifts of two proximal γ bands to the $E_\mathrm{F}$, as shown in Fig. 2(d). In fact, these two proximal bands are unique features of the *Amam* symmetry which denote another bonding-antibonding structure associating with the stagger of the bilayers. This may modulate the charge fluctuation, which we will further discuss in the latter. We also calculate the electronic structures under tensile stress (Fig. S10-S11 of SM) and suggest that the tensile stress along [100] ([001]) plays a similar role with the compression stress along [001] ([100]). Our results show that $n_z^{E_\mathrm{F}}$ abruptly increases when $\varepsilon$ exceeds 0.01 (equivalent to 1.53 GPa of stress), reaches a maximum value at $\varepsilon$ = 0.025 (equivalent to 3.72 GPa of stress), and then gradually decreases.

Since the superconductivity only occur under pressure for $La_3Ni_2O_7$, we can expand the pressure response of $T_c$ along the uniaxial directions. For this reason, we further explore the stress response of the *I*4/*mmm* phase at 15 GPa, with a small $\varepsilon$ = 0.005. The results are presented in Fig. 3, which are calculated under a $\sqrt{2} \times \sqrt{2} \times 1$ supercell to allow direct comparison with our previous results. The band structures in Fig. 3(c-d) indicate that, although barely identify, the effects of these two stresses on electronic structure are unchanged, i.e., hole (electron) dope the $d_{x^2-y^2}$ ($d_{z^2}$) orbital for [100] stress and vice versa for [001] stress. When referring to the DOS, it is clear that $n_z^{E_\mathrm{F}}$ is enhanced for [100] stress but depressed for [001] stress. These suggest that further applying *c* axis compression should depress superconductivity, while applying in-plane pressure should promote superconductivity, which is opposite to that under the ambient pressure.

In Fig. 4, we present the Fermi surfaces of our above discussions, which are at 0 GPa with $\varepsilon$ = 0, 0.02 (upper panel) and at 15 GPa with $\varepsilon$ = 0, 0.005 (lower panel). At 0 GPa, as shown in Fig. 4(b), the Fermi surface is barely affected by the [100] stress as compared to Fig. 4(a). However, for the [001] stress in Fig. 4(c), the situation gets complicated. Careful contrast of Fig. 4(a, c-d) shows that, the [001] stress on one hand brings the γ pocket to the Fermi surface, thus mimics that in Fig. 4(d) but with notable distortion. On the other hand, it significantly reduces the distortions in α pocket, making it approaching that in Fig. 4(d). Combined with the aforementioned analysis of the enhanced charge fluctuation associating with the stagger of the bilayers, this may imply a modulation of the density wave instability from the in-plane to the *c* axis, which can be testified in future experiment. For the lower panel at 15 GPa, similar trends are observed with $\varepsilon$ = 0.005. In this case the γ pocket shows very sensitive dependence on uniaxial pressure and is line with our aforementioned results.

We have shown that, at 0 GPa, the maximal $n_z^{E_\mathrm{F}}$ can be reached by applying the *c* axis compression of $\varepsilon$ = 0.02, in which the resulting $R$ = 3.9% is quite close to that of the $R$ = 4% at 15 GPa without stress. In fact, we find the maxima of $n_z^{E_\mathrm{F}}$ seems always correlates with such $R$ value regardless of the hydrostatic pressures. In Fig. 5, we present the evolution of $n_z^{E_\mathrm{F}}$ as a function of $R$ at different hydrostatic pressures. In each subgraph, the star symbol denotes the case without stress. By applying [001] ([100]) stress, it starts to extend as a curve with solid (open) symbols along the left (right) direction. Following the movement of the star symbol from Fig. 5(a) to Fig. 5(d), one sees that, as increasing pressure, $R$ is gradually decreased, accompanying by the increase of $n_z^{E_\mathrm{F}}$. It eventually stabilizes around $R$ ~ 4% in Fig. 5(c) in which the maxima of $n_z^{E_\mathrm{F}} \approx 2.52$ states/eV/f.u. is reached simultaneously. As further inspecting each subgraph, similar trend can also be observed under the *c* axis compression, namely, the maximum of $n_z^{E_\mathrm{F}}$ can be reached when $R$ approaching ~ 4%. This special $R$ value reflects an ideal self-doping level and could be regarded as a useful quantity in identifying superconductivity in the bilayer nickelates.

Our theoretical DFT study has revealed a close connection between the superconducting $T_c$ and the DOS of Ni-$d_{z^2}$ orbital at the $E_\mathrm{F}$, as both demonstrate a right-triangle shape as a function of pressure. The agreement of these two quantities is in fact quite common in the conventional superconductors where the *s*-wave pairing prevails [69-72]. This leads us to the speculation that a similar mechanism should apply in the unconventional $La_3Ni_2O_7$ system. Here due to localized $s_\pm$-wave pairing structure, the pairing scale $\Delta$ can be maximized by tuning the DOS that involves the pairing at the $E_\mathrm{F}$. This is in stark contrast to the cuprates and iron-based superconductors where the ideal Fermi surface topologies are always needed be considered during variation of the doping level [73]. It is important to note that, however, such consideration is insufficient for a finite $T_c$ without superfluid stiffness [34]. In $La_3Ni_2O_7$ system, the superfluid stiffness is achieved by the hybridization $V$ with the Ni-$d_{x^2-y^2}$ orbital to form the global phase coherence [74]. As we further inspect the evolutions of the hopping parameters under pressure based on our DFT (see Table S2 in SM), we find $V$ is enhanced from 0.14 to 0.23 eV for pressure from 0 to 15 GPa, which is quite significant as compared the in-plane hopping of Ni-$d_{x^2-y^2}$ orbital $t^x$ from 0.43 to 0.49 eV. Taking $V/t^x$ as a relevant quantity, we find it increases from 0.32 to 0.48, which remarkably covers the critical $V_c/t^x$ ~ 0.4 for a finite $T_c$ as revealed in a cluster dynamic mean-field theory study [74]. These highlight the subtle electronic evolutions in the establishment of HTSC in $La_3Ni_2O_7$ system. For the high-pressure side featured by the slope of the right-triangle for P > 15 GPa, further increasing $V$ would lead to an opposite suppression of $T_c$ by disrupting the formation of the pairing scale $\Delta$ [74]. This may explain the slightly larger slope of the $T_c$ curve as compared to that of the $n_z^{E_\mathrm{F}}$ in Fig. 1. Another important result of our study is the plausibility of applying the *c* axis compression of ~2 GPa for a finite $T_c$ at ambient pressure. Regarding the phase coherence, we find that in this case $V/t^x$ = 0.36 is reached, which is quite close to $V_c/t^x$ and should prefer superconductivity. Careful examination of this idea is warranted in future research.

To summarize, we have performed a comprehensive first-principles study of the hydrostatic and uniaxial pressures response of the electronic structure in $La_3Ni_2O_7$ system. Our results reveal a pressure dependence of the Ni-$d_{z^2}$ electron density at $E_\mathrm{F}$ that highly coincides with the right-triangle shape of the superconductivity $T_c$ [22]. We elucidate the physical origin of such coincidence that points to the unique localized $s_\pm$-wave pairing structure in this system. Furthermore, we propose that superconductivity in $La_3Ni_2O_7$ can be achieved by applying ~ 2 GPa of compression along *c* axis, and we exemplify this data from the perspectives of lattice distortion, band structure, Fermi surface and superconducting phase coherence. We also discuss the possible charge modulation under *c* axis compression. Our studies provide helpful guide to the further experimental search of HTSC in RP nickelate.

We thank Meng Wang for the experimental data and discussion and thank Wéi Wú, Wenjie Sun, and Mengwu Huo for many stimulating discussions. This work was supported by National Key R&D Program of China (Nos. 2022YFA1402304 and 2022YFA1402802), National Natural Science Foundation of China (Grants Nos. 12122405, 52072188, 12274169, and 92165204), Program for Science and Technology Innovation Team in Zhejiang (Grant No. 2021R01004), Guangdong Provincial Key Laboratory of Magnetoelectric Physics and Devices (Grant No. 2022B1212010008), and the Fundamental Research Funds for the Central Universities. Some of the calculations were performed at the High Performance Computing Center of Jilin University and using TianHe-1(A) at the National Supercomputer Center in Tianjin.

See Supplemental Material for band structures, DOS, in-plane O-Ni-O average distance, out-plane O-Ni-O average distance, the octahedral regularity parameter *R*, tight-binding parameters and structural information of $La_3Ni_2O_7$ under various compressive.

## References

[1] H. Sun, M. Huo, X. Hu, J. Li, Z. Liu, Y. Han, L. Tang, Z. Mao, P. Yang, B. Wang, J. Cheng, D. Yao, G. Zhang and M. Wang, Signatures of superconductivity near 80 K in a nickelate under high pressure, Nature **621**, 493 (2023).

[2] J. Bednorz and K. Müller, Possbile high $T_c$ superconductivity in the Ba-La-Cu-O system, Z. Phys. B Condens. Matter **64**, 189 (1986).

[3] M. Wu, J. Ashburn, C. Torng, P. Hor, R. Meng, L. Gao, Z. Huang, Y. Wang and C. Chu, Superconductivity at 93 K in a New Mixed-Phase Y-Ba-Cu-O Compound System at Ambient Pressure, Phys. Rev. Lett. **58**, 908 (1987).

[4] B. Keimer, S. Kivelson, M. Norman, S. Uchida and J. Zaanen, From quantum matter to high-temperature superconductivity in copper oxides, Nature **518**, 179 (2015).

[5] P. Lee, N. Nagaosa and X. Wen, Doping a Mott insulator: Physics of high-temperature superconductivity, Rev. Mod. Phys. **78**, 17 (2006).

[6] W. Li, J. Zhao, L. Cao, Z. Hu, Q. Huang, X. Wang, Y. Liu, G. Zhao, J. Zhang, Q. Liu, R. Yu, Y. Long, H. Wu, H. Lin, C. Chen, Z. Li, Z. Gong, Z. Guguchia, J. Kim, G. Stewart, Y. Uemura, S. Uchida and C. Jin, Superconductivity in a unique type of copper oxide, Proc. Natl. Acad. Sci. U.S.A. **116**, 12156 (2019).

[7] Y. Kamihara, T. Watanabe, M. Hirano and H. Hosono, Iron-Based Layered Superconductor $La[O_{1-x}F_x]FeAs$ (x=0.05-0.12) with $T_c = 26$ K, J. Am. Chem. Soc. **130**, 3296 (2008).

[8] H. Takahashi, K. Igawa, K. Arii, Y. Kamihara, M. Hirano and H. Hosono, Superdoncutivity at 43 K in an iron-based layered compound $LaO_{1-x}F_xFeAs$, Nature **453**, 376 (2008).

[9] X. Chen, T. Wu, G. Wu, R. Liu, H. Chen and D. Fang, Superconductivity at 43 K in $SmFeAsO_{1-x}F_x$, Nature **453**, 761 (2008).
[10] H. Wen, G. Mu, L. Fang, H. Yang and X. Zhu, Superconductivity at 25 K in hole-doped $(La_{1-x}Sr_x)OFeAs$, Europhys. Lett. **82**, 17009 (2008).
[11] G. Chen, Z. Li, D. Wu, G. Li, W. Hu, J. Dong, P. Zheng, J. Luo and N. Wang, Superconductivity at 41 K and Its Competition with Spin-Density-Wave Instability in Layered $CeO_{1-x}F_xFeAs$, Phys. Rev. Lett. **100**, 247002 (2008).
[12] M. Rotter, M. Tegel and D. Johrendt, Superconductivity at 38 K in the Iron Arsenide $(Ba_{1-x}K_x)Fe_2As_2$, Phys. Rev. Lett. **101**, 107006 (2008).
[13] D. Li, K. Lee, B. Wang, M. Osada, S. Crossley, H. Lee, Y. Cui, Y. Hikita and H. Hwang, Superconductivity in an infinite-layer nickelate, Nature **572**, 624 (2019).
[14] M. Osada, B. Wang, B. Goodge, S. Harvey, K. Lee, D. Li, L. Kourkoutis and H. Hwang, Nickelate Superconductivity without Rare-Earth Magnetism: $(La,Sr)NiO_2$, Adv. Mater. **33**, 2104083 (2021).
[15] N. Wang, M. Yang, Z. Yang, K. Chen, H. Zhang, Q. Zhang, Z. Zhu, Y. Uwatoko, L. Gu, X. Dong, J. Sun, K. Jin and J. Cheng, Pressure-induced monotonic enhancement of $T_c$ to over 30 K in superconductivity $Pr_{0.82}Sr_{0.18}NiO_2$ thin films, Nat. Commun. **13**, 4367 (2022).
[16] D. Li, B. Wang, K. Lee, S. Harvey, M. Osada, B. Goodge, L. Kourkoutis and H. Hwang, Superconducting Dome in $Nd_{1-x}Sr_xNiO_2$ Infinite Layer Films, Phys. Rev.Lett. **125**, 027001 (2020).
[17] W. Sun, Y. Li, R. Liu, J. Yang, J. Li, W. Wei, G. Jin, S. Yan, H. Sun, W. Guo, Z. Gu, Z. Zhu, Y. Sun, Z. Shi, Y. Deng, X. Wang and Y. Nie, Evidence for Anisotropic Superconductivity Beyond Pauli Limit in Infinite-Layer Lanthanum Nickelates, Adv. Mater. **35**, 2303400 (2023).
[18] J. Hou, P. Yang, Z. Liu, J. Li, P. Shan, L. Ma, G. Wang, N. Wang, H. Guo, J. Sun, Y. Uwatoko, M. Wang, G. Zhang, B. Wang and J. Cheng, Emergence of High-Temperature Superconducting Phase in Pressurized $La_3Ni_2O_7$ Crystals, Chin. Phys. Lett. **40**, 117302 (2023).
[19] Y. Zhang, D. Su, Y. Huang, H. Sun, M. Huo, Z. Shan, K. Ye, Z. Yang, R. Li, M. Smidman, M. Wang, L. Jiao and H. Yuan, High-temperature superconductivity with zero-resistance and strange metal behavior in $La_3Ni_2O_{7-\delta}$, arXiv preprint arXiv:2307.14819 (2023).
[20] Y. Zhou, J. Guo, S. Cai, H. Sun, P. Wang, J. Zhao, J. Han, X. Chen, Q. Wu, Y. Ding, M. Wang, T. Xiang, H. Mao and L. Sun, Evidence of filamentary superconductivity in pressurized $La_3Ni_2O_7$, arXiv preprint arXiv:2311.12361 (2023).
[21] G. Wang, N. Wang, X. Shen, J. Hou, L. Ma, L. Shi, Z. Ren, Y. Gu, H. Ma, P. Yang, Z. Liu, H. Guo, J. Sun, G. Zhang, S. Calder, J. Yan, B. Wang, Y. Uwatoko and J. Cheng, Pressure-Induced Superconductivity In Polycrystalline $La_3Ni_2O_{7-\delta}$, Phys. Rev. X **14**, 011040 (2024).
[22] J. Li, P. Ma, H. Zhang, X. Huang, C. Huang, M. Huo, D. Hu, Z. Dong, C. He, J. Liao, X. Chen, T. Xie, H. Sun and M. Wang, Pressure-driven right-triangle shape superconductivity in bilayer nickelate $La_3Ni_2O_7$, arXiv preprint arXiv:2404.11369 (2024).

[23] Z. Luo, X. Hu, M. Wang, W. Wú and D. Yao, Bilayer Two-Orbital Model of $La_3Ni_2O_7$ under Pressure, Phys. Rev. Lett. **131**, 126001 (2023).
[24] Y. Gu, C. Le, Z. Yang, X. Wu and J. Hu, Effective model and pairing tendency in bilayer Ni-based superconductor $La_3Ni_2O_7$, arXiv preprint arXiv:2306.07275 (2023).
[25] Q. Yang, D. Wang and Q. Wang, Possbile $s_{\pm}$-wave superconductivity in $La_3Ni_2O_7$, Phys. Rev. B **108**, L140505 (2023).
[26] D. Shilenko and I. Leonov, Correlated electronic structure, orbital-selective behavior, and magnetic correlations in double-layer $La_3Ni_2O_7$ under pressure, Phys. Rev. B **108**, 125105 (2023).
[27] X. Chen, P. Jiang, J. Li, Z. Zhong and Y. Lu, Critical charge and spin instabilities in superconducting $La_3Ni_2O_7$, arXiv preprint arXiv:2307.07154 (2023).
[28] Y. Liu, J. Mei, F. Ye, W. Chen and F. Yang, $s^{\pm}$-Wave Pairing and the Destructive Role of Apical-Oxygen Deficencies in $La_3Ni_2O_7$ under Pressure, Phys. Rev. Lett. **131**, 236002 (2023).
[29] V. Christiansson, F. Petocchi and P. Werner, Correlated Electronic Structure of $La_3Ni_2O_7$ under Pressure, Phys. Rev. Lett. **131**, 206501 (2023).
[30] Y. Zhang, L. Lin, A. Moreo and E. Dagotto, Electronic structure, dimer physics, orbital-selective behavior, and magnetic tendencies in the bilayer nickelate superconductor $La_3Ni_2O_7$ under pressure, Phys. Rev. B **108**, L180510 (2023).
[31] Z. Liao, L. Chen, G. Duan, Y. Wang, C. Liu, R. Yu and Q. Si, Electron correlations and superconductivity in $La_3Ni_2O_7$ under pressure tuning, Phys. Rev. B **108**, 214522 (2023).
[32] F. Lechermann, J. Gondolf, S. Bötzel and I. Eremin, Electronic correlations and superconducting instability in $La_3Ni_2O_7$ under high pressure, Phys. Rev. B **108**, L201121 (2023).
[33] Y. Shen, M. Qin and G. Zhang, Effective Bi-Layer Model Hamiltonian and Density-Matrix Renormalization Group Study for the High-$T_c$ Superconductivity in $La_3Ni_2O_7$ under High Pressure, Chin. Phys. Lett. **40**, 127401 (2023).
[34] Y. Yang, G. Zhang and F. Zhang, Interlayer valence bonds and two-component theory for high-$T_c$ superconductivity of $La_3Ni_2O_7$ under pressure, Phys. Rev. B **108**, L201108 (2023).
[35] J. Huang, Z. Wang and T. Zhou, Impurity and vortex states in the bilayer high-temperature superconductivity $La_3Ni_2O_7$, Phys. Rev. B **108**, 174501 (2023).
[36] Q. Qin and Y. Yang, High-$T_c$ superconductivity by mobilizing local spin singlets and possible route to higher $T_c$ in pressurized $La_3Ni_2O_7$, Phys. Rev. B **108**, L140504 (2023).
[37] Z. Luo, B. Lv, M. Wang, W. Wú and D. Yao, High-$T_c$ superconductivity in $La_3Ni_2O_7$ based on the bilayer two-orbital t-J model, arXiv preprint arXiv:2308.16564 (2023).
[38] C. Lu, Z. Pan, F. Yang and C. Wu, Interlayer-Coupling-Driven High-Temperature Superconductivity in $La_3Ni_2O_7$ under Pressure, Phys. Rev. Lett. **132**, 146002 (2024).
[39] Y. Zhang, L. Lin, A. Moreo, T. A. Maier and E. Dagotto, Structural phase transition, $s_{\pm}$-wave pairing, and magnetic stripe order in bilayered superconductor $La_3Ni_2O_7$ under pressure, Nat. Commun. **15**, 2470 (2024).

[40] W. Wú, Z. Luo, D. Yao and M. Wang, Superexchange and charge transfer in the nickelate superconductor $La_3Ni_2O_7$ under pressure, Sci. China-Physics, Mech. **67**, 117402 (2024).
[41] H. Sakakibara, N. Kitamine, M. Ochi and K. Kuroki, Possible High $T_c$ Superconductivity in $La_3Ni_2O_7$ under High Pressure through Manifestation of a Nearly Half-Filled Bilayer Hubbard Model, Phys. Rev. Lett. **132**, 106002 (2024).
[42] X. Qu, D. Qu, J. Chen, C. Wu, F. Yang, W. Li and G. Su, Bilayer $t$-$J$-$J_\perp$ Model and Magnetically Mediated Pairing in the Pressurized Nickelate $La_3Ni_2O_7$, Phys. Rev. Lett. **132**, 036502 (2024).
[43] K. Jiang, Z. Wang and F. Zhang, High-Temperature Superconductivity in $La_3Ni_2O_7$, Chin. Phys. Lett. **41**, 017402 (2024).
[44] Y. Cao and Y. Yang, Flat bands promoted by Hund's rule coupling in the candidate double-layer high-temperature superconductor $La_3Ni_2O_7$ under high pressure, Phys. Rev. B **109**, L081105 (2024).
[45] Y. Tian, Y. Chen, J. Wang, R. He and Z. Lu, Correlation effects and concomitant two-orbital $s_\pm$-wave superconductivity in $La_3Ni_2O_7$ under high pressure, Phys. Rev. B **109**, 165154 (2024).
[46] T. Xie, M. Huo, X. Ni, F. Shen, X. Huang, H. Sun, H. Walker, D. Adroja, D. Yu, B. Shen, L. He, K. Cao and M. Wang, Neutron Scattering Studies on the High-$T_c$ Superconductor $La_3Ni_2O_7$ at Ambient Pressure, arXiv preprint arXiv:2401.12635 (2024).
[47] X. Chen, J. Choi, Z. Jiang, J. Mei, K. Jiang, J. Li, S. Agrestini, M. Garcia-Fernandez, X. Huang, H. Sun, D. Shen, M. Wang, J. Hu, Y. Lu, K. Zhou and D. Feng, Electronic and magnetic excitations in $La_3Ni_2O_7$, arXiv preprint arXiv:2401.12657 (2024).
[48] Y. Zhang, L. Lin, A. Moreo, T. Maier and E. Dagotto, Trends in electronic structures and $s_\pm$-wave pairing for the rare-earth series in bilayer nickelate superconductor $R_3Ni_2O_7$, Phys. Rev. B **108**, 165141 (2023).
[49] B. Geisler, J. Hamlin, G. Stewart, R. Hennig and P. Hirschfeld, Structural transitions, octahedral rotations, and electronic properties of $A_3Ni_2O_7$ rare-earth nickelates under high pressure, npj Quantum Mater. **9**, 38 (2024).
[50] G. Wang, N. Wang, Y. Wang, L. Shi, X. Shen, J. Hou, H. Ma, P. Yang, Z. Liu, H. Zhang, X. Dong, J. Sun, B. Wang, K. Jiang, J. Hu, Y. Uwatoko and J. Cheng, Observation of high-temperature superconductivity in the high-pressure tetragonal phase of $La_2PrNi_2O_{7-\delta}$, arXiv preprint arXiv:2311.08212 (2023).
[51] Y. Zhu, E. Zhang, B. Pan, X. Chen, D. Peng, L. Chen, H. Ren, F. Liu, N. Li, Z. Xing, J. Han, J. Wang, D. Jia, H. Wo, Y. Gu, Y. Gu, L. Ji, W. Wang, H. Guo, Y. Shen, T. Ying, X. Chen, W. Yang, C. Zheng, Q. Zeng, J. Guo and J. Zhao, Superconductivity in trilayer nickelate $La_4Ni_3O_{10}$ single crystals, arXiv preprint arXiv:2311.07353 (2023).
[52] M. Zhang, C. Pei, X. Du, W. Hu, Y. Cao, Q. Wang, J. Wu, Y. Li, H. Liu, C. Wen, Y. Zhao, C. Li, W. Cao, S. Zhu, Q. Zhang, N. Yu, P. Cheng, L. Zhang, Z. Li, J. Zhao, Y. Chen, H. Guo, C. Wu, F. Yang, S. Yan, L. Yang and Y. Qi, Superconductivity in trilayer nickelate $La_4Ni_3O_{10}$ under pressure, arXiv preprint arXiv:2311.07423 (2023).

[53] J. Li, C. Chen, C. Huang, Y. Han, M. Huo, X. Huang, P. Ma, Z. Qiu, J. Chen, X. Hu, L. Chen, T. Xie, B. Shen, H. Sun, D. Yao and M. Wang, Structural transition, electric transport, and electronic structures in the compressed trilayer nickelate $La_4Ni_3O_{10}$, Sci. China-Physics Mech. **67**, 117403 (2024).
[54] Q. Li, Y. Zhang, Z. Xiang, Y. Zhang, X. Zhu and H. Wen, Signature of Superconductivity in Pressurized $La_4Ni_3O_{10}$, Chin. Phys. Lett. **41**, 017401 (2024).
[55] H. Sakakibara, M. Ochi, H. Nagata, Y. Ueki, H. Sakurai, R. Matsumoto, K. Terashima, K. Hirose, H. Ohta, M. Kato, Y. Takano and K. Kuroki, Theoretical analysis on the possibility of superconductivity in the trilayer Ruddlesden-Popper nickelate $La_4Ni_3O_{10}$ under pressure and its experimental examination: Comparison with $La_3Ni_2O_7$, Phys. Rev. B **109**, 144511 (2024).
[56] Q. Yang, K. Jiang, D. Wang, H. Lu and Q. Wang, Effective model and $s_{\pm}$-wave superconductivity in trilayer nickelate $La_4Ni_3O_{10}$, Phys. Rev. B **109**, L220506 (2024).
[57] Q. Qin, J. Wang and Y. Yang, Frustrated Superconductivity in the Trilayer Nickelate $La_4Ni_3O_{10}$, arXiv preprint arXiv:2405.04340 (2024).
[58] C. Chen, Z. Luo, M. Wang, W. Wú and D. Yao, Trilayer multi-orbital models of $La_4Ni_3O_{10}$, Phys. Rev. B **110**, 014503 (2024).
[59] J. Kalikka, X. Zhou, E. Dilcher, S. Wall, J. Li and R. Simpson, Strain-engineered diffusive atomic switching in two-dimensional crystals, Nat. Commun. **7**, 11983 (2016).
[60] B. Chen, N. Gauquelin, D. Jannis, D. M. Cunha, U. Halisdemir, C. Piamonteze, J. H. Lee, J. Belhadi, F. Eltes, S. Abel, Z. Jovanović, M. Spreitzer, J. Fompeyrine, J. Verbeeck, M. Bibes, M. Huijben, G. Rijnders and G. Koster, Strain-Engineered Metal-to-Insulator Transition and Orbital Polarization in Nickelate Superlattices Integrated on Silicon, Adv. Mater. **32**, 2004995 (2020).
[61] Y. Chen, Y. Lei, Y. Li, Y. Yu, J. Cai, M. Chiu, R. Rao, Y. Gu, C. Wang, W. Choi, H. Hu, C. Wang, Y. Li, J. Song, J. Zhang, B. Qi, M. Lin, Z. Zhang, A. Islam, B. Maruyama, S. Dayeh, L. Li, K. Yang, Y. Lo and S. Xu, Strain engineering and epitaxial stabilization of halide perovskites, Nature **577**, 209 (2020).
[62] C. Hicks, D. Brodsky, E. Yelland, A. Gibbs, J. Bruin, M. Barber, S. Edkins, K. Nishimura, S. Yonezawa, Y. Maeno and A. Mackenzie, Strong Increase of $T_c$ of $Sr_2RuO_4$ Under Both Tensile and Compressive Strain, Science **344**, 283 (2014).
[63] A. Steppke, L. Zhao, M. E. Barber, T. Scaffidi, F. Jerzembeck, H. Rosner, A. Gibbs, Y. Maeno, S. Simon, A. Mackenzie and C. Hicks, Strong peak in $T_c$ of $Sr_2RuO_4$ under uniaxial pressure, Science **355**, eaaf9398 (2017).
[64] C. Meingast, O. Kraut, T. Wolf, H. Wühl, A. Erb and G. Müller-Vogt, Large *a*-*b* anisotropy of the expansivity anomaly at $T_c$ in untwinned $YBa_2Cu_3O_{7-\delta}$, Phys. Rev. Lett. **67**, 1634 (1991).
[65] U. Welp, M. Grimsditch, S. Fleshler, W. Nessler, J. Downey, G. Crabtree and J. Guimpel, Effect of uniaxial stress on the superconducting transition in $YBa_2Cu_3O_7$, Phys. Rev. Lett. **69**, 2130 (1992).
[66] H. Kim, S. Souliou, M. Barber, E. Lefrançois, M. Minola, M. Tortora, R. Heid, N. Nandi, R. Borzi, G. Garbarino, A. Bosak, J. Porras, T. Loew, M. König, P. W. Moll, A. Mackenzie, B. Keimer, C. Hicks and M. Le Tacon, Uniaxial pressure control of competing orders in a high-temperature superconductor, Science **362**, 1040 (2018).

[67] B. Geisler and R. Pentcheva, Inducing *n*- and *p*-Type Thermoelectricity in Oxide Superlattices by Strain Tuning of Orbital-Selective Transport Resonances, Phys. Rev. Appl. **11**, 044047 (2019).
[68] J. Yang, H. Sun, X. Hu, Y. Xie, T. Miao, H. Luo, H. Chen, B. Liang, W. Zhu, G. Qu, C. Chen, M. Huo, Y. Huang, S. Zhang, F. Zhang, F. Yang, Z. Wang, Q. Peng, H. Mao, G. Liu, Z. Xu, T. Qian, D. Yao, M. Wang, L. Zhao and X. Zhou, Orbital-dependent electron correlation in double-layer nickelate $La_3Ni_2O_7$, Nat. Commun. **15**, 4373 (2024).
[69] X. Pan, X. Chen, H. Liu, Y. Feng, Z. Wei, Y. Zhou, Z. Chi, L. Pi, F. Yen, F. Song, X. Wan, Z. Yang, B. Wang, G. Wang and Y. Zhang, Pressure-driven dome-shaped superconductivity and electronic structural evolution in tungsten ditelluride, Nat. Commun. **6**, 7805 (2015).
[70] K. S. Fries and S. Steinberg, Fermi-Level Charact eristics of Potential Chalocogenide Superconductors, Chem. Mater. **30**, 2251 (2018).
[71] W. Sano, T. Koretsune, T. Tadano, R. Akashi and R. Arita, Effect of Van Hove singularities on high-$T_c$ superconductivity in $H_3S$, Phys. Rev. B **93**, 094525 (2016).
[72] Y. Ge, F. Zhang, R. Dias, R. Hemley and Y. Yao, Hole-doped room-temperature superconductivity in $H_3S_{1-x}Z_x$ (Z=C,Si), Mater. Today Phys. **15**, 100330 (2020).
[73] J. Hu and H. Ding, Local antiferromagnetic exchange and collaborative Fermi surface as key ingredients of high temperature superconductors, Sci. Rep. **2**, 381 (2012).
[74] Y. Zheng and W. Wú, Superconductivity in the Bilayer Two-orbital Hubbard Model, arXiv preprint arXiv:2312.03605 (2023).

## Figures and Captions

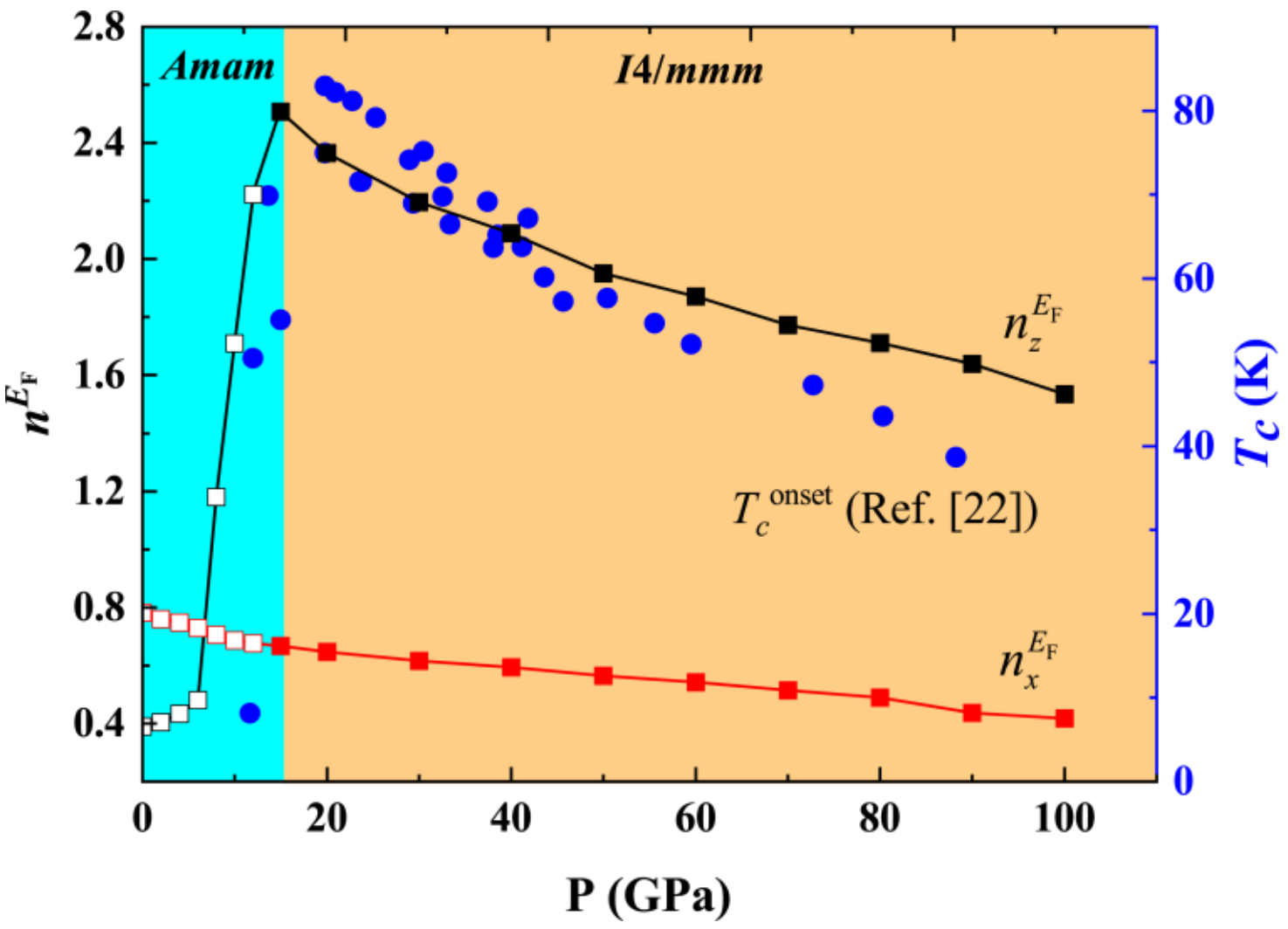

FIG. 1. The calculated Ni-$d_{x^2-y^2}$ and Ni-$d_{z^2}$ DOS at $E_F$, and the experimentally measured $T_c$s as a function of pressure. The solid and open symbols correspond to the data of the *I*4/*mmm* and *Amam* phases, respectively.

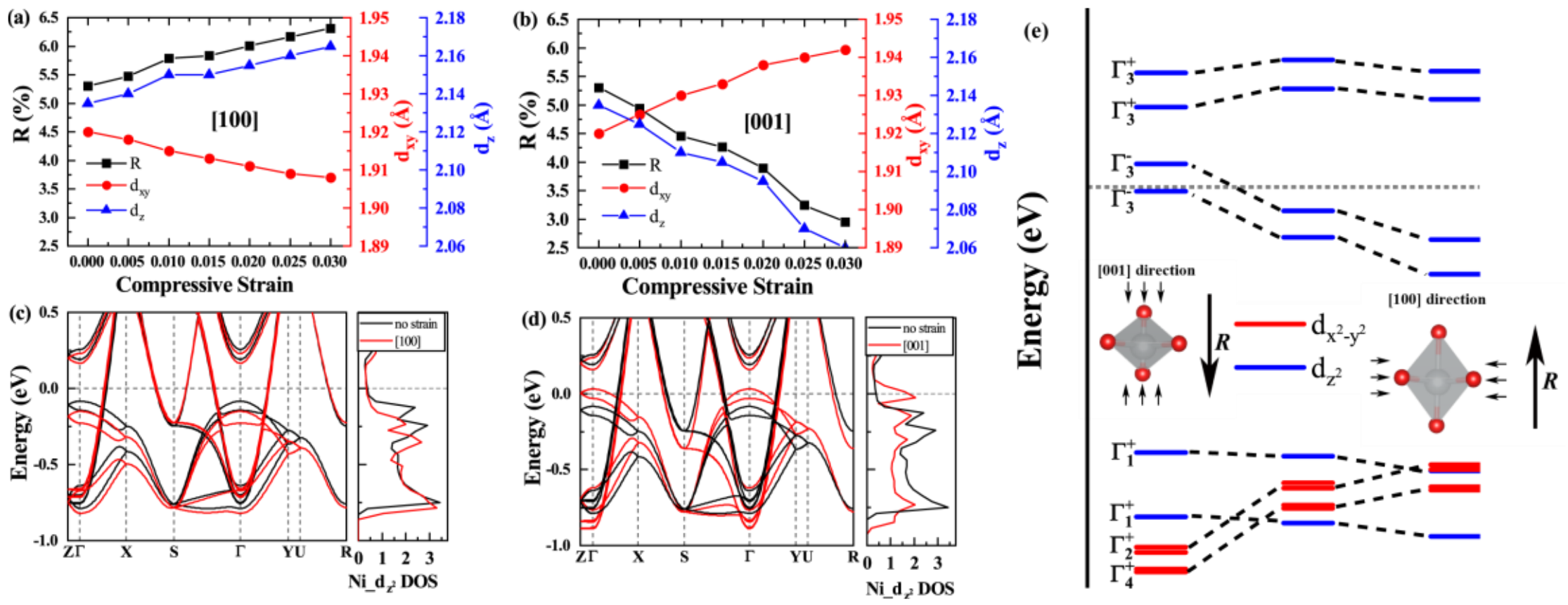

FIG 2. Evolution of *R*, $d_{xy}$, and $d_z$ of the *Amam* phase at ambient pressure as a function of the stress along (a) the [100] and (b) [001] directions. Electronic properties of the *Amam* phase at ambient pressures at $\varepsilon$ = 0 and 0.02 along the (c) [100] and (d) [001] directions. (e) Orbital character and band representations at Γ point of Ni at ambient pressure. The left and right columns denote the energy levels under $\varepsilon$ = 0.02 along the [001] and [100] directions, and the middle column denotes the energy levels without stress. The horizontal gray dashed line represents the Fermi level.

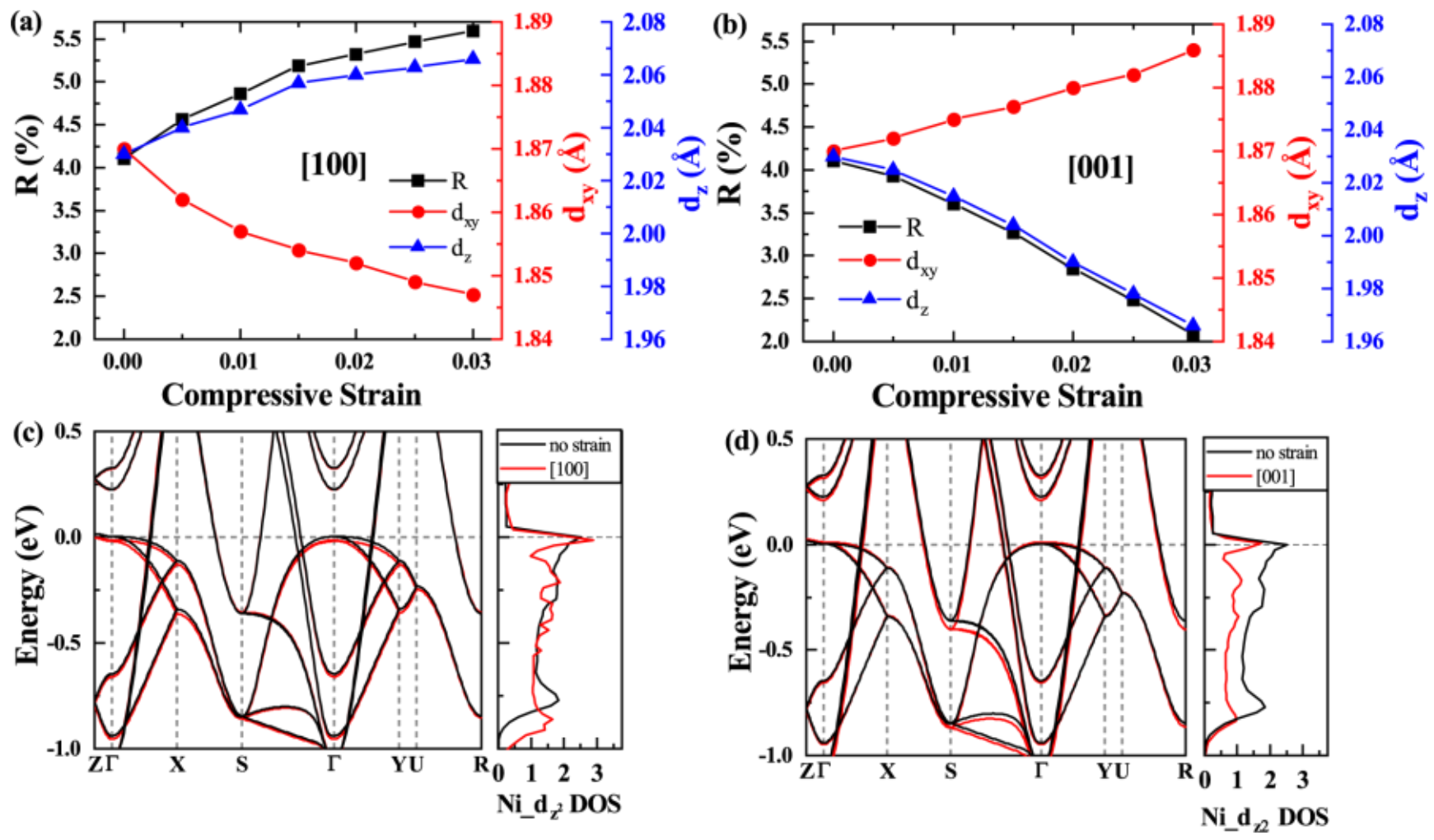

FIG 3. Evolution of $R$, $d_{xy}$, and $d_z$ of the $I4/mmm$ phase at 15 GPa as a function of the compressive strain induced along (a) the [100] and (b) [001] directions. Electronic properties of the $I4/mmm$ phase at 15 GPa with $\varepsilon$ = 0 and 0.005 along the (c) [100] and (d) [001] directions. The horizontal gray dashed line represents the Fermi level.

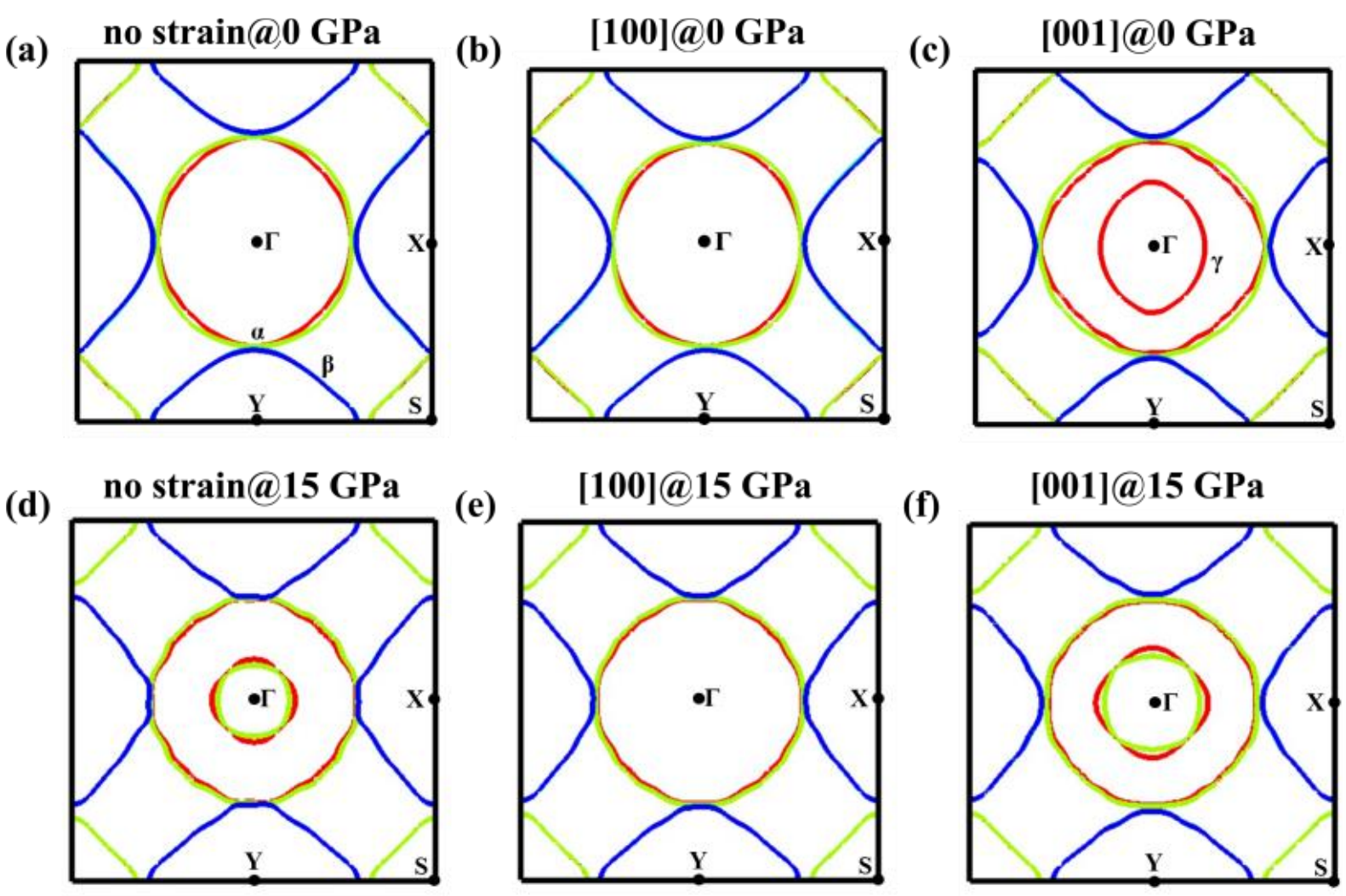

FIG 4. Fermi surface of the *Amam* phase at ambient pressures (a) without strain, (b) with $\varepsilon$ = 0.02 along the [100] direction and (c) $\varepsilon$ = 0.02 along the [001] direction. The calculated two-dimensional Fermi surface of the *I*4/*mmm* phase at 15 GPa (d) without strain, (e) with $\varepsilon$ = 0.005 along the [100] direction, and (f) with $\varepsilon$ = 0.005 along the [001] direction.

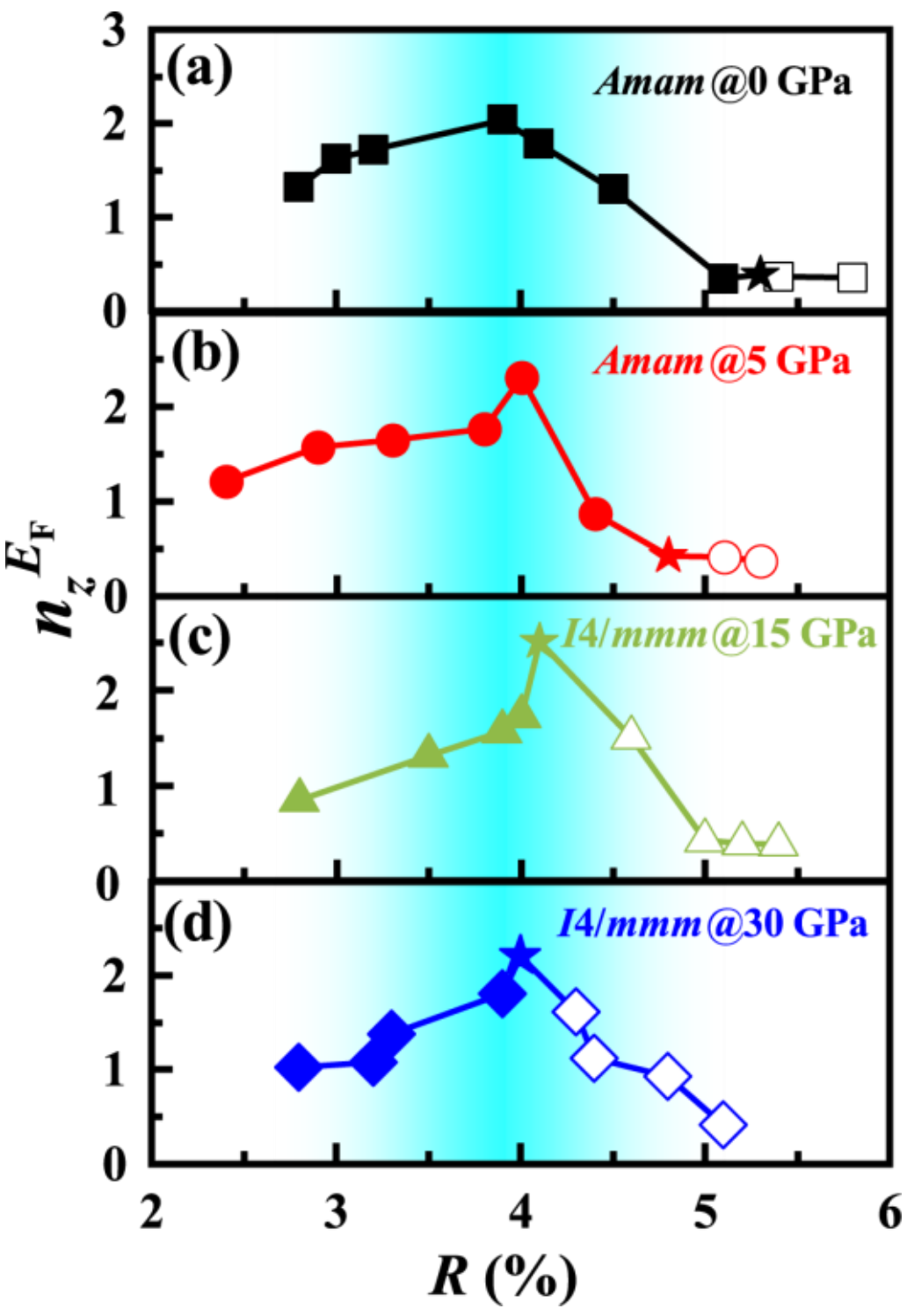

FIG 5. Evolution of the $n_z^{E_\mathrm{F}}$ as a function of $R$. The star symbols represent $R$ in the absence of strain. The solid and open symbols correspond to the induction of compressive strain along the [001] and [100] directions, respectively.

# Supplemental Materials

# Modulation of the Octahedral Structure and Potential Superconductivity of $La_3Ni_2O_7$ through Strain Engineering

Zihao Huo[1,‡], Zhihui Luo[3,‡], Peng Zhang[2], Aiqin Yang[2], Zhengtao Liu[1], Xiangru Tao[2], Zihan Zhang[1], Shumin Guo[1], Qiwen Jiang[1], Wenxuan Chen[1], Dao-Xin Yao[3,*], Defang Duan[1,*],Tian Cui[4,1,*]

[1]*Key Laboratory of Material Simulation Methods & Software of Ministry of Education, State Key Laboratory of Superhard Materials, College of Physics, Jilin University, Changchun 130012, China*

[2]*MOE Key Laboratory for Non-equilibrium Synthesis and Modulation of Condensed Matter, Shaanxi Province Key Laboratory of Advanced Functional Materials and Mesoscopic Physics, School of Physics, Xi'an Jiaotong University, Xi'an 710049, China*

[3]*Center for Neutron Science and Technology, Guangdong Provincial Key Laboratory of Magnetoelectric Physics and Devices, State Key Laboratory of Optoelectronic Materials and Technologies, School of Physics, Sun Yat-Sen University, Guangzhou, 510275, China*

[4]*Institute of High Pressure Physics, School of Physical Science and Technology, Ningbo University, Ningbo 315211, China*

*Corresponding author: duandf@jlu.edu.cn, yaodaox@mail.sysu.edu.cn, cuitian@nbu.edu.cn

[‡]These authors contributed equally: Zihao Huo, Zhihui Luo

# Content

## I. Computational details

We performed first-principles simulations in the framework of density functional theory (DFT) as implemented in the Vienna *ab initio* simulation packages [1]. Exchange and correlations were described using the Perdew-Burke-Ernzerhof formulation of the generalized gradient approximation [2]. The projector-augmented wave potentials [3, 4] were adopted, with $3s^2 3p^6 3d^8 4s^2$, $5s^2 5p^6 5d^1 6s^2$, and $2s^2 2p^4$ shells treated as valence for Ni, La, and O, respectively. A cutoff energy of 700 eV and the Brillouin zone was sampled with a *k*-point mesh of $2\pi \times 0.03$ Å$^{-1}$ to make the enthalpy calculations converge well to less than 1 meV/atom. Static correlation effects were considered within the DFT+$U$ formalism [5, 6]. To determine the $U$ parameter, we tested the $U$ values with 3, 3.5, and 4 eV (Table. S1). The calculated lattice parameters and electronic properties of *Amam* phase at ambient pressure closely matches the experimental data [7, 8] when $U$ is set to 3 eV.

The stress-strain relations were determined by an established method [9-11] with a strain increment of 0.005. At each shear or compressive strain is fixed to determine the shear stress $\sigma_{xz}$ or compressive stress $\sigma_{xx}$, respectively, while the other five independent components of the strain tensors and all atomic positions are simultaneously relaxed with residual forces and stresses less than 0.005 eV/Å and 0.1 GPa, respectively.

Table S1. Calculated lattice parameters (Å) and position of γ band (meV) below the Fermi level

| | a | b | c | position of γ band |
|---|---|---|---|---|
| U = 3 eV | 5.371 | 5.446 | 20.661 | -84.4 |
| U = 3.5 eV | 5.369 | 5.442 | 20.642 | -87.9 |
| U = 4 eV | 5.366 | 5.438 | 20.67 | -92.2 |
| EXP | 5.4[a] | 5.438[a] | 20.455[a] | ~-50[b] |

[a] Ref. [7]

[b] Ref. [8]

## II. Stress response and electronic properties of *Amam* phase under various compressive strain at ambient pressure

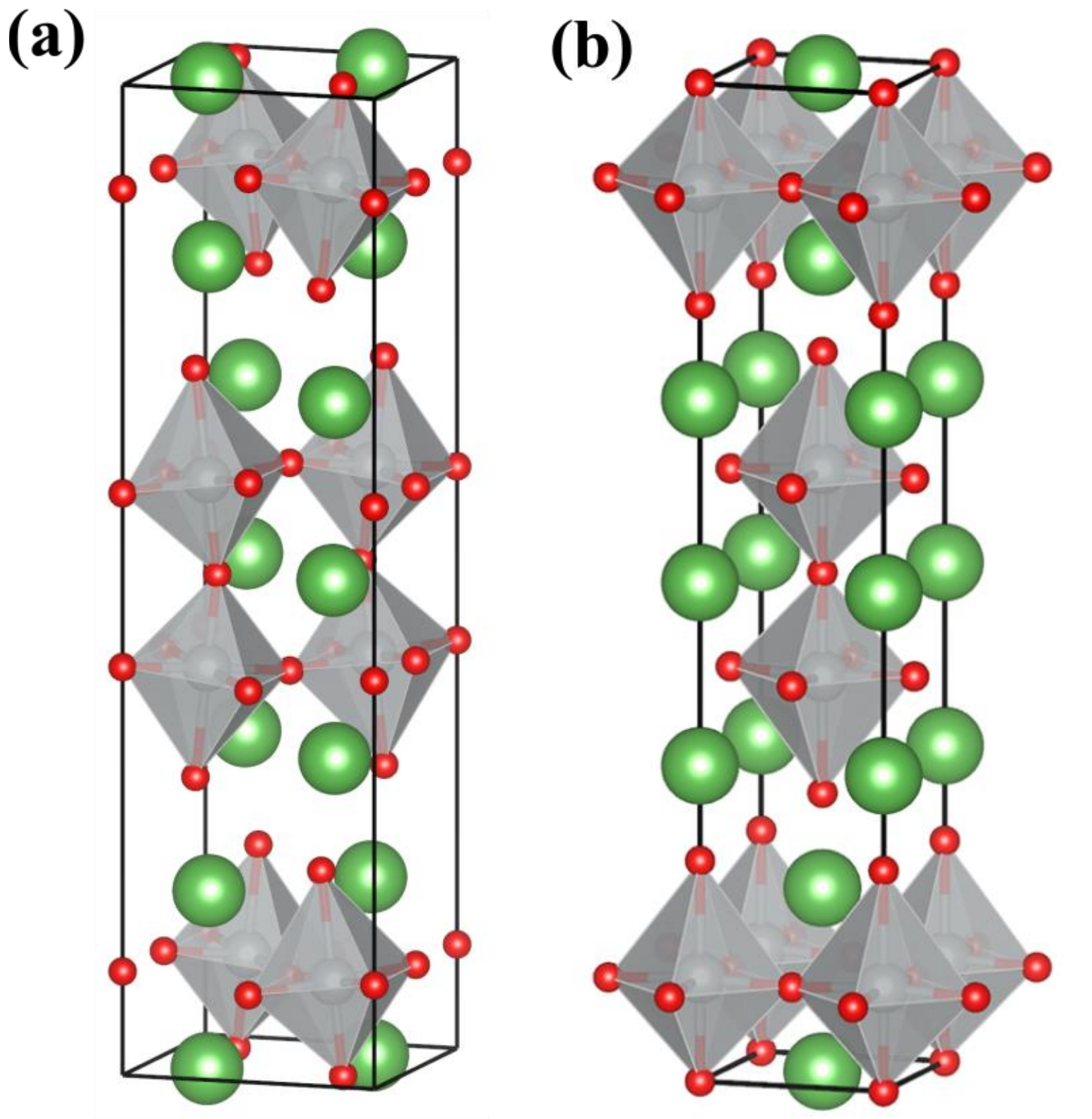

FIG S1. (a-b) Crystal structures of the bilayer Ruddlesden-Popper phase *Amam*-$La_3Ni_2O_7$ and *I4/mmm*-$La_3Ni_2O_7$. The green, grey, and red spheres represent La, Ni, and O atoms, respectively.

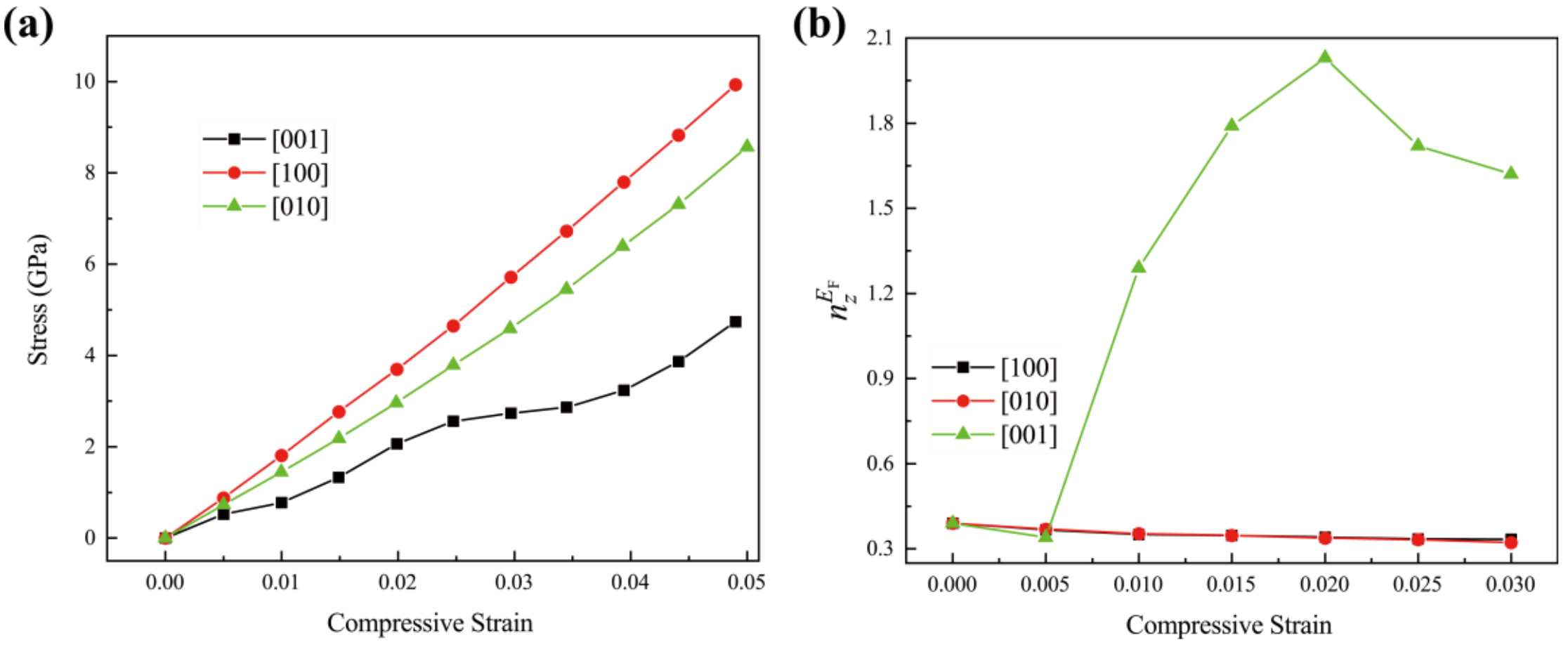

FIG S2. (a) First-principles-determined stress responses to compressive strain of *Amam* phase along [100], [010], and [001] directions at ambient pressure. (b) Evolution of the Ni 3$d_{z^2}$ DOS at fermi level and the compressive strain of *Amam* phase at ambient pressure along the above selected loading paths.

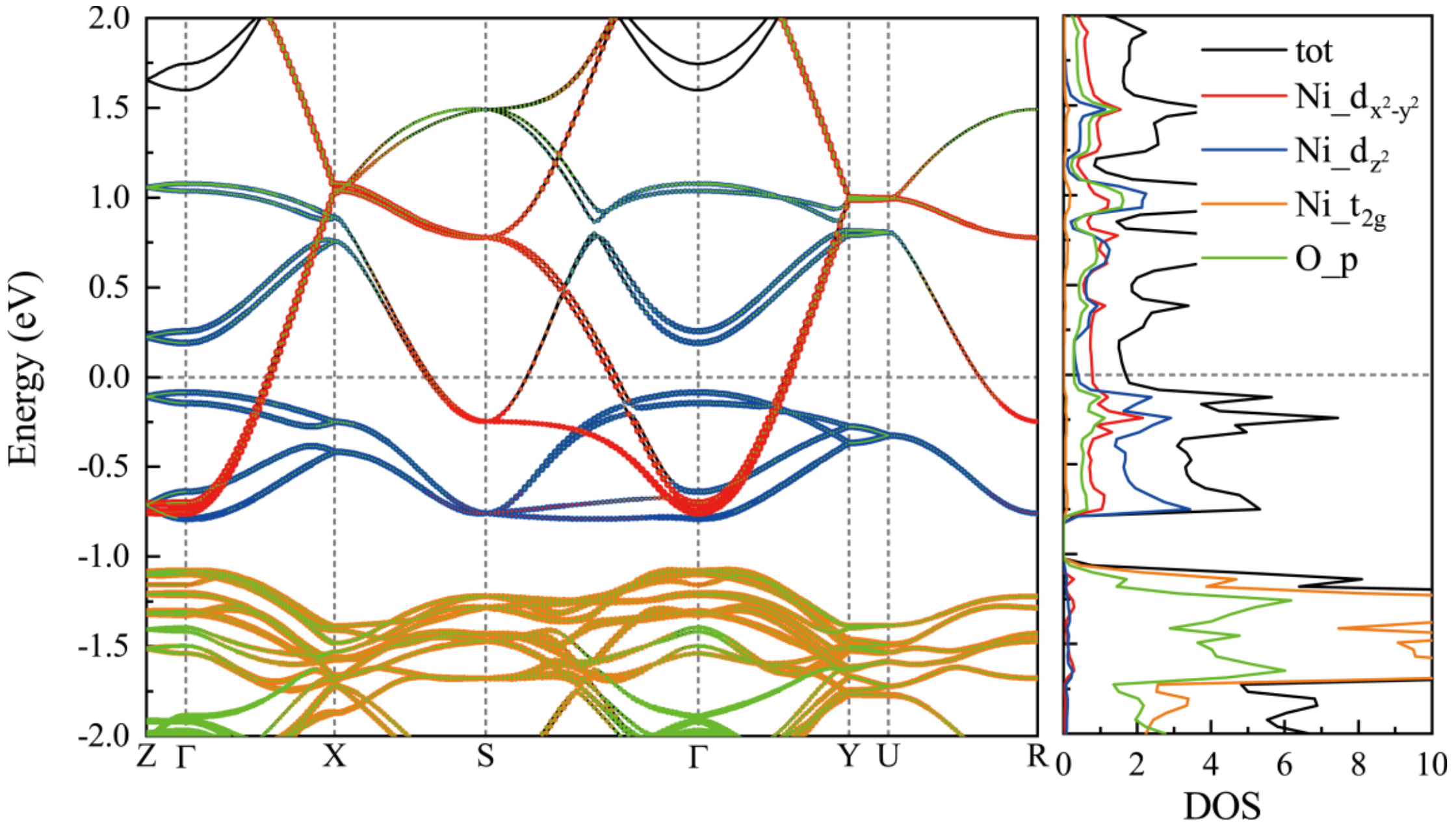

FIG S3. Projected electronic band structures of Ni cations and O anions of *Amam* phase without strain at ambient pressure. The corresponding DOS are shown on the right. The horizontal grey dash line represents the Fermi level.

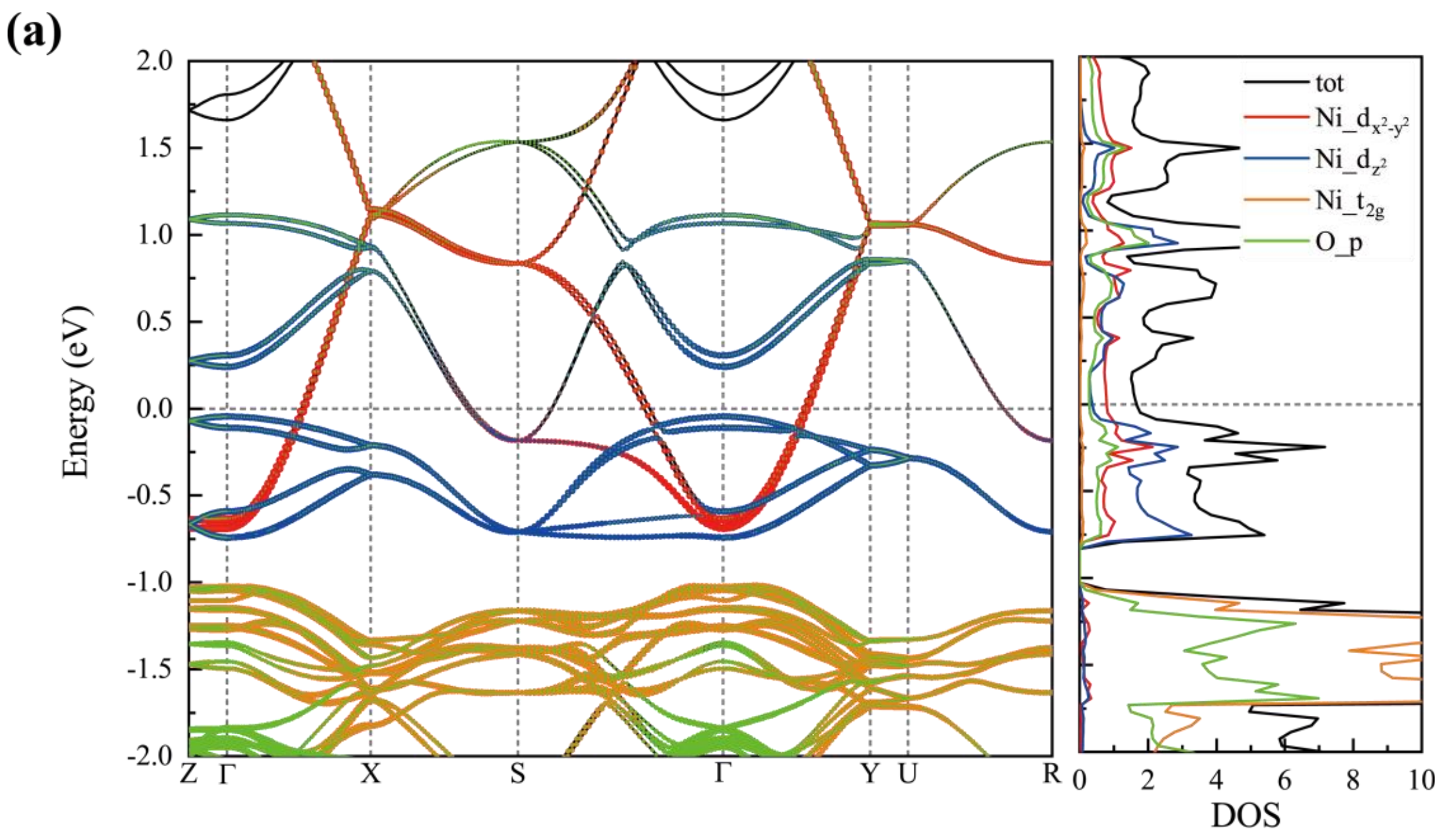

FIG S4. Projected electronic band structures of Ni cations and O anions of *Amam* phase at ambient pressure under compressive strain of (a) 0.005 and (b) 0.01 along [100] direction. The corresponding DOS are shown on the right. The horizontal grey dash line means the Fermi level.

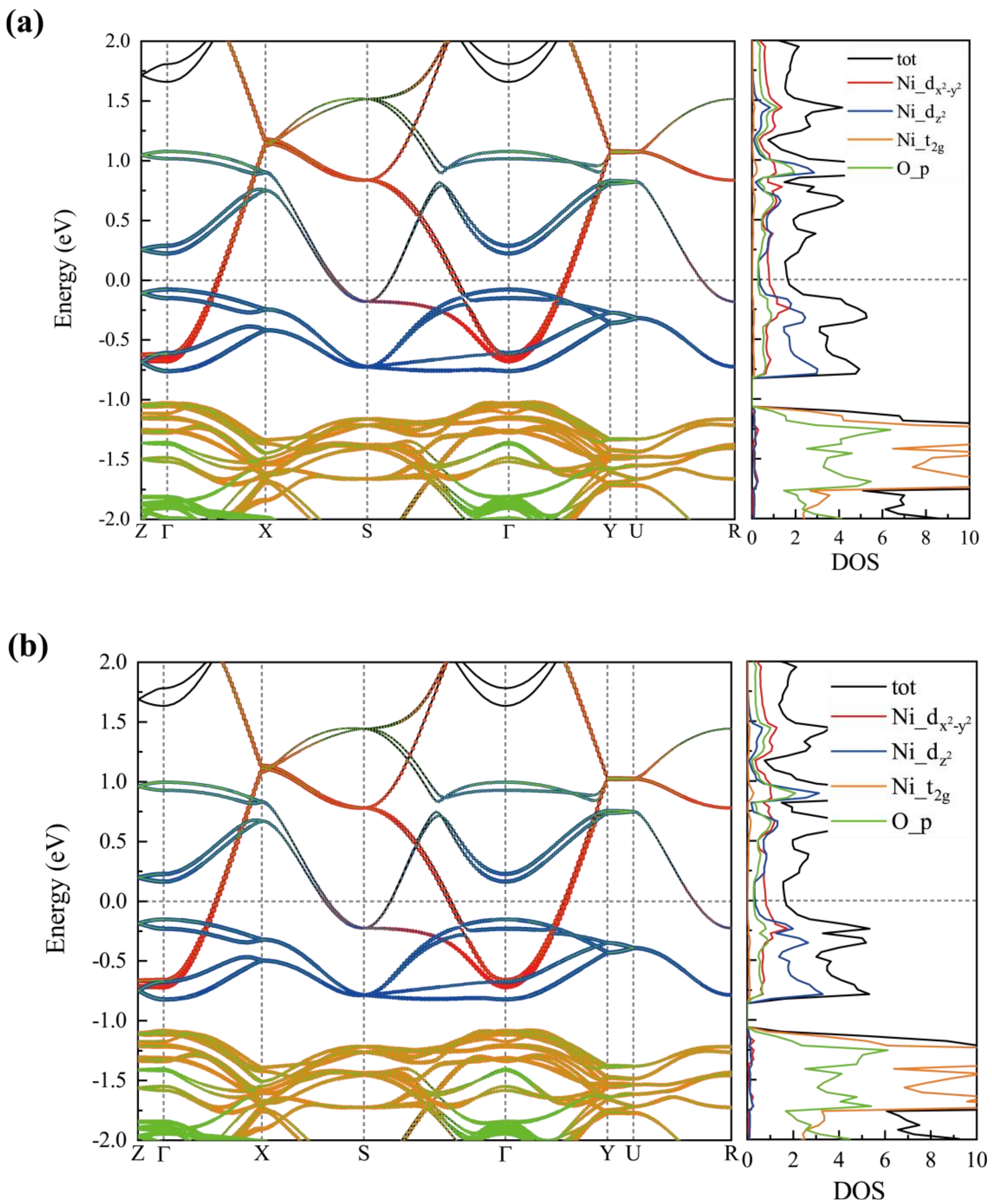

FIG S5. Projected electronic band structures of Ni cations and O anions of *Amam* phase at ambient pressure under compressive strain of (a) 0.015 and (b) 0.02 along [100] direction. The corresponding DOS are shown on the right. The horizontal grey dash line means the Fermi level.

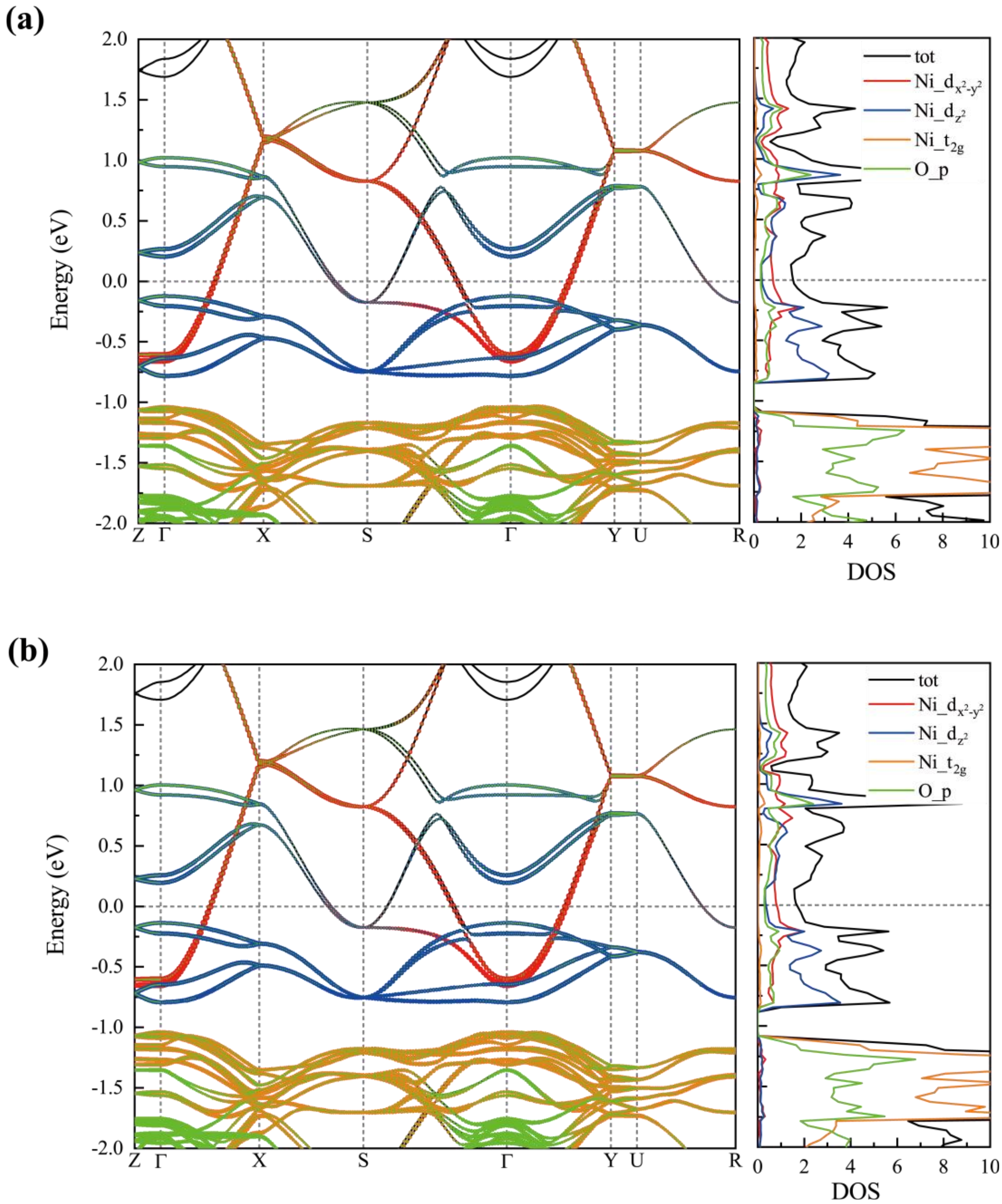

FIG S6. Projected electronic band structures of Ni cations and O anions of *Amam* phase at ambient pressure under compressive strain of (a) 0.025 and (b) 0.03 along [100] direction. The corresponding DOS are shown on the right. The horizontal grey dash line means the Fermi level.

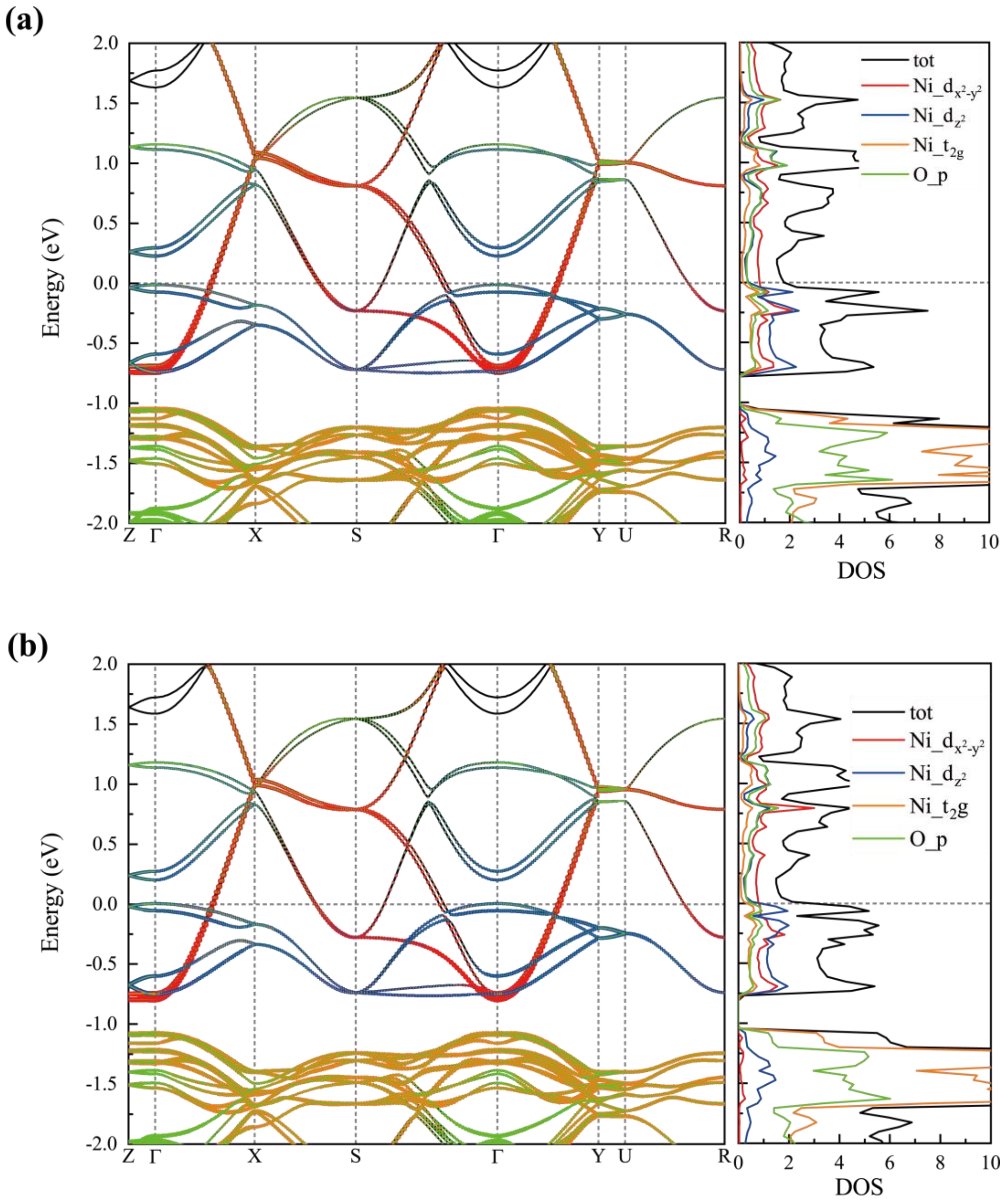

FIG S7. Projected electronic band structures of Ni cations and O anions of *Amam* phase at ambient pressure under compressive strain of (a) 0.005 and (b) 0.01 along [001] direction. The corresponding DOS are shown on the right. The horizontal grey dash line means the Fermi level.

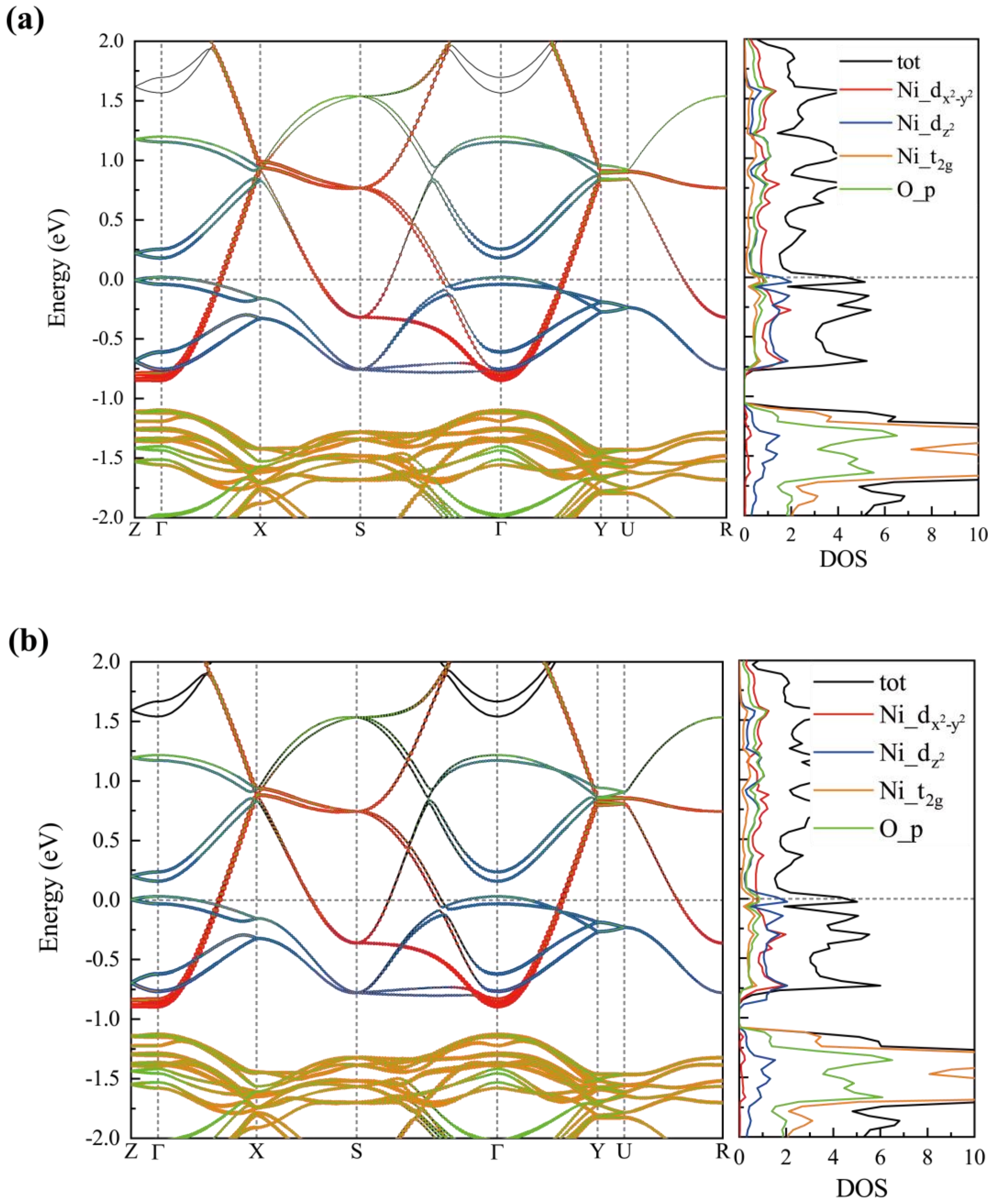

FIG S8. Projected electronic band structures of Ni cations and O anions of *Amam* phase at ambient pressure under compressive strain of (a) 0.015 and (b) 0.02 along [001] direction. The corresponding DOS are shown on the right. The horizontal grey dash line means the Fermi level.

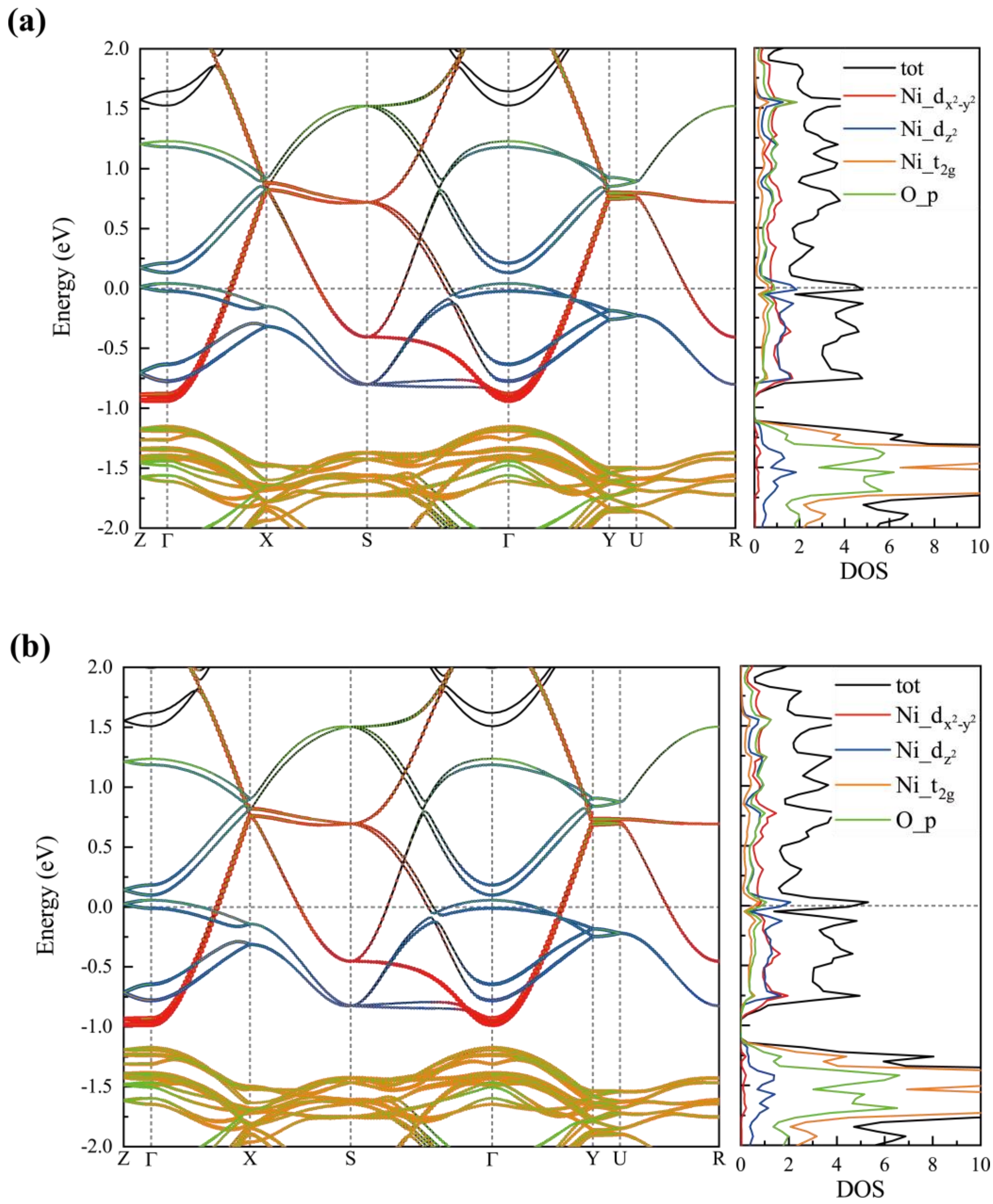

FIG S9. Projected electronic band structures of Ni cations and O anions of *Amam* phase at ambient pressure under compressive strain of (a) 0.025 and (b) 0.03 along [001] direction. The corresponding DOS are shown on the right. The horizontal grey dash line means the Fermi level.

## III. Stress response and electronic properties of *Amam* phase under various tensile strain at ambient pressure

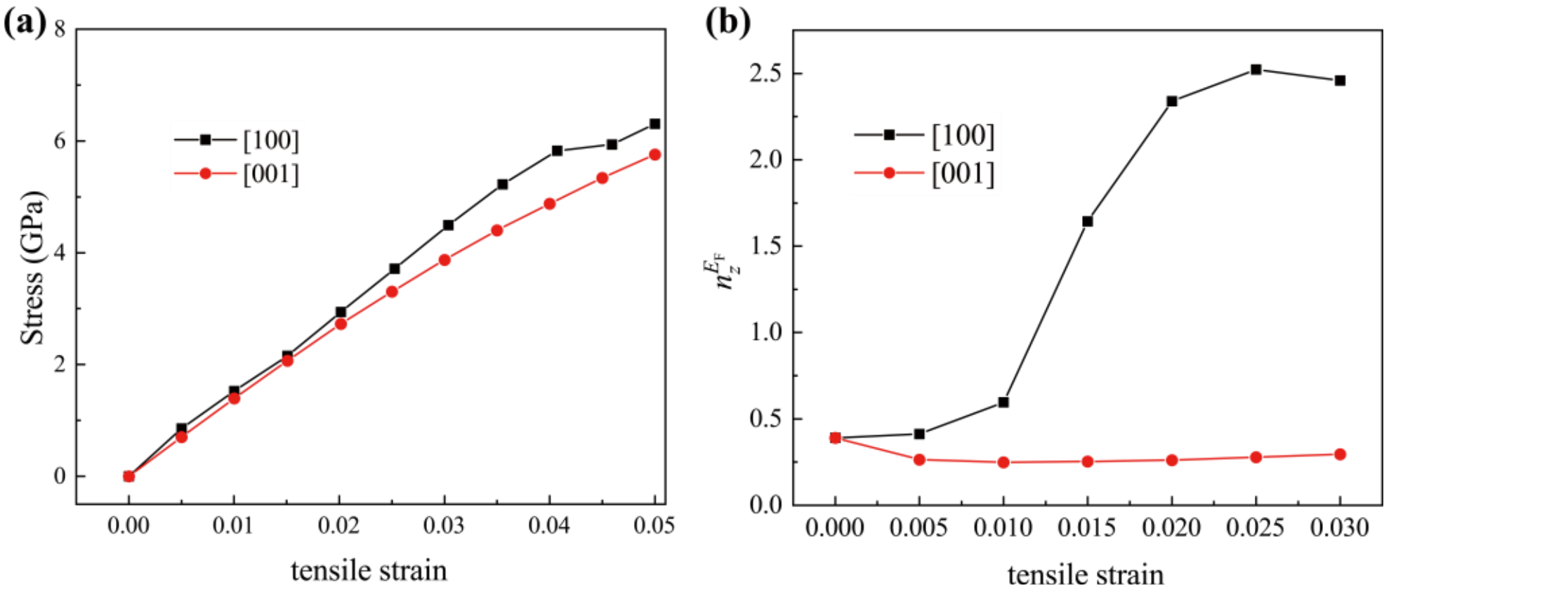

FIG S10. (a) First-principles-determined stress responses to tensile strain of *Amam* phase along [100] and [001] directions at ambient pressure. (b) Evolution of the Ni 3$d_{z^2}$ DOS at fermi level and the tensile strain of *Amam* phase at ambient pressure along the above selected loading paths.

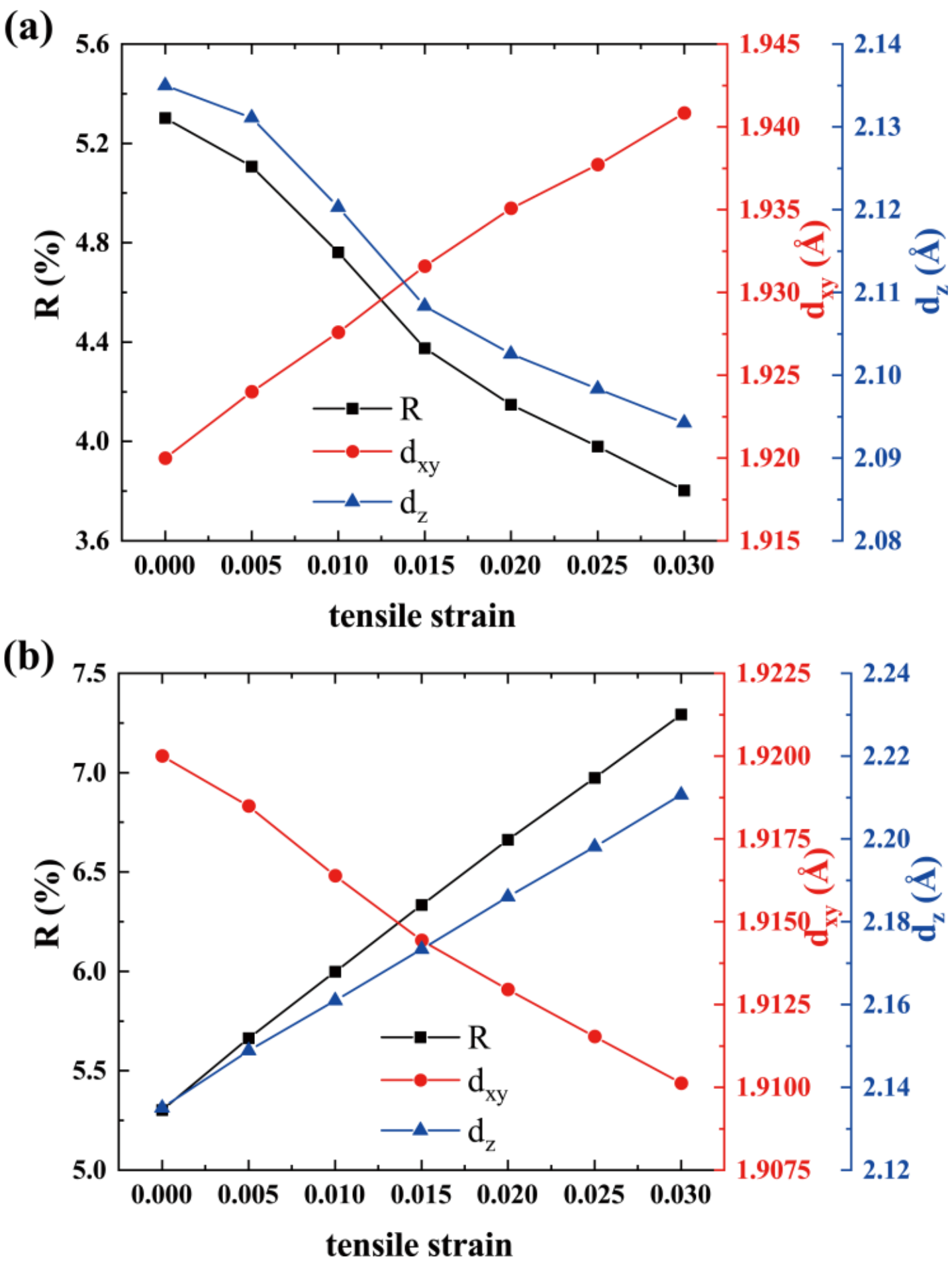

FIG S11. Evolution of the *R*, $d_{xy}$, and $d_z$ of *Amam* phase at ambient pressure with different tensile strain along (a) [100] and (b) [001] direction.

**IV. Stress response and electronic properties of *Amam* phase under various compressive strain at 5 GPa**

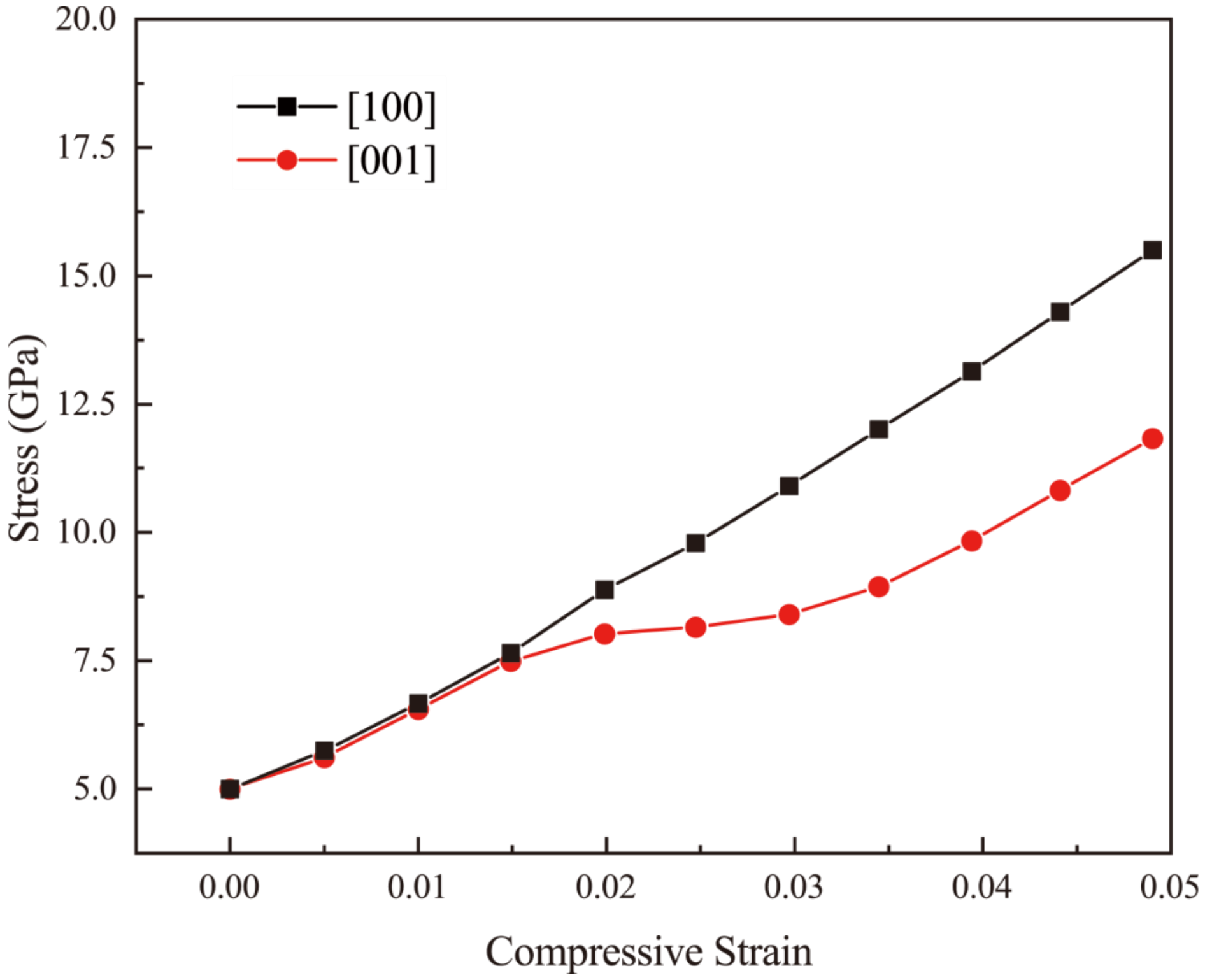

FIG S12. First-principles-determined stress responses to compressive strain of *Amam* phase along [001] and [100] directions at 5 GPa.

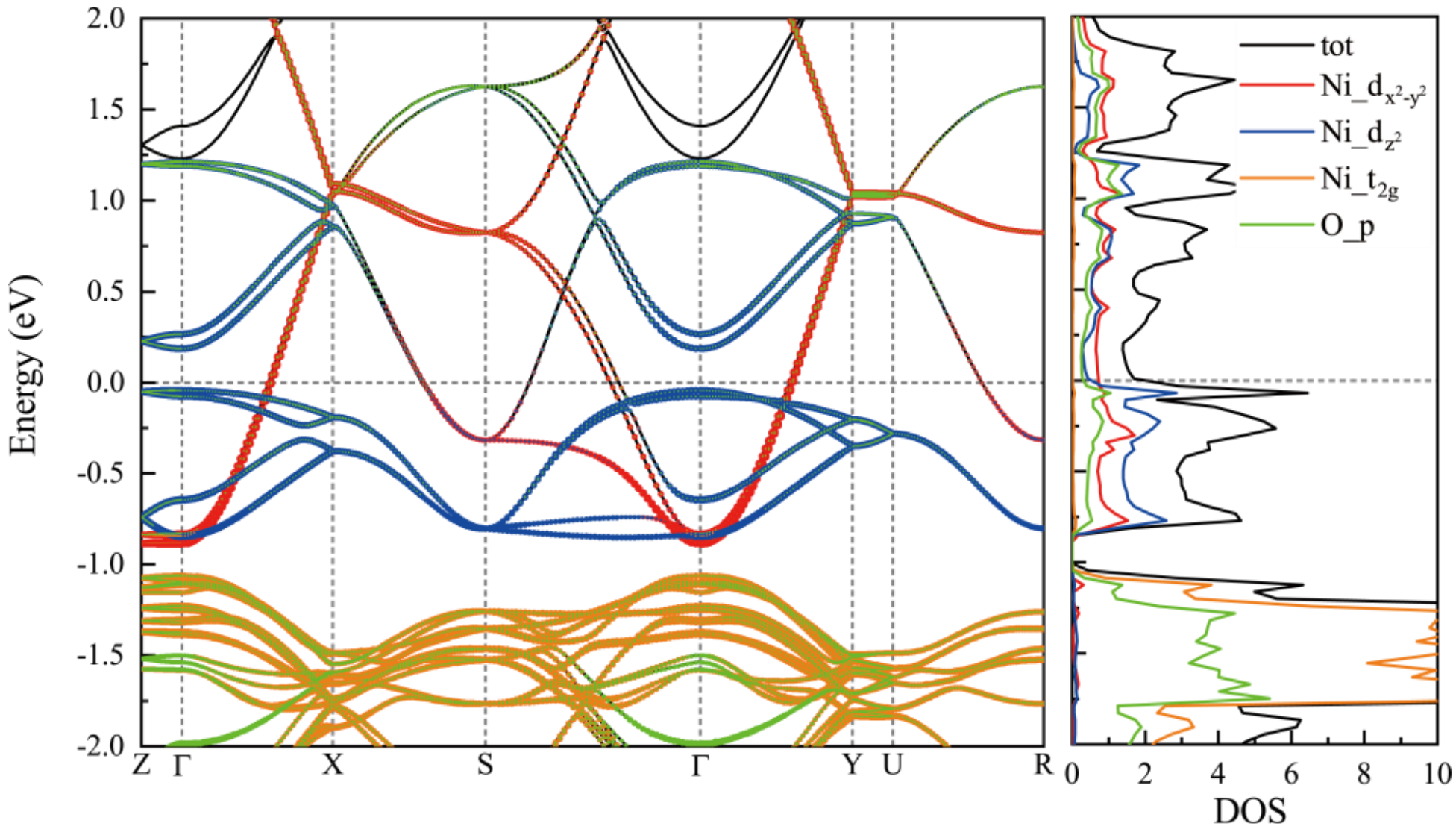

FIG S13. Projected electronic band structures of Ni cations and O anions of *Amam* phase without strain at 5 GPa. The corresponding DOS are shown on the right. The horizontal grey dash line represents the Fermi level.

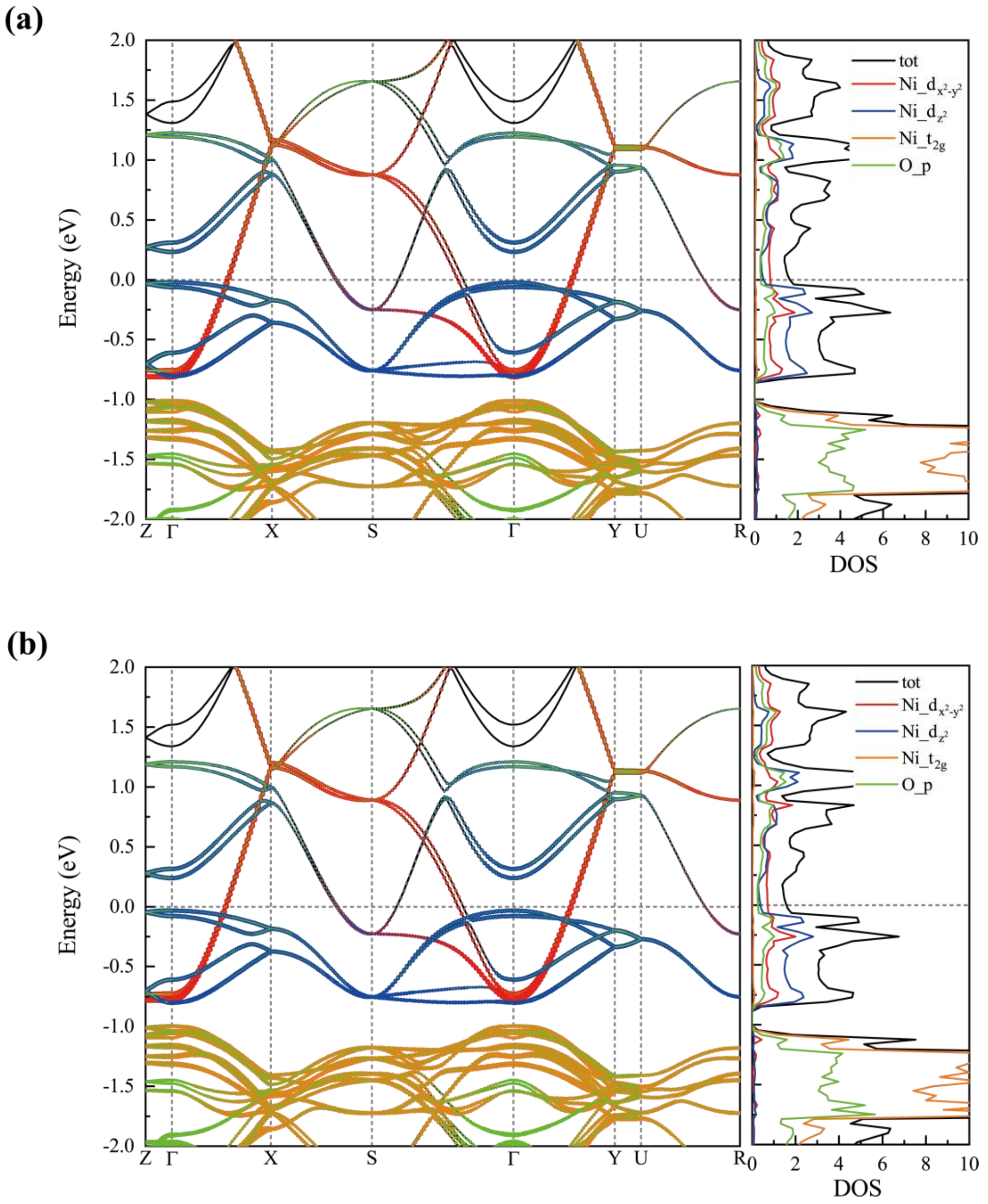

FIG S14. Projected electronic band structures of Ni cations and O anions of *Amam* phase at 5 GPa under compressive strain of (a) 0.005 and (b) 0.01 along [100] direction. The corresponding DOS are shown on the right. The horizontal grey dash line means the Fermi level.

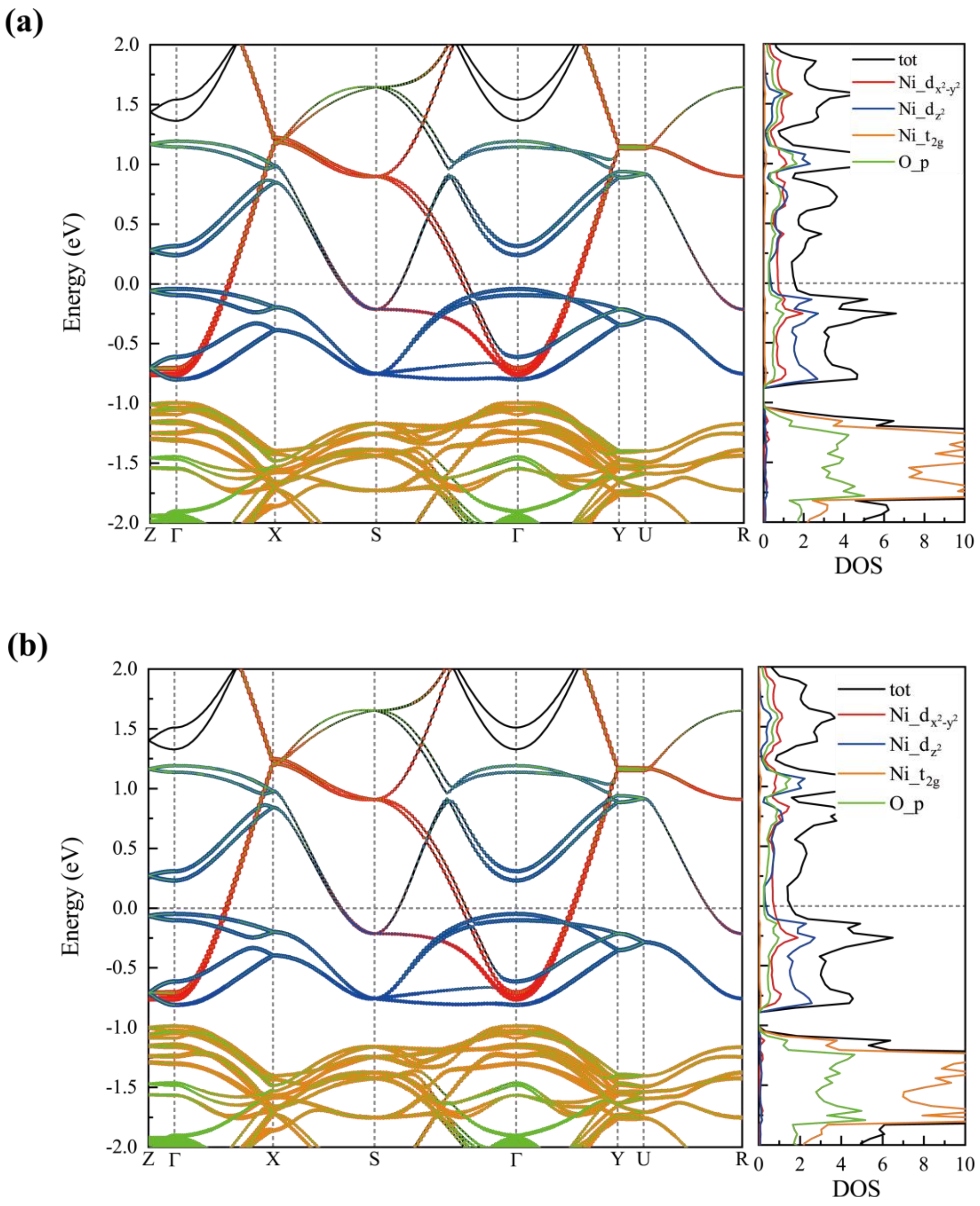

FIG S15. Projected electronic band structures of Ni cations and O anions of *Amam* phase at 5 GPa under compressive strain of (a) 0.015 and (b) 0.02 along [100] direction. The corresponding DOS are shown on the right. The horizontal grey dash line means the Fermi level.

(a)

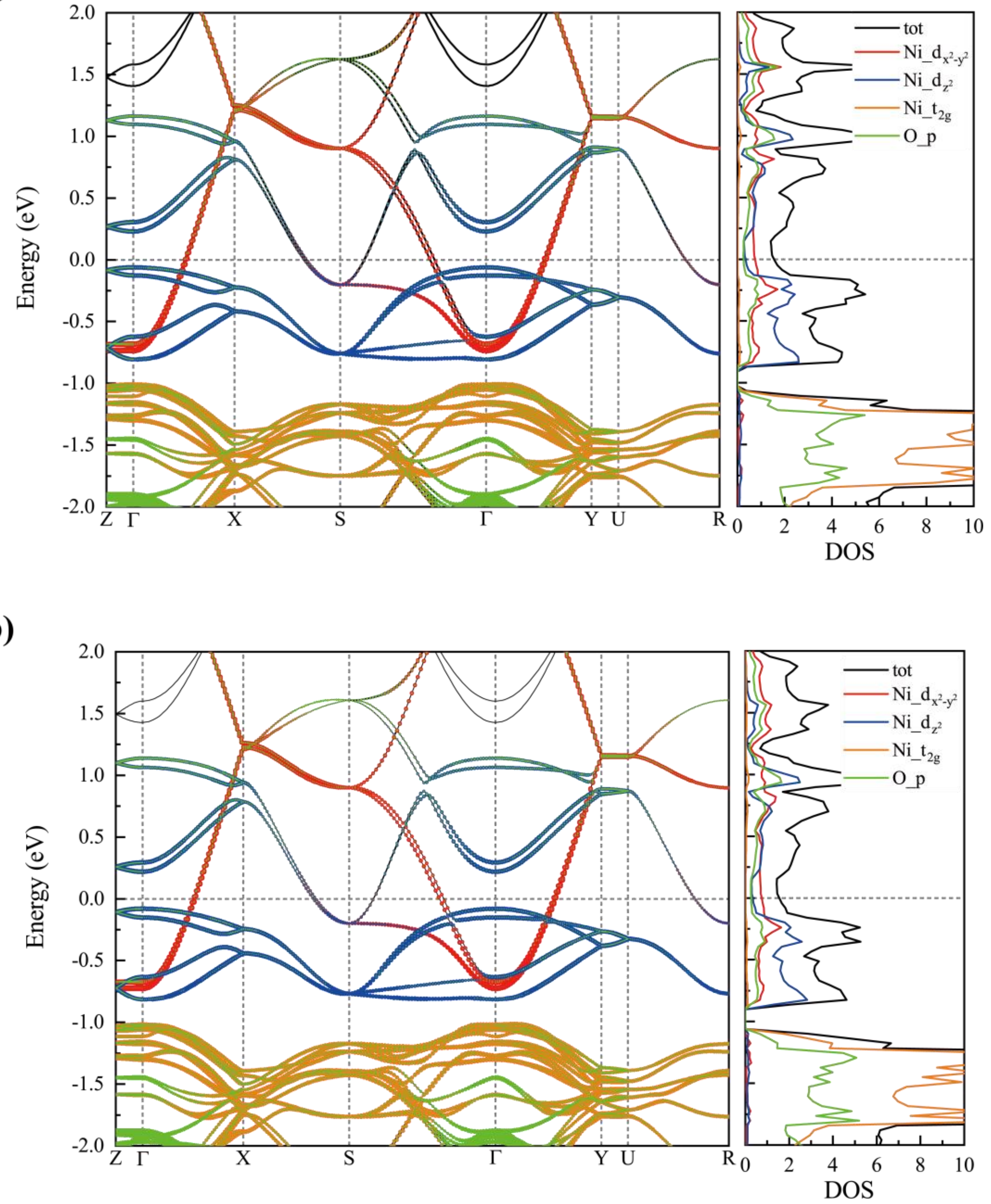

FIG S16. Projected electronic band structures of Ni cations and O anions of *Amam* phase at 5 GPa under compressive strain of (a) 0.025 and (b) 0.03 along [100] direction. The corresponding DOS are shown on the right. The horizontal grey dash line means the Fermi level.

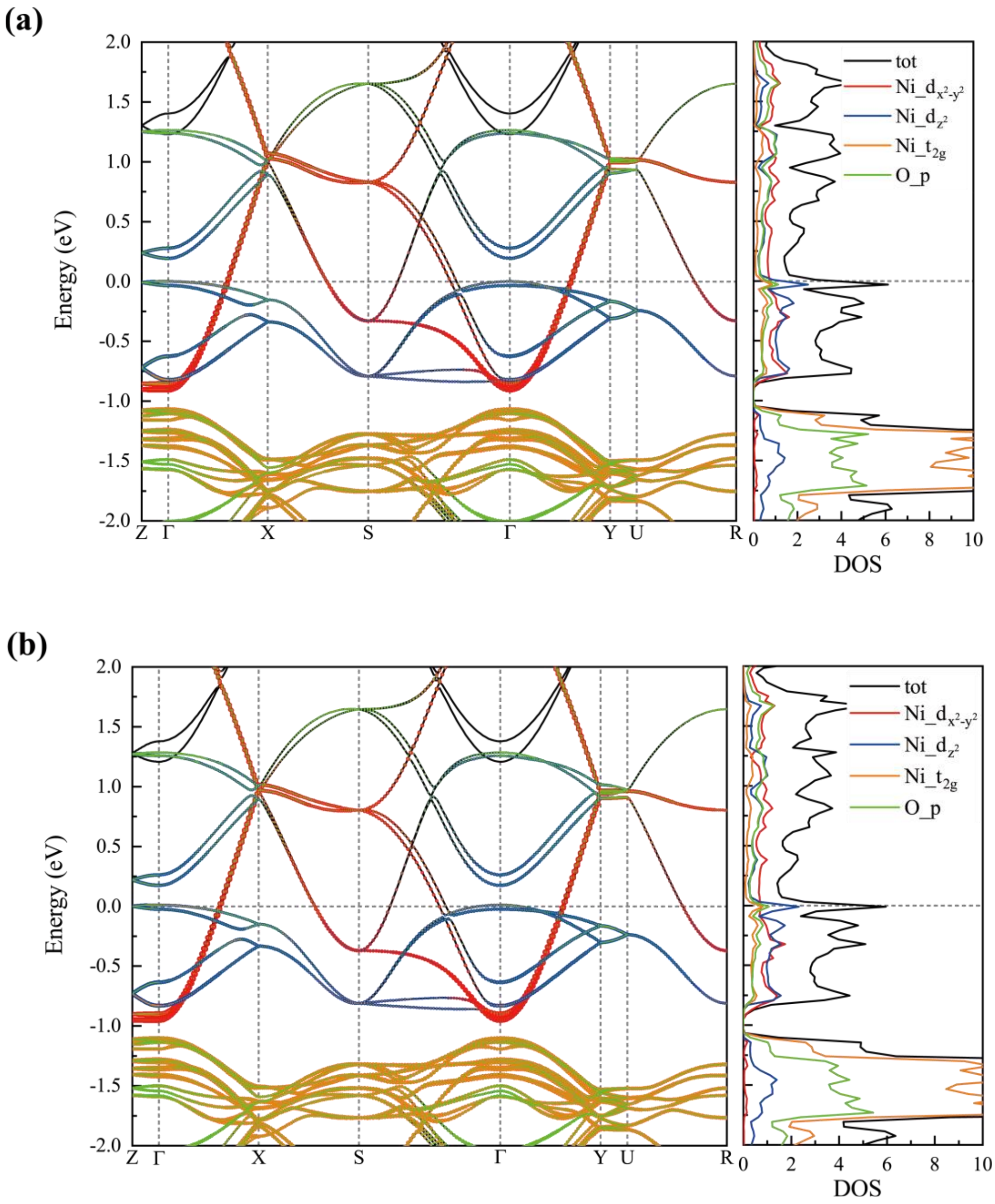

FIG S17. Projected electronic band structures of Ni cations and O anions of *Amam* phase at 5 GPa under compressive strain of (a) 0.005 and (b) 0.01 along [001] direction. The corresponding DOS are shown on the right. The horizontal grey dash line means the Fermi level.

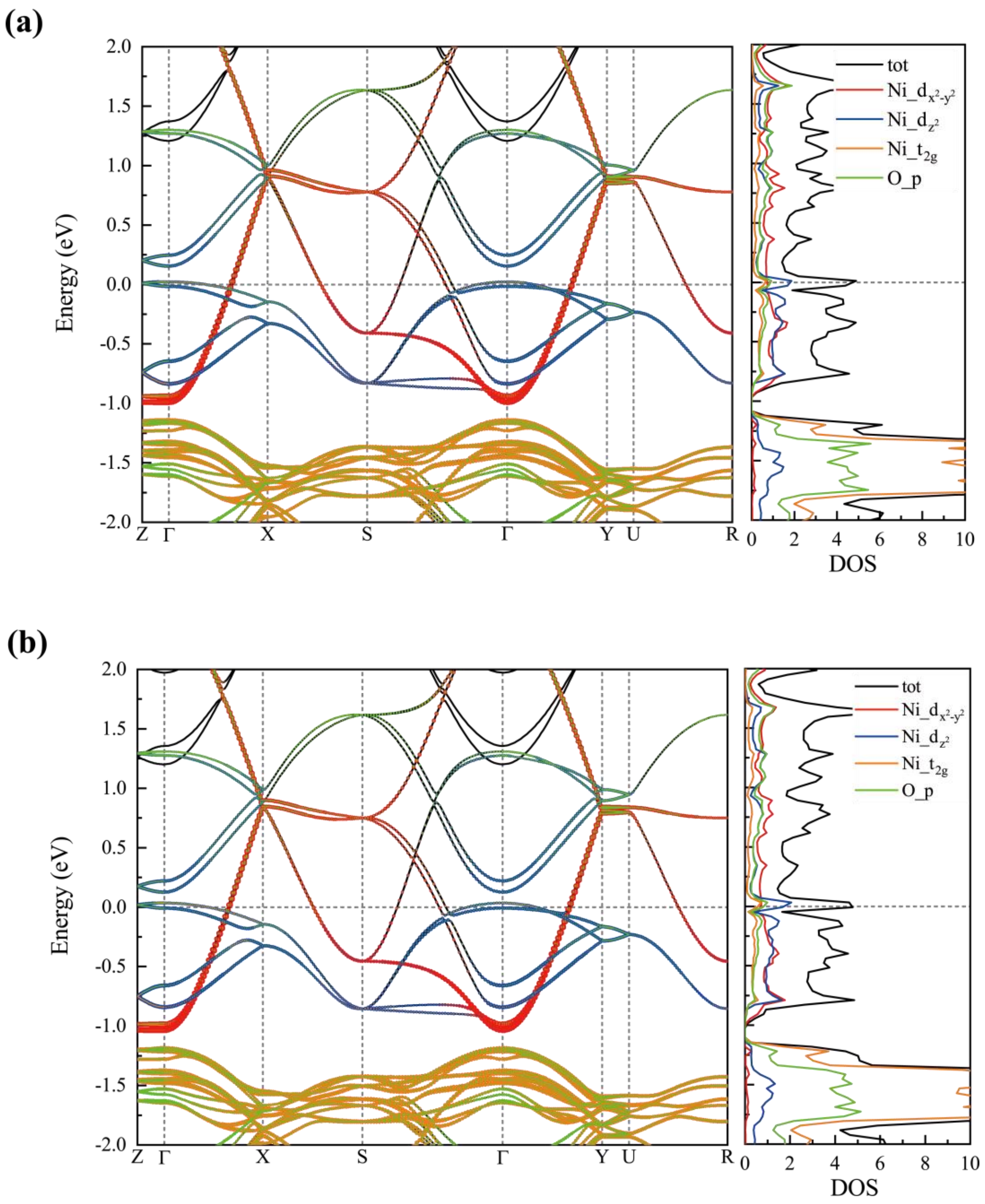

FIG S18. Projected electronic band structures of Ni cations and O anions of *Amam* phase at 5 GPa under compressive strain of (a) 0.015 and (b) 0.02 along [001] direction. The corresponding DOS are shown on the right. The horizontal grey dash line means the Fermi level.

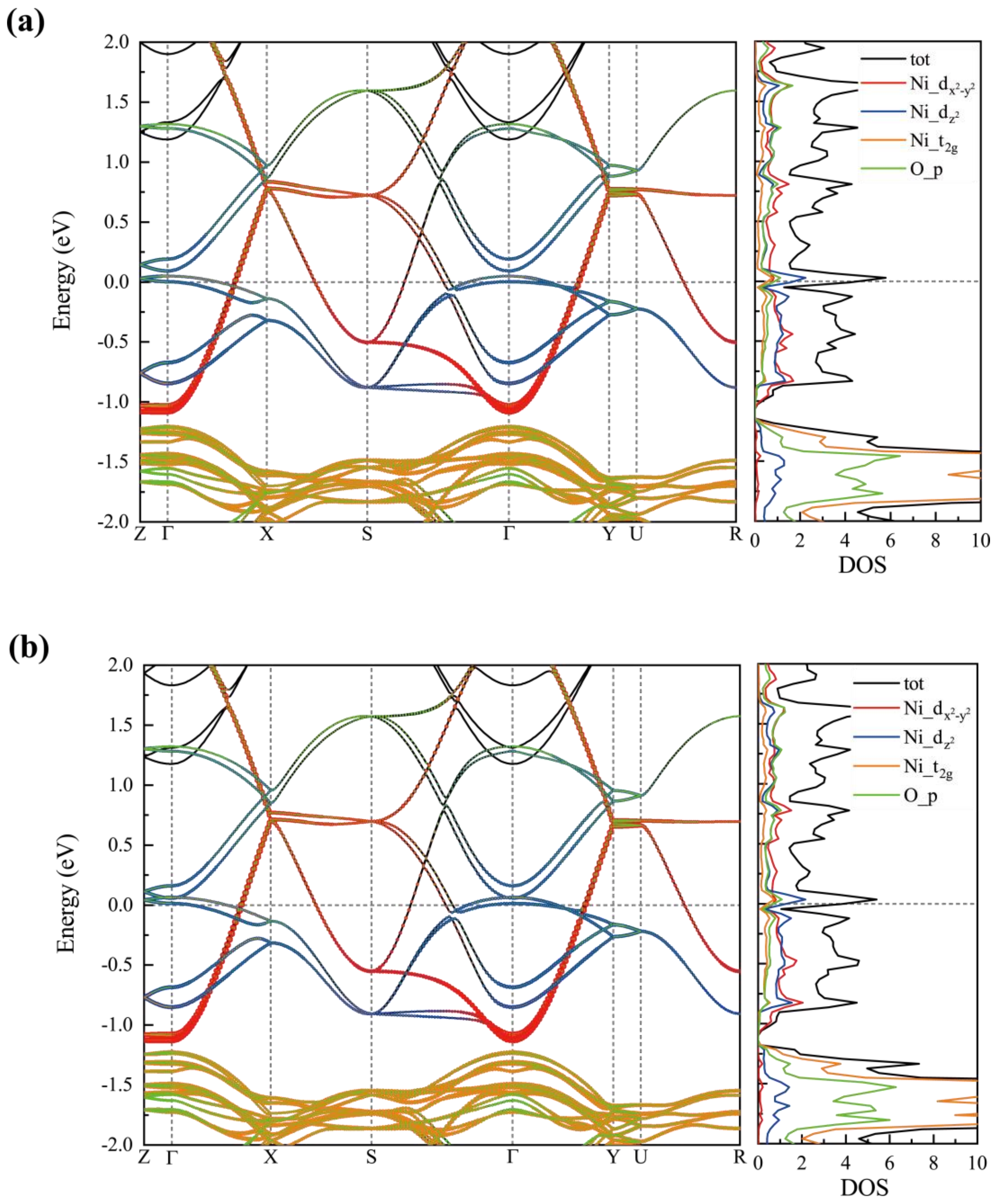

FIG S19. Projected electronic band structures of Ni cations and O anions of *Amam* phase at 5 GPa under compressive strain of (a) 0.025 and (b) 0.03 along [001] direction. The corresponding DOS are shown on the right. The horizontal grey dash line means the Fermi level.

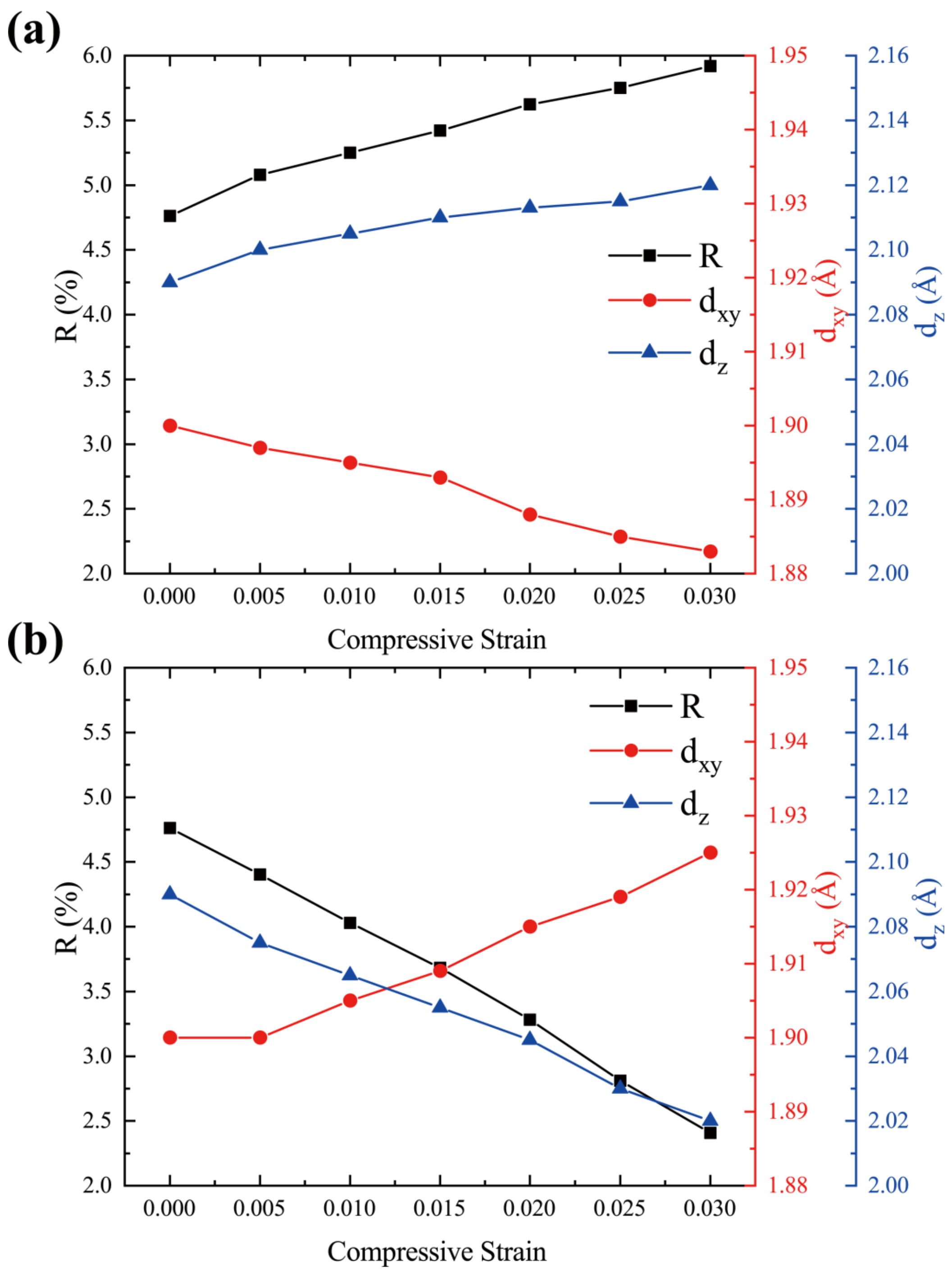

FIG S20. Evolution of the $R$, $d_{xy}$, and $d_z$ of *Amam* phase at 5 GPa with different compressive strain along (a) [100] and (b) [001] direction.

**V. Stress response and electronic properties of *I*4/*mmm* phase under various compressive strain at 15 GPa**

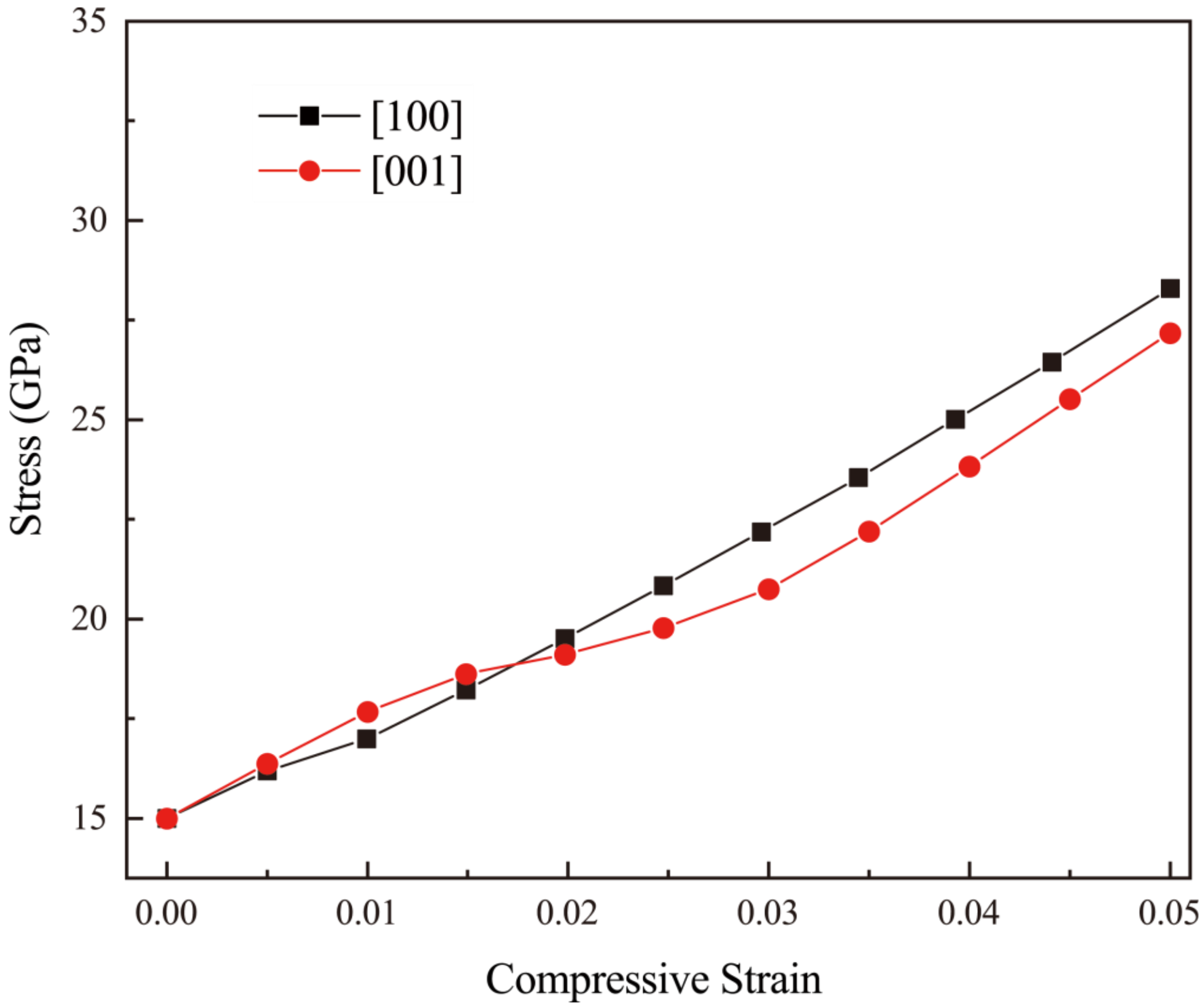

FIG S21. First-principles-determined stress responses to compressive strain of *I*4/*mmm* phase along [001] and [100] directions at 15 GPa.

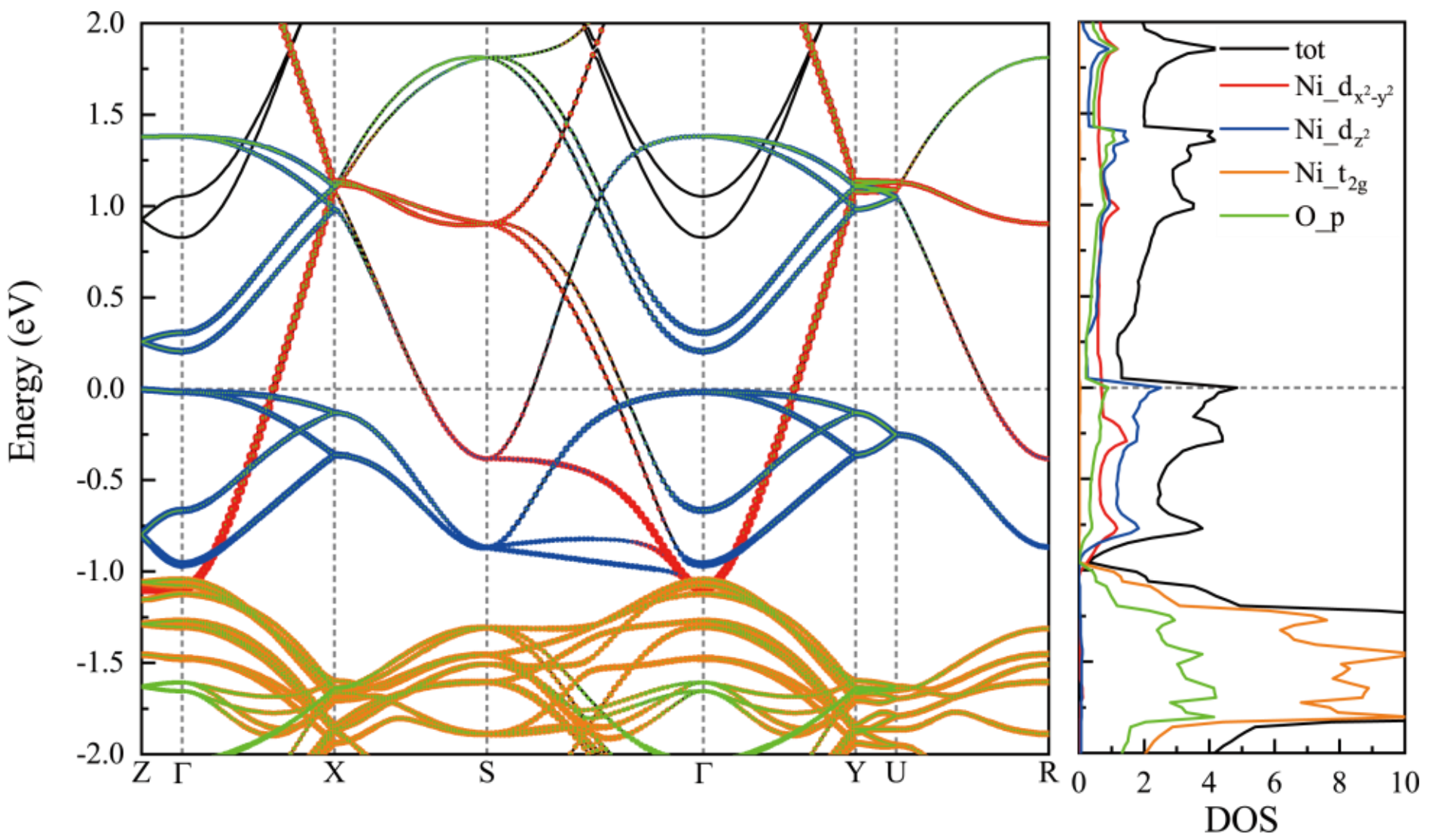

FIG S22. Projected electronic band structures of Ni cations and O anions of *I*4/*mmm* phase without strain at 15 GPa. The corresponding DOS are shown on the right. The horizontal grey dash line represents the Fermi level.

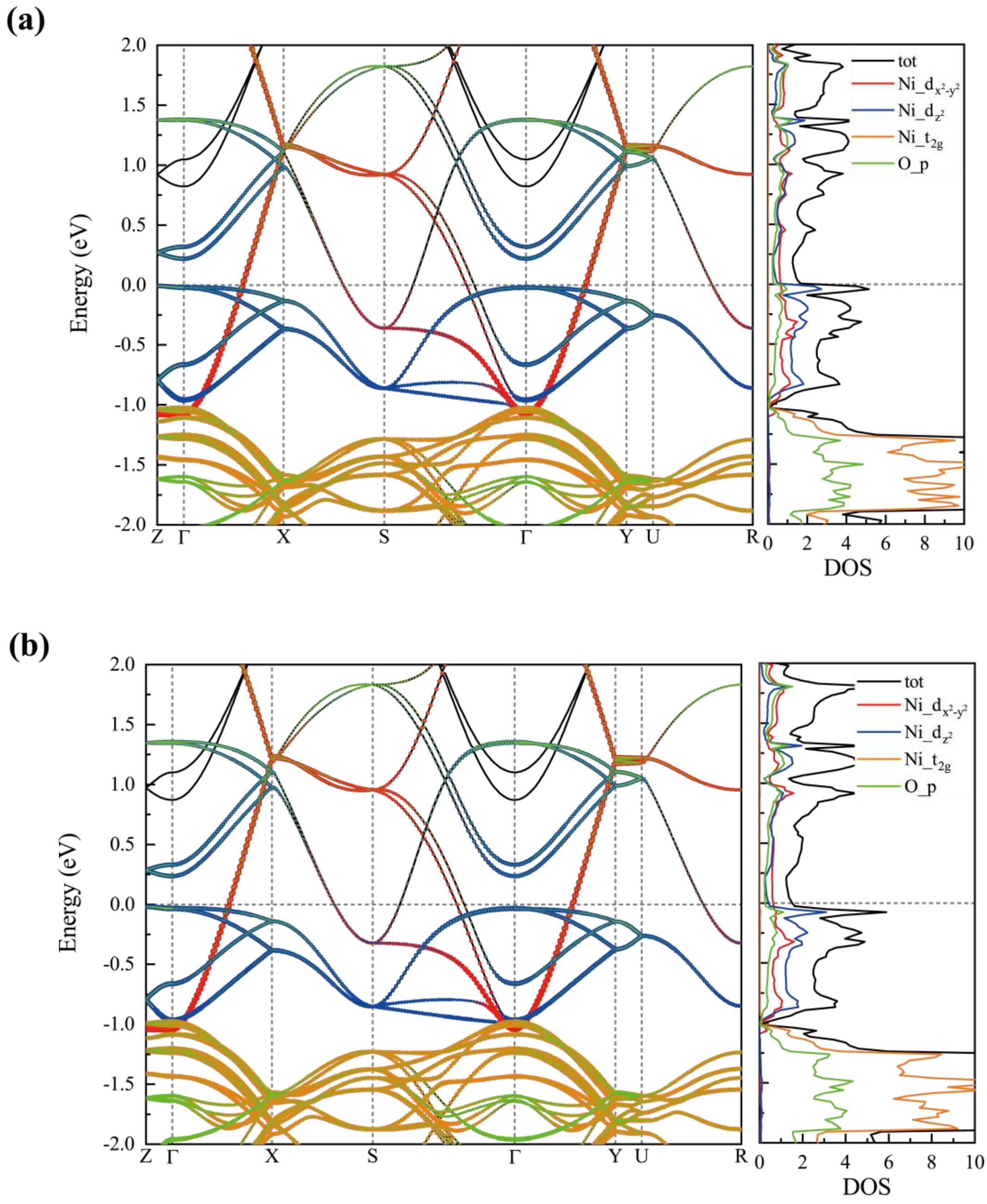

FIG S23. Projected electronic band structures of Ni cations and O anions of *I*4/*mmm* phase at 15 GPa under compressive strain of (a) 0.005 and (b) 0.01 along [100] direction. The corresponding DOS are shown on the right. The horizontal grey dash line means the Fermi level.

**(a)**

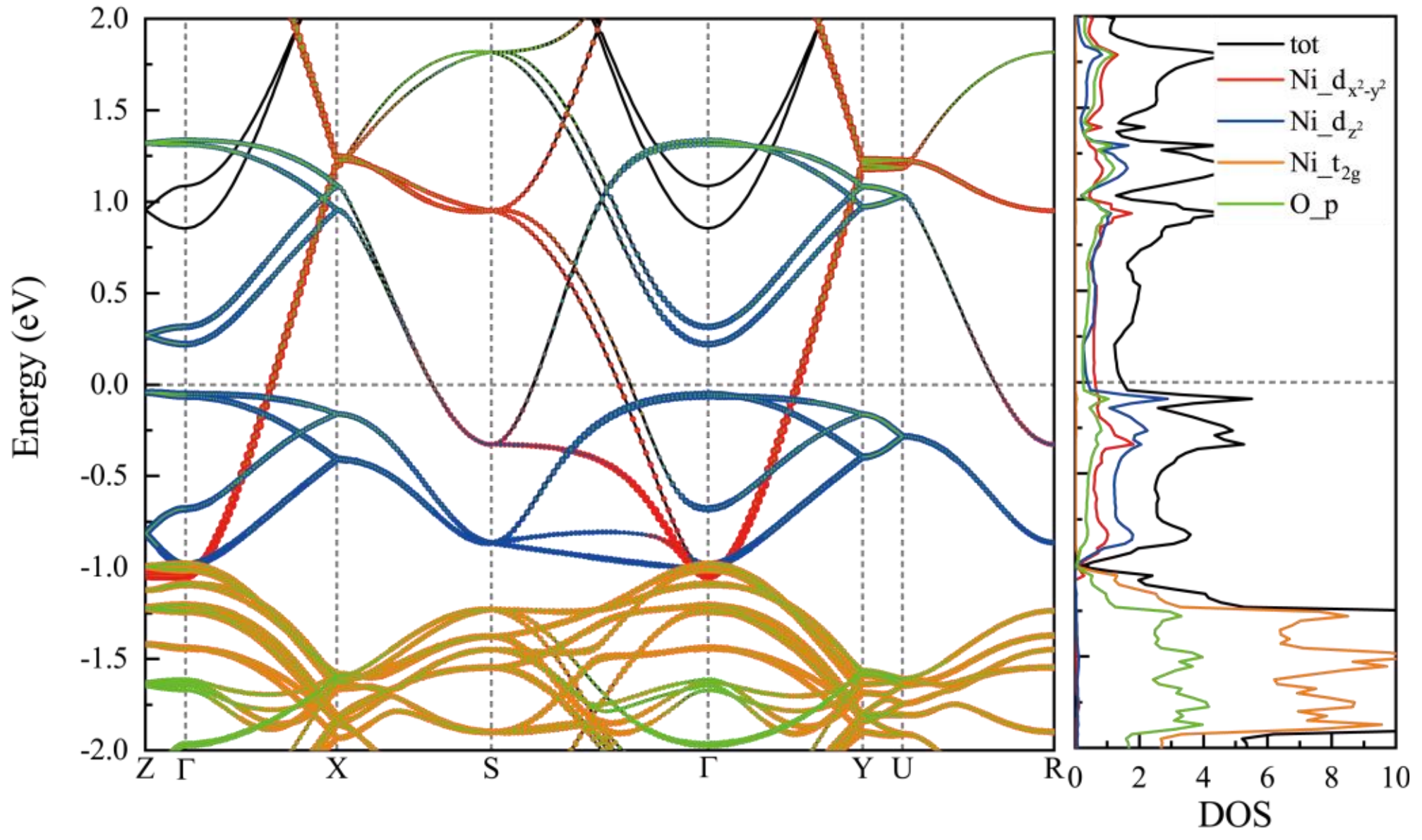

**(b)**

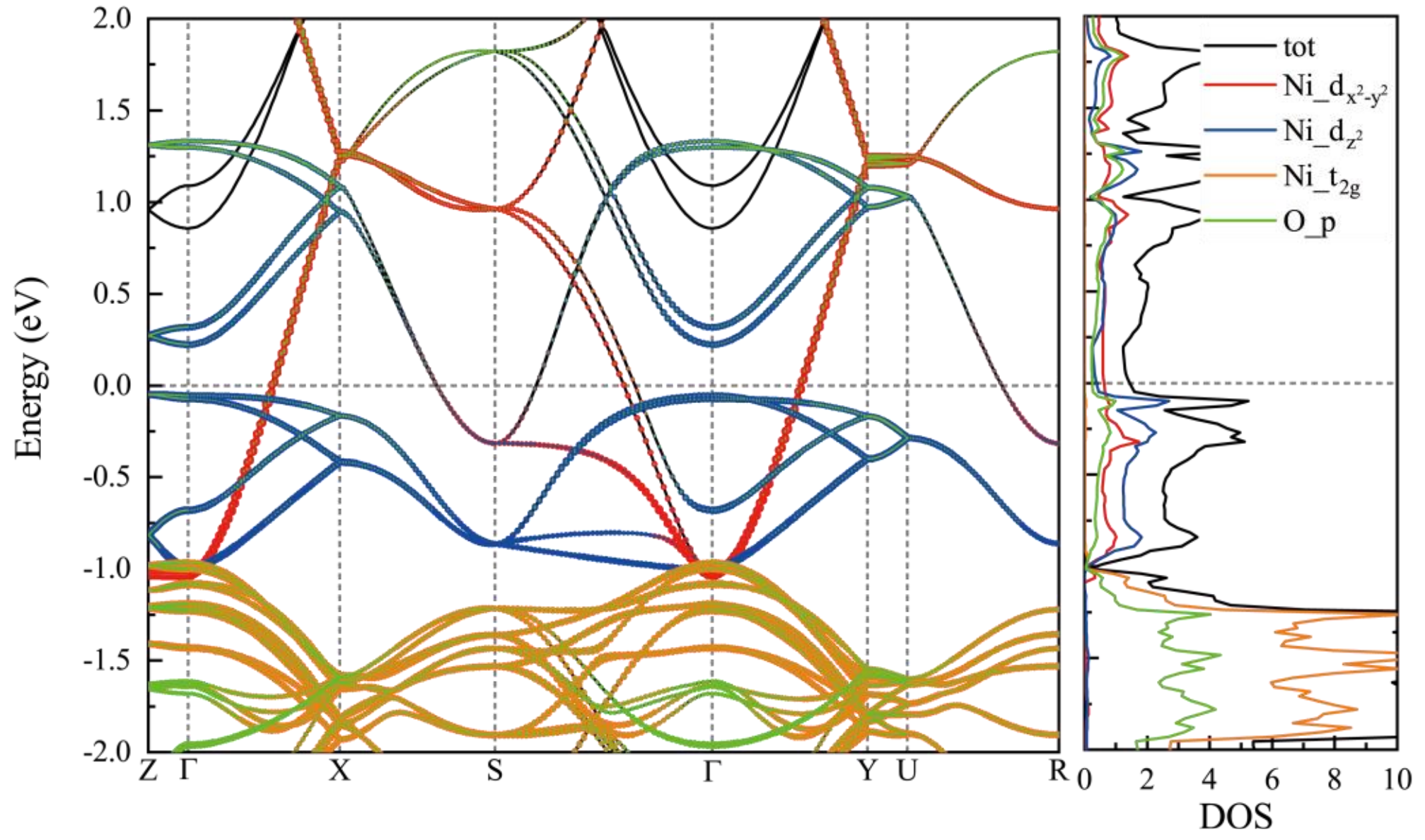

FIG S24. Projected electronic band structures of Ni cations and O anions of *I*4/*mmm* phase at 15 GPa under compressive strain of (a) 0.015 and (b) 0.02 along [100] direction. The corresponding DOS are shown on the right. The horizontal grey dash line means the Fermi level.

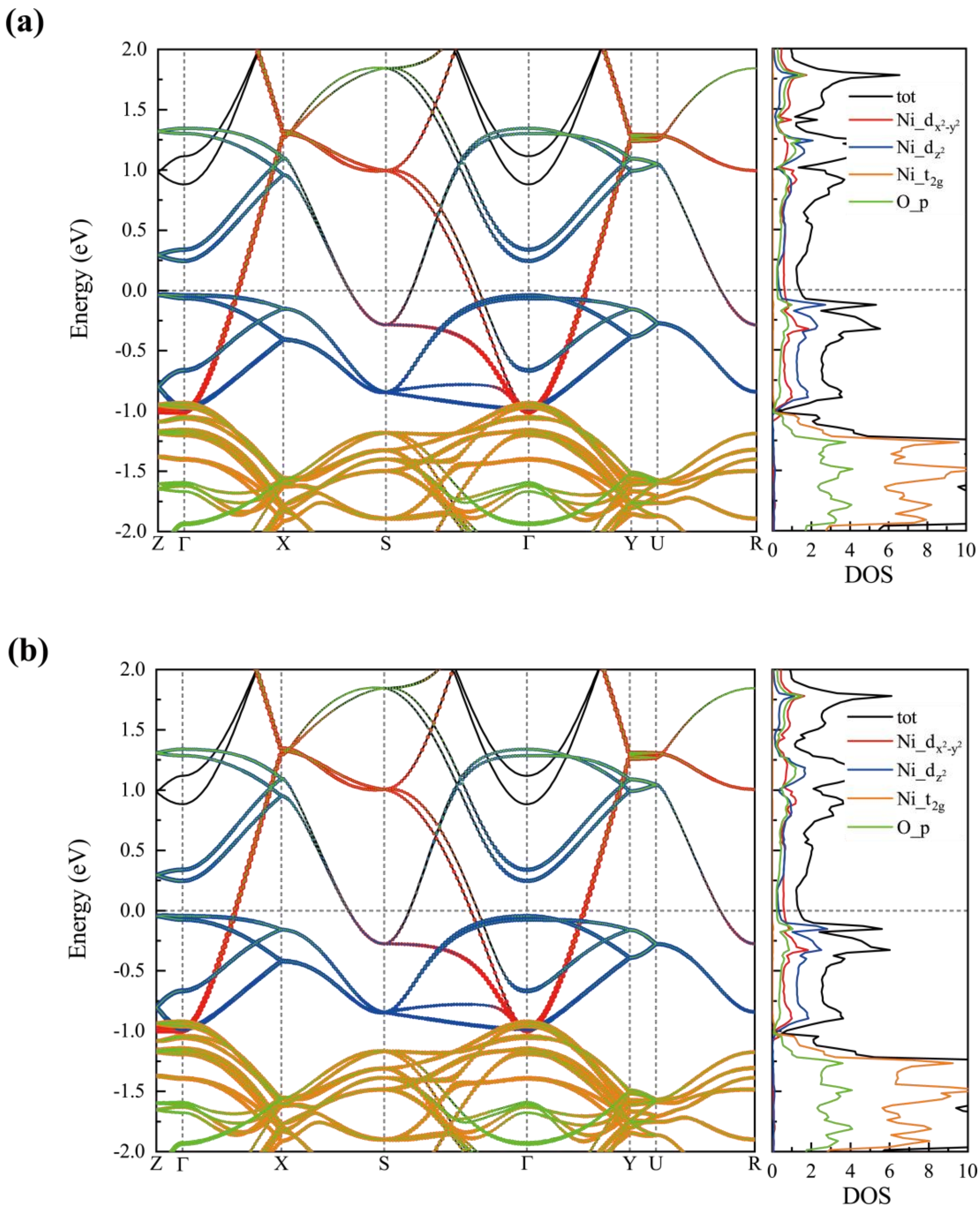

FIG S25. Projected electronic band structures of Ni cations and O anions of *I*4/*mmm* phase at 15 GPa under compressive strain of (a) 0.025 and (b) 0.03 along [100] direction. The corresponding DOS are shown on the right. The horizontal grey dash line means the Fermi level.

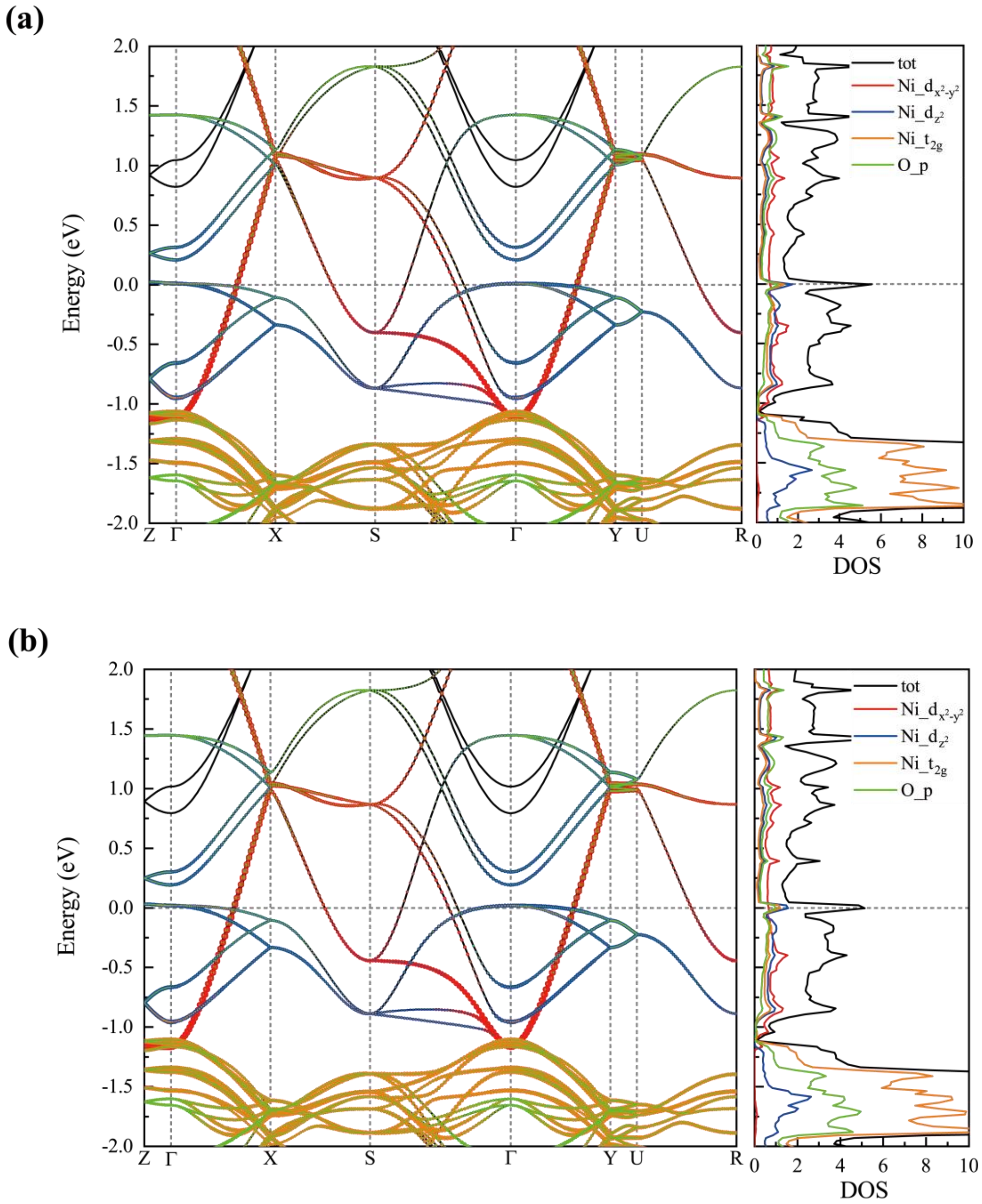

FIG S26. Projected electronic band structures of Ni cations and O anions of *I*4/*mmm* phase at 15 GPa under compressive strain of (a) 0.005 and (b) 0.01 along [001] direction. The corresponding DOS are shown on the right. The horizontal grey dash line means the Fermi level.

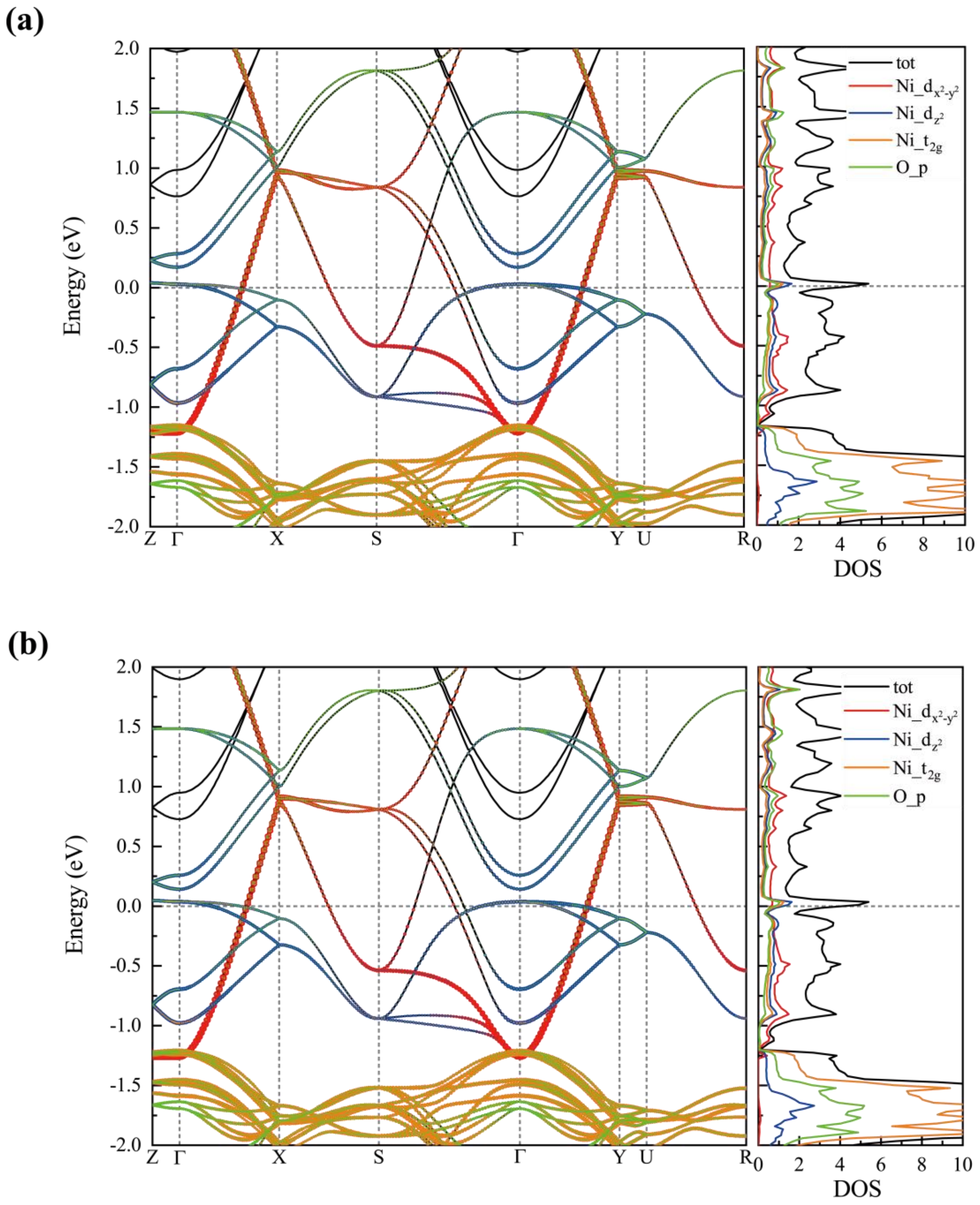

FIG S27. Projected electronic band structures of Ni cations and O anions of *I*4/*mmm* phase at 15 GPa under compressive strain of (a) 0.015 and (b) 0.02 along [001] direction. The corresponding DOS are shown on the right. The horizontal grey dash line means the Fermi level.

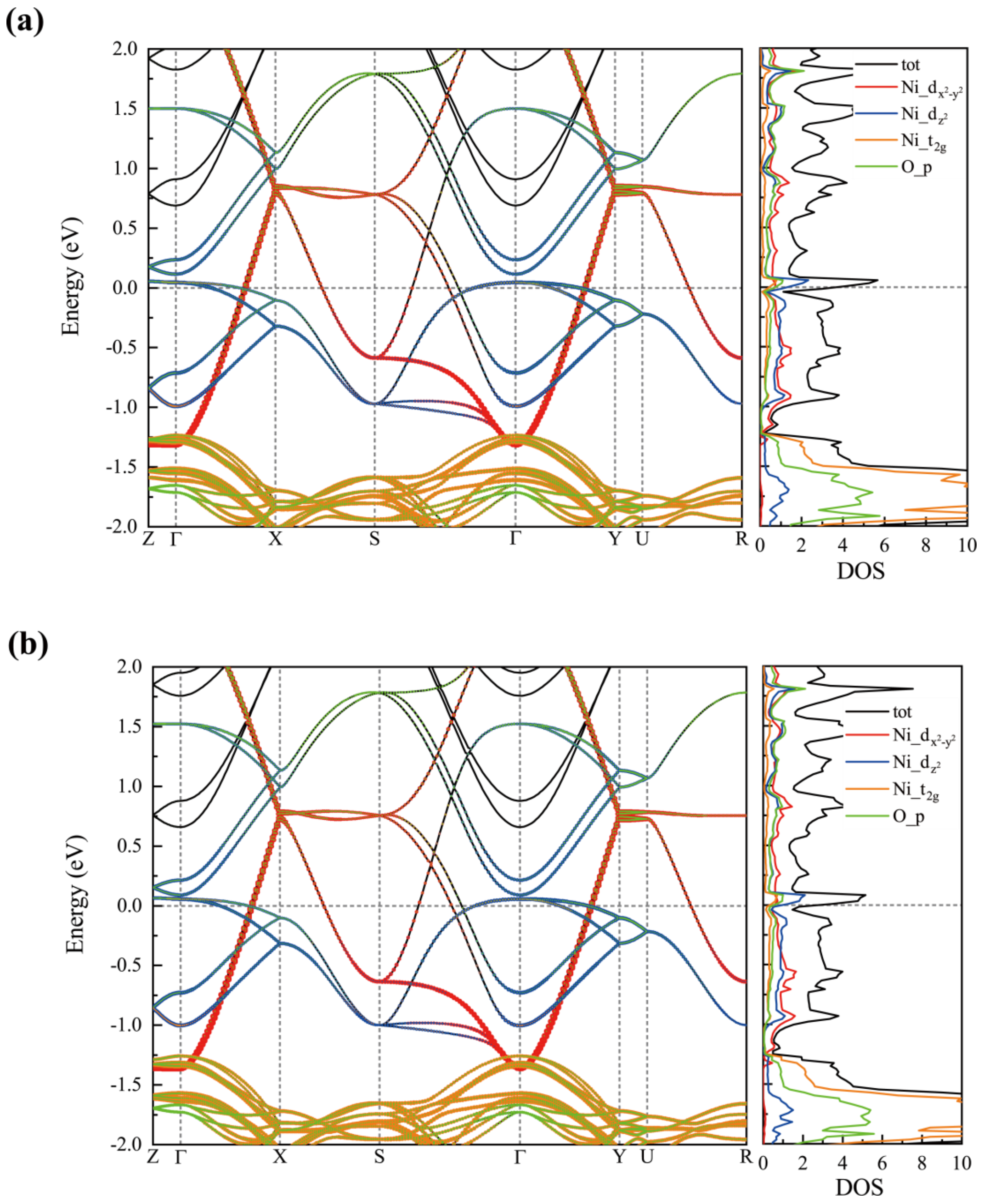

FIG S28. Projected electronic band structures of Ni cations and O anions of *I*4/*mmm* phase at 15 GPa under compressive strain of (a) 0.025 and (b) 0.03 along [001] direction. The corresponding DOS are shown on the right. The horizontal grey dash line means the Fermi level.

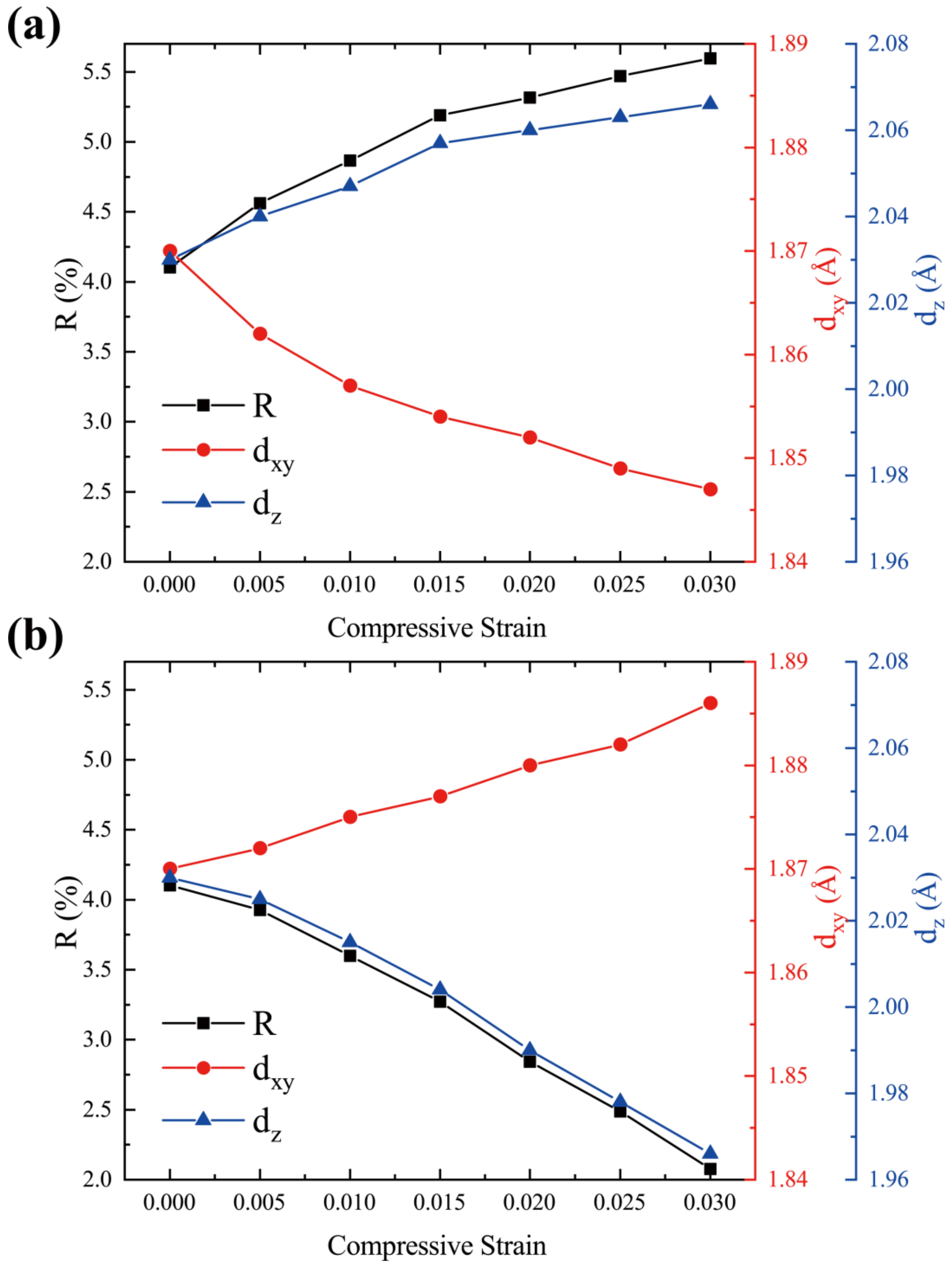

FIG S29. Evolution of the $R$, $d_{xy}$, and $d_z$ of $I4/mmm$ phase at 15 GPa with different compressive strain along (a) [100] and (b) [001] direction.

**VI. Stress response and electronic properties of *I*4/*mmm* phase under various compressive strain at 30 GPa**

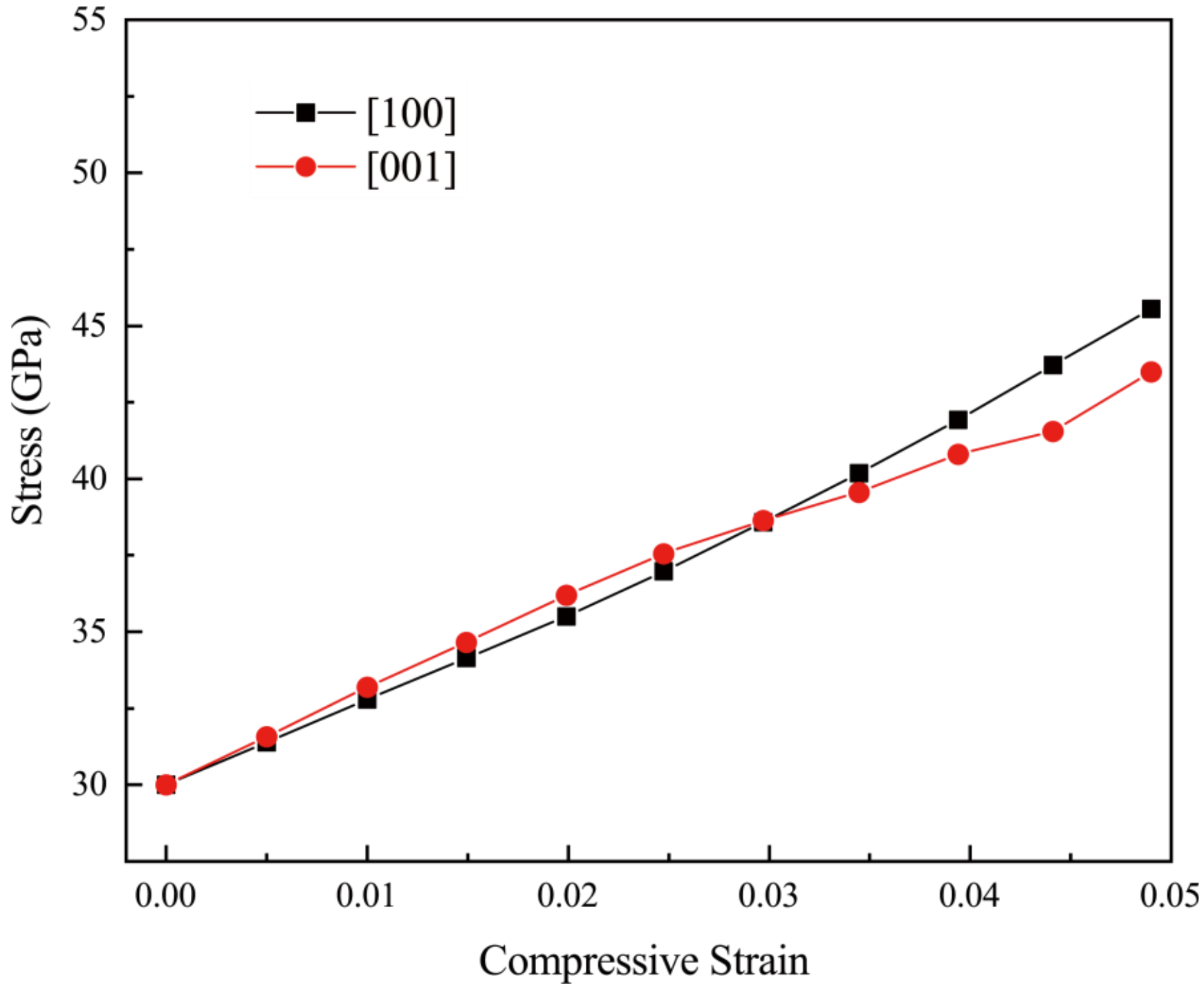

FIG S30. First-principles-determined stress responses to compressive strain of *I*4/*mmm* along [001] and [100] directions at 30 GPa.

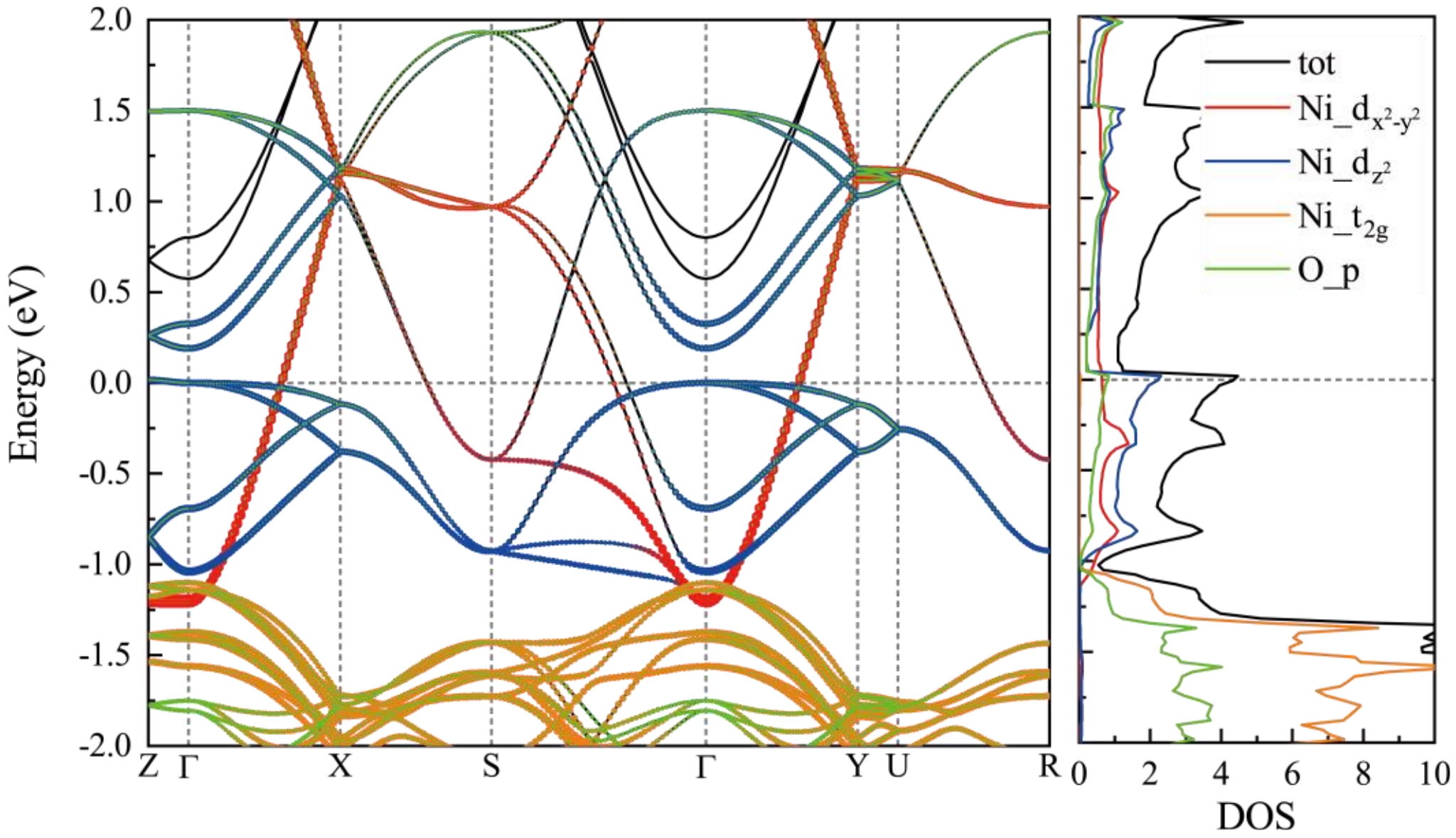

FIG S31. Projected electronic band structures of Ni cations and O anions of *I*4/*mmm* phase without strain at 30 GPa. The corresponding DOS are shown on the right. The horizontal grey dash line represents the Fermi level.

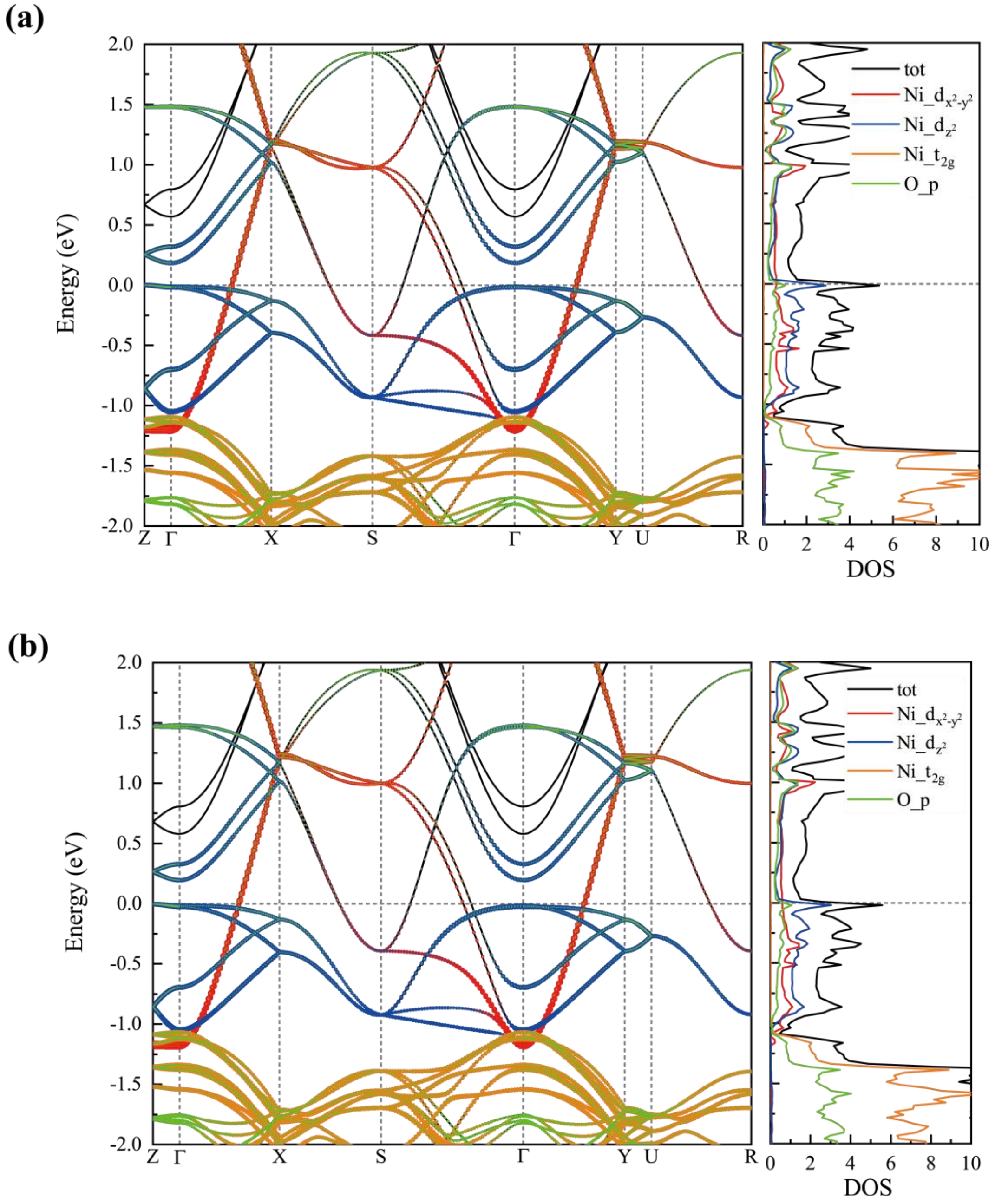

FIG S32. Projected electronic band structures of Ni cations and O anions of *I*4/*mmm* phase at 30 GPa under compressive strain of (a) 0.005 and (b) 0.01 along [100] direction. The corresponding DOS are shown on the right. The horizontal grey dash line means the Fermi level.

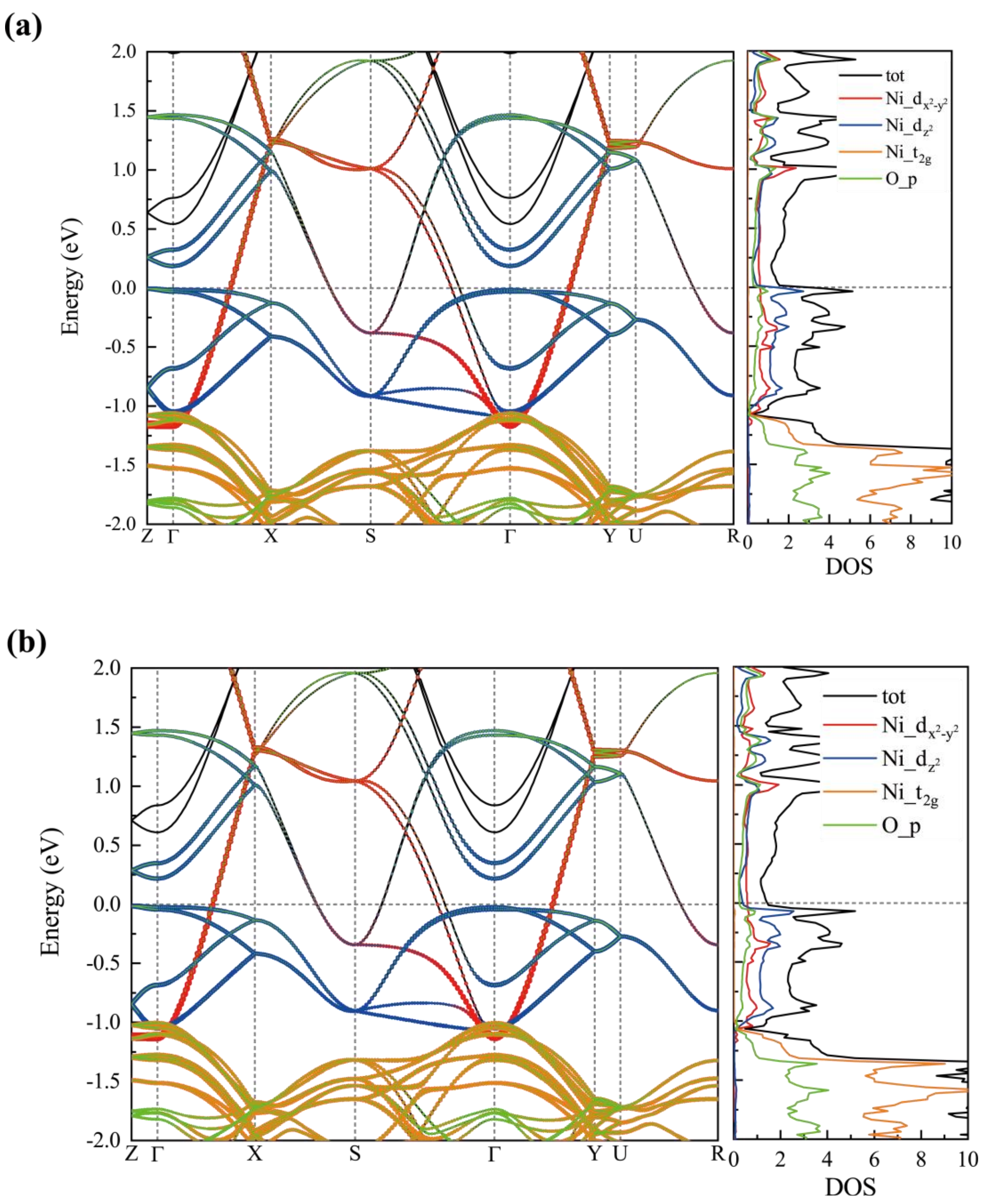

FIG S33. Projected electronic band structures of Ni cations and O anions of *I*4/*mmm* phase at 30 GPa under compressive strain of (a) 0.015 and (b) 0.02 along [100] direction. The corresponding DOS are shown on the right. The horizontal grey dash line means the Fermi level.

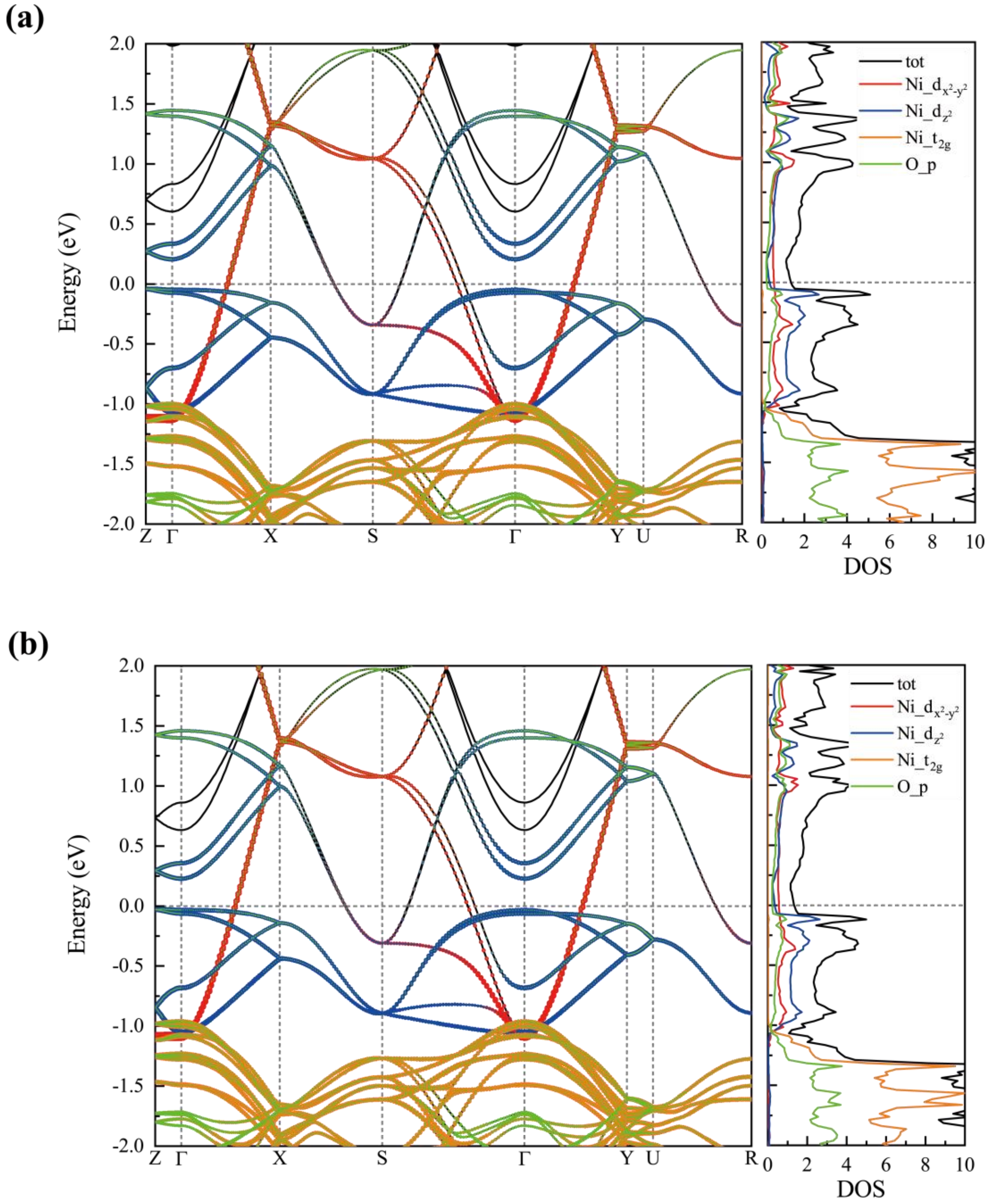

FIG S34. Projected electronic band structures of Ni cations and O anions of *I*4/*mmm* phase at 30 GPa under compressive strain of (a) 0.025 and (b) 0.03 along [100] direction. The corresponding DOS are shown on the right. The horizontal grey dash line means the Fermi level.

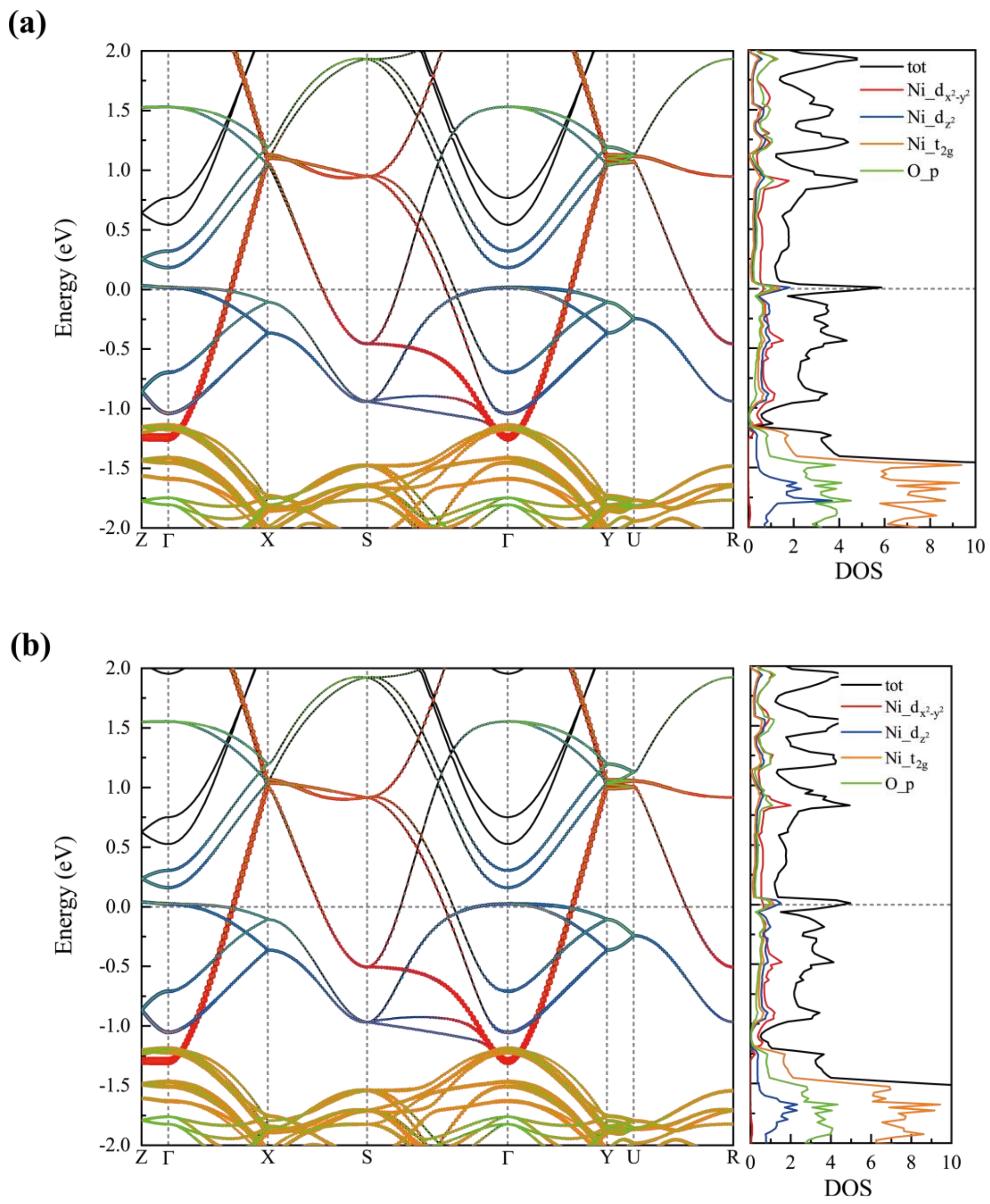

FIG S35. Projected electronic band structures of Ni cations and O anions of *I*4/*mmm* phase at 30 GPa under compressive strain of (a) 0.005 and (b) 0.01 along [001] direction. The corresponding DOS are shown on the right. The horizontal grey dash line means the Fermi level.

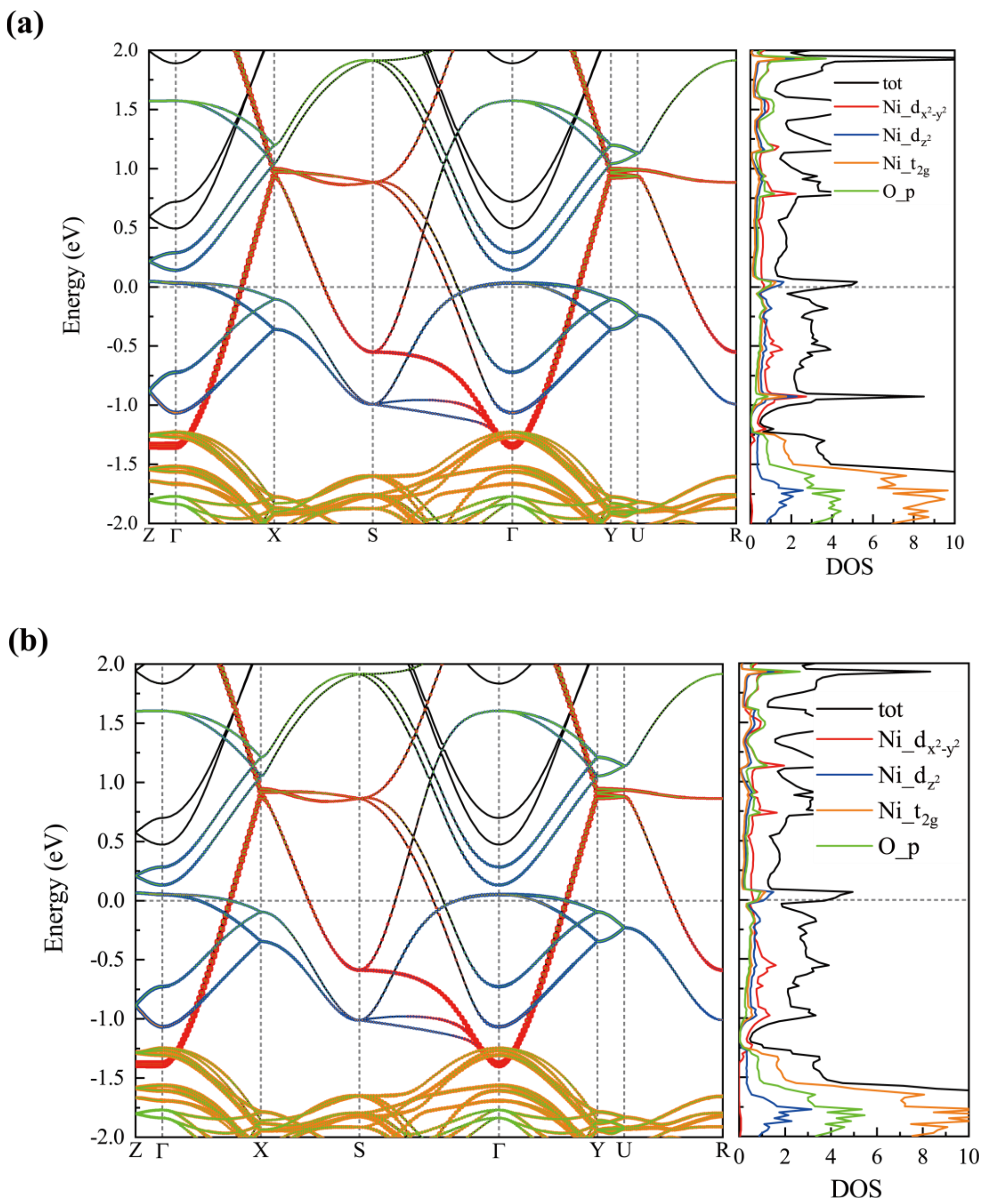

FIG S36. Projected electronic band structures of Ni cations and O anions of *I*4/*mmm* phase at 30 GPa under compressive strain of (a) 0.015 and (b) 0.02 along [100] direction. The corresponding DOS are shown on the right. The horizontal grey dash line means the Fermi level.

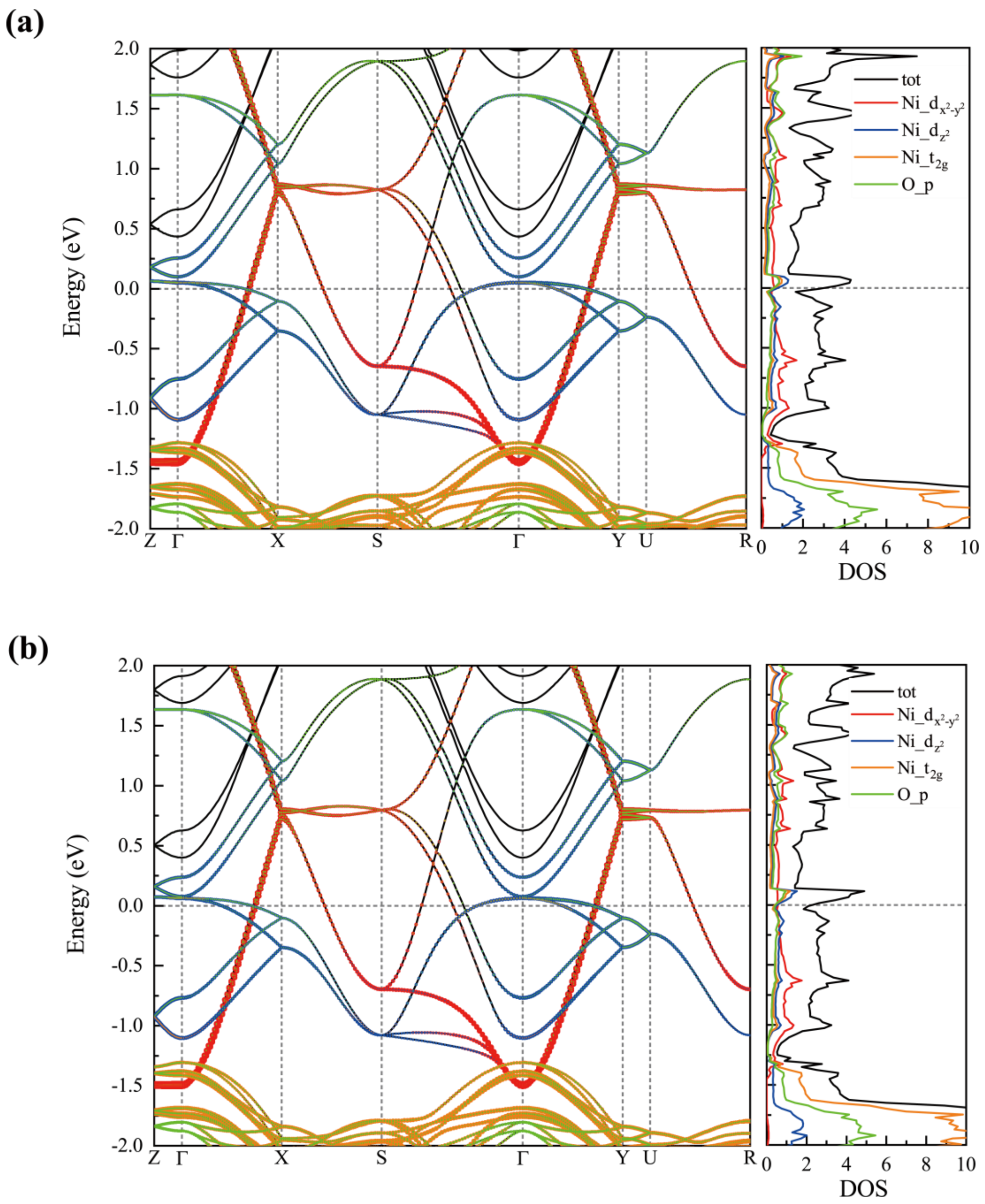

FIG S37. Projected electronic band structures of Ni cations and O anions of *I*4/*mmm* phase at 30 GPa under compressive strain of (a) 0.025 and (b) 0.03 along [001] direction. The corresponding DOS are shown on the right. The horizontal grey dash line means the Fermi level.

## VII. Bilayer two-orbital model

We fit our DFT band structure to the bilayer two-orbital model as proposed in Ref.[12]. The results are presented in Table S2. Here we follow the notation of the tight-binding parameters in Ref.[12], in which $t^{x/z}$ denotes the in-plane nearest-neighbor-hopping of Ni-$d_{x^2-y^2}/d_{z^2}$ orbital, $t_{\perp}^{z}$ denotes the inter-layer hopping of Ni-$d_{z^2}$ orbital, and $\varepsilon^{x/z}$ is the site-energy of Ni-$d_{x^2-y^2}/d_{z^2}$ orbital. Also, $t_3^{xz}$ is the in-plane nearest-neighbor hopping between $d_{x^2-y^2}$ and $d_{z^2}$ orbitals, which corresponds to the hybridization $V$ in our main text.

Table S2. Tight-binding parameters of the bilayer two-orbital model. The units are eV.

| Pressure (GPa) | $t_1^x$ | $t_1^z$ | $t_2^x$ | $t_2^z$ | $t_{\perp}^z$ | $t_{\perp}^x$ | $\varepsilon^x$ | $\varepsilon^z$ | $t_3^{xz}$ | $t_4^{xz}$ |
|---|---|---|---|---|---|---|---|---|---|---|
| 0 | -0.43 | -0.09 | 0.1 | 0.0 | -0.56 | 0.0 | 0.5 | 0.05 | 0.14 | -0.01 |
| 2 | -0.44 | -0.1 | 0.1 | 0.0 | -0.58 | 0.0 | 0.51 | 0.1 | 0.15 | -0.01 |
| 4 | -0.45 | -0.1 | 0.1 | -0.01 | -0.6 | 0.01 | 0.52 | 0.15 | 0.16 | -0.01 |
| 6 | -0.46 | -0.11 | 0.1 | -0.02 | -0.62 | 0.01 | 0.51 | 0.2 | 0.17 | -0.01 |
| 8 | -0.48 | -0.12 | 0.11 | -0.02 | -0.64 | 0.02 | 0.52 | 0.22 | 0.18 | -0.01 |
| 10 | -0.49 | -0.12 | 0.11 | -0.03 | -0.67 | 0.01 | 0.52 | 0.26 | 0.19 | -0.01 |
| 12 | -0.49 | -0.12 | 0.07 | -0.04 | -0.69 | 0.01 | 0.64 | 0.36 | 0.23 | -0.02 |
| 15 | -0.49 | -0.12 | 0.07 | -0.04 | -0.69 | 0.01 | 0.7 | 0.38 | 0.23 | -0.02 |
| 20 | -0.49 | -0.12 | 0.06 | -0.04 | -0.71 | 0.02 | 0.7 | 0.4 | 0.25 | -0.02 |
| 30 | -0.5 | -0.13 | 0.05 | -0.04 | -0.75 | 0.02 | 0.7 | 0.42 | 0.27 | -0.02 |
| 40 | -0.5 | -0.13 | 0.03 | -0.05 | -0.79 | 0.03 | 0.69 | 0.44 | 0.28 | -0.03 |
| 50 | -0.51 | -0.14 | 0.02 | -0.05 | -0.81 | 0.03 | 0.69 | 0.46 | 0.3 | -0.03 |
| 60 | -0.51 | -0.15 | 0.03 | -0.05 | -0.83 | 0.03 | 0.67 | 0.47 | 0.3 | -0.03 |
| 70 | -0.52 | -0.15 | 0.02 | -0.05 | -0.86 | 0.04 | 0.61 | 0.48 | 0.3 | -0.03 |
| 80 | -0.52 | -0.16 | 0.02 | -0.05 | -0.88 | 0.04 | 0.59 | 0.49 | 0.31 | -0.03 |
| 90 | -0.53 | -0.16 | 0.01 | -0.05 | -0.91 | 0.04 | 0.59 | 0.5 | 0.32 | -0.04 |
| 100 | -0.53 | -0.17 | 0.01 | -0.05 | -0.93 | 0.04 | 0.57 | 0.52 | 0.32 | -0.04 |
| [001] | -0.42 | -0.11 | 0.1 | 0.0 | -0.56 | 0.0 | 0.38 | 0.17 | 0.15 | -0.01 |

## VIII. Structural information

Table S3. Detailed structural information of deformed *Amam* phase at ambient pressure.

| Compression strain | Direction | Lattice Parameters (Å) | Atomic Coordinates |
|---|---|---|---|
| 2% | [100] | a=5.265<br>b=5.489<br>c=20.75<br>α=β=γ=90 | Ni 0.250 0.746 0.405<br>Ni 0.750 0.254 0.595<br>Ni 0.250 0.746 0.595<br>Ni 0.750 0.254 0.405<br>Ni 0.250 0.246 0.905<br>Ni 0.750 0.754 0.095<br>Ni 0.250 0.246 0.095<br>Ni 0.750 0.754 0.905<br>O 0.250 0.796 0.294<br>O 0.750 0.204 0.706<br>O 0.250 0.796 0.706<br>O 0.750 0.204 0.294<br>O 0.250 0.296 0.794<br>O 0.750 0.704 0.206<br>O 0.250 0.296 0.206<br>O 0.750 0.704 0.794<br>O 0.250 0.704 0.500<br>O 0.750 0.296 0.500<br>O 0.250 0.204 0.000<br>O 0.750 0.796 0.000<br>O 0.000 0.000 0.413<br>O 0.000 0.000 0.587<br>O 0.500 0.000 0.587<br>O 0.500 0.000 0.413<br>O 0.000 0.500 0.913<br>O 0.000 0.500 0.087<br>O 0.500 0.500 0.087<br>O 0.500 0.500 0.913<br>O 0.500 0.500 0.395<br>O 0.500 0.500 0.605<br>O 0.000 0.500 0.605<br>O 0.000 0.500 0.395<br>O 0.500 0.000 0.895<br>O 0.500 0.000 0.105<br>O 0.000 0.000 0.105<br>O 0.000 0.000 0.895<br>La 0.750 0.762 0.320<br>La 0.250 0.238 0.680<br>La 0.750 0.762 0.680<br>La 0.250 0.238 0.320<br>La 0.750 0.262 0.820<br>La 0.250 0.738 0.180<br>La 0.750 0.262 0.180<br>La 0.250 0.738 0.820<br>La 0.250 0.246 0.500 |
| | | | La 0.750 0.754 0.500<br>La 0.250 0.746 0.000<br>La 0.750 0.254 0.000 |
| 2% | [001] | a=5.405<br>b=5.492<br>c=20.201<br>α=β=γ=90 | Ni 0.250 0.748 0.404<br>Ni 0.750 0.252 0.596<br>Ni 0.250 0.748 0.596<br>Ni 0.750 0.252 0.404<br>Ni 0.250 0.248 0.904<br>Ni 0.750 0.752 0.096<br>Ni 0.250 0.248 0.096<br>Ni 0.750 0.752 0.904<br>O 0.250 0.793 0.295<br>O 0.750 0.207 0.705<br>O 0.250 0.793 0.705<br>O 0.750 0.207 0.295<br>O 0.250 0.293 0.795<br>O 0.750 0.707 0.205<br>O 0.250 0.293 0.205<br>O 0.750 0.707 0.795<br>O 0.250 0.704 0.500<br>O 0.750 0.296 0.500<br>O 0.250 0.204 0.000<br>O 0.750 0.796 0.000<br>O 1.000 0.000 0.412<br>O 0.000 0.000 0.588<br>O 0.500 0.000 0.588<br>O 0.500 0.000 0.412<br>O 1.000 0.500 0.912<br>O 0.000 0.500 0.088<br>O 0.500 0.500 0.088<br>O 0.500 0.500 0.912<br>O 0.500 0.500 0.395<br>O 0.500 0.500 0.605<br>O 1.000 0.500 0.605<br>O 0.000 0.500 0.395<br>O 0.500 0.000 0.895<br>O 0.500 0.000 0.105<br>O 1.000 0.000 0.105<br>O 0.000 0.000 0.895<br>La 0.750 0.761 0.321<br>La 0.250 0.239 0.679<br>La 0.750 0.761 0.679<br>La 0.250 0.237 0.321<br>La 0.750 0.261 0.821<br>La 0.250 0.739 0.179<br>La 0.750 0.261 0.179<br>La 0.250 0.739 0.821<br>La 0.250 0.251 0.500<br>La 0.750 0.749 0.500<br>La 0.250 0.751 0.000 |

| | |
|---|---|
| | La 0.750 0.249 0.000 |

## IX. References

[1] G. Kresse and J. Furthmüller, Efficiency of ab-initio total energy calculations for metals and semiconductors unsing a plane-wave basis set, Comput. Mater. Sci. **6**, 15 (1996).

[2] J. Perdew, K. Burke and M. Ernzerhof, Generalized Gradient Approximation Made Simple, Phys. Rev. Lett. **77**, 3865 (1996).

[3] P. Blöchl, Projector augmented-wave method, Phys. Rev. B **50**, 17953 (1994).

[4] G. Kresse and D. Joubert, From ultrasoft pseudopotentials to the projector augmented-wave method, Phys. Rev. B **59**, 1758 (1999).

[5] A. Liechtenstein, V. Anisimov and J. Zaanen, Density-functional theory and strong interactions: Orbital ordering in Mott-Hubbard insulators, Phys. Rev. B **52**, R5467 (1995).

[6] S. Dudarev, G. Botton, S. Savrasov, C. Humphreys and A. Sutton,Electron-energy-loss spectra and the structural stability of nickel oxide: An LSDA+U study, Phys. Rev. B **57**, 1505 (1998).

[7] Z. Liu, H. Sun, M. Huo, X. Ma, Y. Ji, E. Yi, L. Li, H. Liu, J. Yu, Z. Zhang, Z. Chen, F. Liang, H. Dong, H. Guo, D. Zhong, B. Shen, S. Li and M. Wang, Evidence for charge and spin density waves in single crystal of $La_3Ni_2O_7$ and $La_3Ni_2O_6$, Sci. China Phys. Mech. **66**, 217411 (2022).

[8] J. Yang, H. Sun, X. Hu, Y. Xie, T. Miao, H. Luo, H. Chen, B. Liang, W. Zhu, G. Qu, C. Chen, M. Huo, Y. Huang, S. Zhang, F. Zhang, F. Yang, Z. Wang, Q. Peng, H. Mao, G. Liu, Z. Xu, T. Qian, D. Yao, M. Wang, L. Zhao and X. Zhou, Orbital-dependent electron correlation in double-layer nickelate $La_3Ni_2O_7$, Nat. Commun. **15**, 4373 (2024).

[9] Y. Zhang, H. Sun and C. Chen, Superhard Cubic $BC_2N$ Compared to Diamond, Phys. Rev. Lett. **93**, 195504 (2004).

[10] Z. Pan, H. Sun and C. Chen, Colossal Shear-Strength Enhancement of Low-Density Cubic $BC_2N$ by Nanoindentation, Phys. Rev. Lett. **98**, 135505 (2007).

[11] Z. Pan, H. Sun and C. Chen, Indenter-angle-sensitive fracture modes and stress response at incipient plasticity, Phys. Rev. B **79**, 104102 (2009).

[12] Z. Luo, X. Hu, M. Wang, W. Wú and D. Yao, Bilayer Two-Orbital Model of $La_3Ni_2O_7$ under Pressure, Phys. Rev. Lett. **131**, 126001 (2023).